\documentclass[seceq,preprint]{ptptex}
\usepackage{amsmath,amssymb,mathtools}
\usepackage{wrapft}

\newcommand{\mspd}[1]{[#1]_G}
\newcommand{\epsi}{\eta\Psi}
\newcommand{\qpsi}{Q_G\Psi}
\newcommand{\qxi}{Q_G\Xi}
\newcommand{\eLambda}{\eta\Lambda_1}
\newcommand{\qLambda}{Q_G\Lambda_{\frac{1}{2}}}
\newcommand{\eLambdaf}{\eta\Lambda_{\frac{3}{2}}}
\newcommand{\eLambdax}{\eta\tilde{\Lambda}_{\frac{1}{2}}}

\newcommand{\llangle}{\langle\!\langle}
\newcommand{\rrangle}{\rangle\!\rangle}
\notypesetlogo                       
\preprintnumber[3cm]{
YITP-15-56}

\title{
Symmetries and Feynman Rules for Ramond Sector\\ in Heterotic String Field Theory
}

\author{
Hiroshi \textsc{Kunitomo}\footnote{%
E-mail:\  {\tt kunitomo@yukawa.kyoto-u.ac.jp}}
}

\inst{
Yukawa Institute for Theoretical Physics, Kyoto University, \\
Kyoto 606-8502, Japan
}

\abst{
Examining the symmetries of the pseudo-action, we propose a prescription
for the new Feynman rules for the Ramond sector of the WZW-like heterotic
string field theory.
The new rules are an analog of that recently proposed for the open superstring 
field theory and respect all the gauge symmetries including those provided
we impose the constraint after transformation.
It is shown that the new rules 
reproduce the well-known on-shell tree-level amplitudes for four and five
external strings including fermions.
}

\begin{document}

\maketitle

\section{Introduction}

In the previous paper\cite{Kunitomo:2014qla}, we examined the gauge symmetries
of the pseudo-action, the action supplemented by the constraint, in
Wess-Zumino-Witten (WZW)-like open superstring field 
theory.\cite{Berkovits:1995ab,Berkovits:2001im} It was found that the pseudo-action
has a new kind of symmetry provided we impose the constraint
after the transformation. We proposed a prescription for the new Feynman rules
for the Ramond (R) sector so as to respect all these symmetries.
It was shown that the new rules reproduce the well-known 
on-shell tree-level amplitudes in the case of four and five 
external states, including those that cannot be reproduced by the self-dual 
Feynman rules which had already been 
proposed.\cite{Michishita:2004by,Michishita:Riken, Michishita:2012ku}
The aim of this paper is to extend these arguments to the heterotic
string field theory and to propose a similar prescription providing
the new Feynman rules.

Similar to the open superstring field theory, the heterotic string 
field theory can also be constructed utilizing 
the large Hilbert space,\cite{Okawa:2004ii,Berkovits:2004xh} 
which is WZW-like in the sense that the Neveu-Schwarz (NS) action 
is constructed as a WZW-type action.\cite{Berkovits:2004xh} 
In spite of this success in the NS sector, 
it is difficult to construct a covariant action including the R sector, 
which is a disadvantage of the formulation. 
Without introducing any extra degrees of freedom,
only the equations of motion have been constructed in a covariant 
manner.\cite{Kunitomo:2013mqa,Kunitomo:2014hba}
Alternatively, however, we can define the pseudo-action by
introducing an auxiliary R string field. 
The pseudo-action of the heterotic string field theory
is non-polynomial in both the NS and R string fields, 
which is required so as to reproduce the correct 
amplitudes,\cite{Saadi:1989tb,Kugo:1989aa,Kugo:1989tk,Zwiebach:1992ie}
and was constructed at some lower order in the fermion expansion,
the expansion with respect to the number of the R string 
fields.\cite{Kunitomo:2013mqa}
The self-dual Feynman rules were also proposed in a parallel way to 
the open superstring case and shown to reproduce the on-shell
four-point amplitudes.\cite{Kunitomo:2013mqa} 
It was pointed out, however, that these rules contain some ambiguity,
which appears when we calculate the amplitudes with five or more
external states including the fermions.

We will examine, in this paper, the gauge symmetries of the pseudo-action
in detail. It will be found, at some lower order in $\Psi$, 
that the missing gauge symmetries, which have been considered the symmetries 
of only the equations of motion, are realized as a new kind of symmetry 
under which the pseudo-action is transformed into the form proportional 
to the constraint. We will then improve the self-dual Feynman rules 
to those which respect all these gauge symmetries and have no ambiguity.
We will show that the new Feynman rules reproduce the correct on-shell 
amplitudes at the tree level, at least for the case of four and five 
external states including fermions.

This paper is organized as follows.
In \S \ref{sec2}, we will first summarize the known basic properties
of the WZW-like heterotic string field theory. After fixing the linearized
gauge symmetries, we will introduce the self-dual Feynman rules proposed
previously. Then the symmetries of the pseudo-action will be studied at
lower-order levels in the fermion expansion.
It will be found that the pseudo-action is invariant under the missing gauge
symmetries if we suppose it to be subject to the constraint after 
the transformation. The new Feynman rules will be proposed without ambiguity 
so as to respect all the gauge symmetries. 
The on-shell tree-level amplitudes 
for the case of the four and the five external states including fermions
will be calculated in \S \ref{sec3} and shown to agree with those obtained
in the first quantized formulation. The final section \S \ref{sec4}
is devoted to the conclusion and discussion.
Some lengthy results of the missing gauge symmetries at a higher order
will be given in Appendix. 
The higher-order corrections to the constraint, which do not exist 
in the case of the open superstring, first become important at this order.

\section{WZW-like heterotic string field theory and the self-dual Feynman rules}\label{sec2}

In this section, after introducing the WZW-like heterotic string field theory 
including the R sector, we will recall the self-dual Feynman rules.
Examining the gauge symmetries of the pseudo-action, we will propose a prescription
for the new Feynman rules, which respects all the gauge symmetries.

\subsection{WZW-like heterotic string field theory}

We denote the Neveu-Schwarz (NS) string as $V$, which is Grassmann odd
and has the ghost and picture numbers $(G,P)=(1,0)$. The action for the NS sector
of the heterotic string field theory is given by a WZW-type action,
\begin{equation}
 S_{NS} =\ \int^1_0 dt\langle \eta V, G(tV)\rangle,\label{NS action}
\end{equation}
where the pure-gauge string field $G(tV)$ is  defined as
\begin{equation}
G(tV) =\ tQV+\frac{\kappa}{2}t^2[V,QV]+\frac{\kappa^2}{3!}t^3\left(
[V,(QV)^2]+[V,[V,QV]]\right)+\cdots,
\end{equation}
by integrating the gauge transformation of the \textit{bosonic} closed string 
field theory.\cite{Berkovits:2004xh}
The BRST charge $Q$ and the string products satisfy the algebraic 
relation\cite{Zwiebach:1992ie}
\begin{align}
0 =&\ Q[B_1,B_2,\cdots,B_n] 
+ \sum_{i=1}^n (-1)^{B_1+\cdots+B_{i-1}}[B_1,\cdots,QB_i,\cdots,B_n]
\nonumber\\
& + \sum_{\{i_l,j_k\} \atop l+k=n} \sigma(i_l,j_k) 
[B_{i_1},\cdots,B_{i_l},[B_{j_1},\cdots,B_{j_k}]],
\label{fundamental}
\end{align}
where $\sigma(i_l,j_k)$ is a sign factor defined to be the sign
picked up when one rearranges the sequence $\{Q,B_1,\cdots,B_n\}$
into the order $\{B_{i_1},\cdots,B_{i_l},Q,B_{j_1},\cdots,B_{j_k}\}$.
The arbitrary variation of the integrand of the action becomes the total derivative, 
and is integrated as 
\begin{equation}
 \delta S_{NS} =\ -\langle B_\delta(V), \eta G(V)\rangle,
\label{variation}
\end{equation}
where $B_\delta(V)$ is a function of $V$ and $\delta V$ defined by a solution of 
some specific ordinary differential equation,\cite{Berkovits:2001im} whose 
first few terms are given by\footnote{
This relation is invertible and solved by $\delta V$ as
\begin{equation}
\delta V(B_\delta) =\ B_\delta - \frac{\kappa}{2}[V,B_\delta]
-\frac{\kappa^2}{12}\left(4[V,QV,B_\delta]-[V,[V,B_\delta]]\right)+\cdots.
\nonumber
\end{equation}
}
\begin{equation}
B_\delta(V) =\ \delta V +\frac{\kappa}{2}[V,\delta V]
+\frac{\kappa^2}{6}\left(2[V,QV,\delta V]+[V,[V,\delta V]]\right)+\cdots.
\label{NS variation}
\end{equation}
The pseudo-action for the R sector is constructed by introducing two R strings, 
$\Psi$ and $\Xi$, which are both Grassmann odd and have the ghost and picture 
numbers $(G,P)=(1,1/2)$ and $(1,-1/2)$, respectively.
The fermion bilinear term of the pseudo-action is then given 
by a straightforward extension of that of the open superstring field theory as
\begin{equation}
 S_{R[2]}=-\frac{1}{2}\langle \epsi, \qxi\rangle,\label{R bilinear}
\end{equation}
where the shifted BRST charge $Q_G$ is defined by the operator acting on a general
string field $B$ as
\begin{equation}
Q_GB =\ QB + \sum_{m=1}^\infty \frac{\kappa^m}{m!}[G(V)^m,B].
\end{equation}
From simple consideration, however, one can easily see that the pseudo-action
has to be non-polynomial not only in the NS string field but also 
in the R string fields to reproduce the on-shell 
fermion amplitudes.\cite{Kunitomo:2013mqa}  
The explicit form of such a pseudo-action can in principle be obtained order 
by order in the fermions, the number of the R string fields,
\begin{equation}
S_R=\sum_{n=1}^\infty S_{R[2n]},\label{R action} 
\end{equation}
starting from (\ref{R bilinear}), 
where each $S_{R[2n]}$ contains $n$ $\Psi$ and $n$ $\Xi$.
In particular, the next-leading (four-fermion) action,
which is necessary for calculating the four- and five-point amplitudes 
in the next section, is given by
\begin{equation}
 S_{R[4]}=\frac{\kappa^2}{4!}\langle\epsi,\mspd{\Psi,(\qxi)^2} \rangle.
\end{equation}
Here the shifted string product $[\cdot]_G$ is defined by
\begin{equation}
[B_1,\cdots,B_n]_G =\ \sum_{m=0}^\infty\frac{\kappa^m}{m!} [G(V)^m,B_1,\cdots,B_n],
\end{equation}
for general $n$ string fields $\{B_1,\cdots,B_n\}$.
The equations of motion derived from the variation of $S=S_{NS}+S_R$ 
agree with those obtained without introducing the auxiliary 
field,\cite{Kunitomo:2013mqa, Kunitomo:2014hba} 
if we impose the constraint
\begin{align}
 Q_G\Xi =&\ \Omega,\label{constraint}
\end{align}
where\footnote{
This $\Omega$ is denoted as $B_{-1/2}$ in Ref.~\citen{Kunitomo:2014hba},
which can be determined order by order in $\Psi$.
}
\begin{equation}
 \Omega =\ \eta\Psi+\frac{\kappa^2}{3!}\mspd{\Psi,(\eta\Psi)^2}+\cdots.
\label{corr}
\end{equation}
In this sense, the pseudo-action (\ref{R action}) describes 
the R sector of the heterotic string field theory.

\subsection{Gauge fixing and the self-dual Feynman rules}

Let us next explain how tree-level amplitudes are calculated in this formulation.
For the NS sector, the Feynman rules can be derived from the action (\ref{NS action})
in a conventional way.
Expanding the action in the power of the coupling constant $\kappa$,
\begin{equation}
 S_{NS} =\ \sum_{n=0}^\infty S_{NS}^{(n)},
\end{equation}
the kinetic term of the NS string is given by
\begin{equation}
 S^{(0)}_{NS} =\ \frac{1}{2}\langle \eta V, QV\rangle.
\label{NS quadra}
\end{equation}
Since this is invariant under the gauge transformations
\begin{equation}
\delta V =\ Q\Lambda_0+\eta\Lambda_1,\label{linearized tf NS} 
\end{equation}
we have to fix these symmetries to obtain the propagator.
If we impose the simplest gauge conditions,
\begin{equation}
 b_0^+V =\ \xi_0V =\ 0,\label{gauge condition} 
\end{equation}
the NS propagator is given by
\begin{align}
\overbracket[0.5pt]{\!\!VV\!\!}\ \equiv\ \Pi_{NS}=&\
 \xi_0\frac{b_0^-b_0^+}{L_0^+}\delta(L_0^-)
\nonumber\\
=&\ \int_0^\infty \!\!dT\!\!\int_0^{2\pi}\frac{d\theta}{2\pi}
(\xi_0b_0^-b_0^+)e^{-TL^+_0-i\theta L^-_0}.
\label{NS propagator}
\end{align}
The three and four NS string vertices, which are necessary for
the calculation in the next section, are given by
\begin{align}
 S_{NS}^{(1)} =&\ \frac{\kappa}{3!}\langle \eta V, [V, QV]\rangle,\label{NS three}\\
 S_{NS}^{(2)} =&\ \frac{\kappa^2}{4!}\langle\eta V, [V, (QV)^2]\rangle
+\frac{\kappa^2}{4!}\langle\eta V, [V, [V, QV]]\rangle.\label{NS four}
\end{align}
Note that the first term in the four-point vertices (\ref{NS four})
contains the integration over two parameters (moduli) realized
by the restricted tetrahedron,\cite{Saadi:1989tb,Kugo:1989aa} 
and corresponding anti-ghost insertion.\cite{Kugo:1989tk,Zwiebach:1992ie}
The second term in (\ref{NS four}), on the other hand, is integrated
over one parameter, the twist angle of the collapsed propagator.

For the R sector, however, the Feynman rules cannot be uniquely derived 
from the pseudo-action (\ref{R action}) since it is not the true action.
We can only propose some plausible Feynman rules and confirm whether 
they reproduce the correct physical on-shell amplitudes. 
In the previous paper, we proposed the Feynman rules, 
which we refer to as the self-dual Feynman rules and 
confirmed that they actually reproduce the well-known four-point amplitudes
with external fermions.\cite{Kunitomo:2013mqa} 
We first summarize the self-dual Feynman rules.
Similar to the NS case, we can expand the pseudo-action in the power of the coupling
constant $\kappa$ as
\begin{equation}
 S_R =\ \sum_{n=0}^\infty S_{R[2]}^{(n)} + \sum_{n=0}^\infty S_{R[4]}^{(n)} + \cdots.
\end{equation}
The kinetic term of the R string,
\begin{equation}
S_{R[2]}^{(0)} =\ -\frac{1}{2}\langle \eta\Psi, Q\Xi\rangle,\label{R quadra}
\end{equation}
is invariant under the gauge transformations
\begin{equation}
\delta\Psi =\ Q\Lambda_{\frac{1}{2}}+\eta\Lambda_{\frac{3}{2}},\qquad
\delta\Xi =\ Q\Lambda_{-\frac{1}{2}}+\eta\tilde{\Lambda}_{\frac{1}{2}}.
\label{linearized tf R}
\end{equation}
Fixing them by the same gauge conditions as for the NS string, 
(\ref{gauge condition}),
\begin{equation}
 b_0^+\Psi =\ \xi_0\Psi =\ 0,\qquad
 b_0^+\Xi =\ \xi_0\Xi =\ 0,\label{simplest gauge}
\end{equation}
the propagator of the R sector in this gauge is given by
\begin{align}
 \overbracket[0.5pt]{\!\!\Psi\Xi\!\!}\ =\ \overbracket[0.5pt]{\!\!\Xi\Psi\!\!}\
\equiv&\ \Pi_R \nonumber\\
=&\ -2\xi_0\frac{b_0^-b_0^+}{L_0^+}\delta(L_0^-)= -2\Pi_{NS}.\label{naive R propagator}
\end{align}
For the R sector, in addition, 
the constraint (\ref{constraint}) has to be taken into account.
For the on-shell external states, 
this is naturally implemented by simply restricting them to those satisfying
the linearized constraint, $Q\Xi=\eta\Psi$. 
In contrast, however, the prescription for the off-shell (propagating) states is 
not unique. The self-dual Feynman rules are defined
by adopting a prescription in which only the \textit{self-dual} part 
$\omega=(Q\Xi+\eta\Psi)/2$ of the R strings propagates through the effective propagator
\begin{align}
\overbracket[0.5pt]{\!\!\omega\omega\!\!}\ =&\ \frac{1}{4}(Q\Pi_R\eta+\eta\Pi_RQ)\nonumber\\
=& -\frac{1}{2}(Q\Pi_{NS}\eta+\eta\Pi_{NS}Q).
\end{align}
Although the fermion interaction vertices can be obtained by replacing 
the R string fields with their self-dual part, we need some preparation 
since, unlike the case of the open superstring field theory, 
the R string fields do not appear only in the form of $Q\Xi$ or $\eta\Psi$. 
For example, the terms with three, four and five string fields needed in the next
section are given as
\begin{subequations} \label{bare vertices}
 \begin{align}
 S_{R[2]}^{(1)} =&\ -\frac{\kappa}{2}\langle\eta\Psi, [QV, \Xi]\rangle,
\label{two-one}\\
 S_{R[2]}^{(2)} =&\ -\frac{\kappa^2}{4}\langle\eta\Psi, [(QV)^2, \Xi]\rangle
-\frac{\kappa^2}{4}\langle\eta\Psi, [[V,QV], \Xi]\rangle,\label{two-two}\\
 S_{R[4]}^{(2)} =&\ \frac{\kappa^2}{4!}\langle\eta\Psi, [\Psi, (Q\Xi)^2]\rangle,
\label{four-zero}\\
 S_{R[2]}^{(3)} =&\ -\frac{\kappa^3}{12}\langle\eta\Psi, [(QV)^3, \Xi]\rangle
-\frac{\kappa^3}{12}\langle\eta\Psi, [[V,(QV)^2],\Xi]\rangle,\nonumber\\
 &-\frac{\kappa^3}{4}\langle\eta\Psi, [[V,QV],QV,\Xi]\rangle
-\frac{\kappa^3}{12}\langle\eta\Psi,  [[V, [V, QV]], \Xi]\rangle,
\label{two-three}\\
 S_{R[4]}^{(3)} =&\ \frac{\kappa^3}{4!}\langle\eta\Psi, [QV, \Psi, (Q\Xi)^2]\rangle
+\frac{\kappa^3}{12}\langle\eta\Psi, [\Psi, Q\Xi, [QV, \Xi]]\rangle,
\label{four-one}
\end{align}
\end{subequations}
by expanding the pseudo-action $S_R$, 
where both the $\Psi$ and $\Xi$ appear in the form not accompanied by $\eta$
and $Q$, respectively. 
Nevertheless, if we assume that the field redefinition
\begin{align}
 \tilde{\Xi} =&\ \Xi - \kappa[V, \Xi]
-\frac{\kappa^2}{2}[V, QV, \Xi]+\frac{\kappa^2}{2}[V, [V, \Xi]]
-\frac{\kappa^3}{3!}[V, (QV)^2, \Xi]+\frac{\kappa^3}{3}[V, [V, QV, \Xi]]
\nonumber\\
&
+\frac{\kappa^3}{3!}[V, QV, [V, \Xi]]-\frac{\kappa^3}{3!}[[V, QV], V, \Xi]
-\frac{\kappa^3}{3!}[V, [V, [V, \Xi]]]+\cdots
\label{xi redefinition}
\end{align}
does not affect the on-shell physical amplitudes, as with the point
transformation in the conventional quantum field theory,
we can rewrite (\ref{bare vertices}) so that
the $\tilde{\Xi}$ always appears in the form of $Q\tilde{\Xi}$ 
thanks to the relation 
\begin{align}
 Q_G\Xi =&\
Q\tilde{\Xi}+\kappa[V, Q\tilde{\Xi}]+\frac{\kappa^2}{2}[V, QV, Q\tilde{\Xi}]
+\frac{\kappa^2}{2}[V, [V, Q\tilde{\Xi}]]
\nonumber\\
&
+\frac{\kappa^3}{3!}[V, (QV)^2, Q\tilde{\Xi}]+\frac{\kappa^3}{3!}[V, [V, QV, Q\tilde{\Xi}]]
+\frac{\kappa^3}{3}[V, QV, [Q, Q\tilde{\Xi}]]
\nonumber\\
&
+\frac{\kappa^3}{3!}[[V, QV], V, Q\tilde{\Xi}]+\frac{\kappa^3}{3!}[V, [V, [V, Q\tilde{\Xi}]]]
+\cdots.
\end{align}
Then we can replace $Q\tilde{\Xi}$ with $\omega$ in the alternative expression.
Contrary to this, the prescription for $\Psi$ is not unique but depends on the gauge condition
in general. In the simplest gauge (\ref{simplest gauge}), we can replace $\Psi$ with 
$\xi_0\omega$ since $\Psi=\{\eta,\xi_0\}\Psi=\xi_0\eta\Psi$. 
However, we have two choices in replacing $\eta\Psi$;
either we simply replace it with $\omega$, or $\eta(\xi_0\omega)$ in accordance with the above 
prescription for $\Psi$. Since $\omega\ne\eta\xi_0\omega$ for the off-shell states,
this is an ambiguity in the self-dual Feynman rules, which does not appear in the
four-point amplitudes. If we take the former choice, 
the interaction vertices for the self-dual rules become
\begin{align}
 \tilde{S}_{R[2]}^{(1)} =&\ -\frac{\kappa}{2}\langle\omega,[V, \omega]\rangle,\\
 \tilde{S}_{R[2]}^{(2)} =&\ -\frac{\kappa^2}{4!}\langle\omega, [V, QV, \omega]\rangle,\\
 \tilde{S}_{R[4]}^{(2)} =&\ \frac{\kappa^2}{4!}\langle\xi_0\omega, [\omega^3]\rangle,\\
 \tilde{S}_{R[2]}^{(3)} =&\ -\frac{\kappa^3}{12}\langle\omega, [V, (QV)^2, \omega]\rangle
 -\frac{\kappa^3}{12}\langle\omega, [V, [V, QV, \omega]]\rangle
-\frac{\kappa^3}{6}\langle\omega, [V, QV, [V, \omega]]\rangle
\nonumber\\
&-\frac{\kappa^3}{12}\langle\omega, [V, \omega, [V, QV]]\rangle
-\frac{\kappa^3}{12}\langle\omega, [V, [V, [V, \omega]]]\rangle,\\
 \tilde{S}_{R[4]}^{(3)} =&\ \frac{\kappa^3}{4!}\langle\xi_0\omega, [QV, \omega^3]\rangle
+\frac{\kappa^3}{12}\langle\xi_0\omega, [\omega^2, [V, \omega]]\rangle,
\end{align}
after the replacements.
It was shown that these self-dual Feynman rules reproduce the well-known on-shell
tree-level amplitudes for the case of four external states including 
the fermions.\cite{Kunitomo:2013mqa}

\subsection{Gauge symmetries and the new Feynman rules}\label{2-3}

In order to revise the Feynman rules, let us examine the gauge symmetries in detail.
As was pointed out in Ref.~\citen{Kunitomo:2013mqa},
the total action, $S=S_{NS}+S_R$, is invariant under the gauge 
transformations
\begin{equation}
 B_\delta =\ Q_G\Lambda_0,\qquad
\delta\Psi =\ 0,\qquad
\delta\Xi =\ Q_G\Lambda_{-\frac{1}{2}}
\label{symmetries}
\end{equation}
by construction. The self-dual Feynman rules respect these symmetries 
since both of the $Q_G\Xi$ and $\Omega$ are invariant under (\ref{symmetries}).
However, they do not include all the gauge symmetries at the linearized
level, (\ref{linearized tf NS}) and (\ref{linearized tf R}), which 
have to be fixed to invert the kinetic terms, 
(\ref{NS quadra}) and (\ref{R quadra}). 
We can show, at some lower order in $\Psi$, 
that the missing symmetries are realized as those provided we
impose the constraint after transformation.

Let us first consider the transformation generated by $\Lambda_1$ 
in (\ref{linearized tf NS}), which is extended to the nonlinear form
\begin{equation}
 B_{\delta_{\Lambda_1}}^{[0]} =\ \eLambda 
\end{equation}
at the leading (zeroth) order of $\Psi$ so as to keep the NS action 
(\ref{NS action}) invariant:
\begin{equation}
 \delta^{[0]}S_{NS} =\ 0.
\end{equation}
We can define the next-order transformation,
\begin{align}
 \delta^{[0]}_{\Lambda_1}\Psi =\ &-\kappa\mspd{\Psi,\eLambda},\qquad
 \delta^{[0]}_{\Lambda_1}\Xi =\ -\kappa\mspd{\Xi,\eLambda}, \\
 &\quad B_{\delta_{\Lambda_1}}^{[2]}=\
\frac{\kappa^2}{2}\mspd{\Psi,\qxi,\eLambda},
\end{align}
so that the total action is invariant up to the higher-order 
corrections:
\begin{equation}
 \delta^{[2]}S_{NS}+\delta^{[0]}S_{R[2]} =\ 0.
\end{equation}
At the next-next-order, however, we cannot keep the action invariant.
Instead, we can find the transformations,
\begin{align}
\delta^{[2]}_{\Lambda_1}\Psi=&\ \frac{\kappa^3}{6}\mspd{\Psi,\qpsi,\qxi,\eLambda}
-\frac{\kappa^3}{4}\mspd{\Psi,\mspd{\Psi,\qxi,\eLambda}}
\nonumber\\
&+\frac{\kappa^3}{4}\mspd{\mspd{\Psi,\qxi},\Psi,\eLambda},\\
\delta^{[2]}_{\Lambda_1}\Xi=&\ \frac{\kappa^3}{6}\mspd{\Psi,(\qxi)^2,\eLambda}
-\frac{\kappa^3}{2}\mspd{\Xi,\mspd{\Psi,\qxi,\eLambda}},\\
B_{\delta_{\Lambda_1}}^{[4]}=&\
-\frac{\kappa^4}{4!}\mspd{\Psi,\qpsi,(\qxi)^2,\eLambda}
+\frac{\kappa^4}{4!}\mspd{\Psi,\mspd{\Psi,(\qxi)^2,\eLambda}}
\nonumber\\
&
+\frac{\kappa^4}{8}\mspd{\Psi,\qpsi,\mspd{\Psi,\qxi,\eLambda}}
-\frac{\kappa^4}{8}\mspd{\mspd{\Psi,\qxi},\Psi,\qxi,\eLambda}
\nonumber\\
&
-\frac{\kappa^4}{4!}\mspd{\mspd{\Psi,(\qxi)^2},\Psi,\eLambda},
\end{align}
by which the pseudo-action is transformed
to the form proportional to the constraint (\ref{constraint}):
\begin{align}
\delta^{[4]}_{\Lambda_1}S_{NS}+&\delta^{[2]}_{\Lambda_1}S_{[2]}
+\delta^{[0]}_{\Lambda_1}S_{R[4]} 
\nonumber\\
=&\
\frac{\kappa^3}{4!}\langle\eLambda,\mspd{\Psi,\qxi,\mspd{\epsi,\qxi}}\rangle
-\frac{\kappa^3}{4!}\langle\eLambda,\mspd{\Psi,\epsi,\mspd{(\qxi)^2}}\rangle
\nonumber\\
&
+\frac{\kappa^3}{4!}\langle\epsi,\mspd{(\qxi)^2,\mspd{\Psi,\epsi}}\rangle
-\frac{\kappa^3}{4!}\langle\epsi,\mspd{\epsi,\qxi,\mspd{\Psi,\qxi}}\rangle.
\label{tf lambda1}
\end{align}
The right-hand side vanishes, up to the higher-order corrections,
if we impose the constraint (\ref{constraint}). 
We can also construct the nonlinear transformation generated by $\Lambda_{1/2}$.
The leading-order transformation, 
\begin{align}
 B^{[0]}_{\delta_{\Lambda_{1/2}}} =\ 0,\qquad
 \delta^{[0]}_{\Lambda_{1/2}}\Psi =&\ \qLambda,\qquad
 \delta^{[0]}_{\Lambda_{1/2}}\Xi =\ 0,
\end{align}
is first combined with
\begin{equation}
  B^{[2]}_{\delta_{\Lambda_{1/2}}} =\ -\frac{\kappa}{2}\mspd{\Xi,\qLambda},
\end{equation}
which keeps the pseudo-action invariant at $\mathcal{O}(\Psi^2)$:
\begin{equation}
 \delta^{[2]}_{\Lambda_{1/2}}S_{NS}+\delta^{[0]}_{\Lambda_{1/2}}S_{R[2]} =\ 0.
\end{equation}
This can be extended to the next order as
\begin{align}
 \delta^{[2]}_{\Lambda_{1/2}}\Psi =&\ -\frac{\kappa^2}{3!}\mspd{\Psi,\qxi,\qLambda},\\
 \delta^{[2]}_{\Lambda_{1/2}}\Xi =&\ -\frac{\kappa^2}{3!}\mspd{\Xi,\qxi,\qLambda}
+\frac{\kappa^2}{3!}\mspd{\Xi,\mspd{\Xi,\qLambda}},\\
 B^{[4]}_{\delta_{\Lambda_{1/2}}} =&\ \frac{\kappa^3}{4!}\mspd{\Psi,(\qxi)^2,\qLambda},
\end{align}
which transforms the pseudo-action in the form proportional to the constraint as
\begin{align}
\delta^{[4]}_{\Lambda_{1/2}}S_{NS}+&\delta^{[2]}_{\Lambda_{1/2}}S_{R[2]}
+\delta^{[0]}_{\Lambda_{1/2}}S_{R[4]}
\nonumber\\
=&\ \frac{\kappa^2}{12}\langle\qLambda, \mspd{\qxi,\mspd{\Xi,\epsi}}\rangle
-\frac{\kappa^2}{12}\langle\qLambda, \mspd{\epsi,\mspd{\Xi,\qxi}}.
\label{tf lambda1half}
\end{align}
The remaining two gauge symmetries in (\ref{linearized tf R}) 
generated by $\Lambda_{3/2}$ and $\tilde{\Lambda}_{1/2}$ can similarly 
be found order by order in $\Psi$.
The transformation
\begin{equation}
B^{[0]}_{\delta_{\Lambda_{3/2}}} =\  B^{[2]}_{\delta_{\Lambda_{3/2}}} =\ 0,\qquad 
\delta^{[0]}_{\Lambda_{3/2}}\Psi =\ \eLambdaf,\qquad
 \delta^{[0]}_{\Lambda_{3/2}}\Xi =\ 0,
\end{equation}
can be improved by combining with the corrections
\begin{align}
 \delta^{[2]}_{\Lambda_{3/2}}\Psi =&\ \frac{\kappa^2}{3!}\mspd{\Psi,\qxi,\eLambdaf},\qquad
 \delta^{[2]}_{\Lambda_{3/2}}\Xi =\ 0,\\
 B^{[4]}_{\delta_{\Lambda_{3/2}}} =&\ 
-\frac{\kappa^3}{4!}\mspd{\Psi,(Q_G\Xi)^2,\eta\Lambda_{\frac{3}{2}}},
\end{align}
so as to keep the pseudo-action invariant up to $\mathcal{O}(\Psi^4)$:
\begin{align}
 \delta_{\Lambda_{3/2}}^{[2]}S_{NS}+\delta_{\Lambda_{3/2}}^{[0]}S_{R[2]} =&\ 0,\\
 \delta_{\Lambda_{3/2}}^{[4]}S_{NS}+\delta_{\Lambda_{3/2}}^{[2]}S_{R[2]}
+\delta_{\Lambda_{3/2}}^{[0]}S_{R[4]} =&\ 0.
\end{align}
This is also a new kind of symmetry, which is shown in Appendix
by constructing the next-order correction.
The last gauge transformation, defined at the linearized level by
\begin{equation}
B^{[0]}_{\delta_{\tilde{\Lambda}_{1/2}}} =\ 0,\qquad
 \delta^{[0]}_{\tilde{\Lambda}_{1/2}}\Psi =\ 0,\qquad
 \delta^{[0]}_{\tilde{\Lambda}_{1/2}}\Xi =\ \eLambdax,
\end{equation}
can be improved by the next-order correction
\begin{equation}
 B^{[2]}_{\delta_{\tilde{\Lambda}_{1/2}}} =\ \frac{\kappa}{2}\mspd{\Psi,\eLambdax},
\end{equation}
to make the action invariant at $\mathcal{O}(\Psi^2)$:
\begin{equation}
 \delta_{\tilde{\Lambda}_{1/2}}^{[2]}S_{NS}+\delta_{\tilde{\Lambda}_{1/2}}^{[0]}S_{R[2]} =\ 0.
\end{equation}
We can find the next-order transformation,
\begin{align}
\delta^{[2]}_{\tilde{\Lambda}_{1/2}}\Psi =&\ \frac{\kappa^2}{6}\mspd{\Psi,\qpsi,\eLambdax}
-\frac{\kappa^2}{3}\mspd{\Psi,\mspd{\Psi,\eLambdax}}\\
 \delta^{[2]}_{\tilde{\Lambda}_{1/2}}\Xi =&\ \frac{\kappa^2}{3}\mspd{\Psi,\qxi,\eLambdax}
-\frac{\kappa^2}{2}\mspd{\Xi,\mspd{\Psi,\eLambdax}},\\
 B^{[4]}_{\delta_{\tilde{\Lambda}_{1/2}}} =&\ -\frac{\kappa^3}{12}\mspd{\Psi,\qpsi,\qxi,\eLambdax}
  +\frac{\kappa^3}{12}\mspd{\Psi,\mspd{\Psi,\qxi,\eLambdax}}\nonumber\\
 &+\frac{\kappa^3}{6}\mspd{\Psi,\qxi,\mspd{\Psi,\eLambdax}}
  -\frac{\kappa^3}{6}\mspd{\mspd{\Psi,\qxi},\Psi,\eLambdax},
\end{align}
so as to transform the pseudo-action 
in the form proportional to the constraint at $\mathcal{O}(\Psi^4)$:
\begin{align}
& \delta^{[4]}_{\tilde{\Lambda}_{1/2}}S_{NS}+\delta^{[2]}_{\tilde{\Lambda}_{1/2}}S_{R[2]}
+\delta^{[0]}_{\tilde{\Lambda}_{1/2}}S_{R[4]} 
\nonumber\\
&\hspace{37mm}
=\frac{\kappa^2}{12}\langle\eLambdax,\mspd{\qxi,\mspd{\Psi,\epsi}}\rangle
-\frac{\kappa^2}{12}\langle\eLambdax,\mspd{\epsi,\mspd{\Psi,\qxi}}\rangle.
\label{tf tildelambda}
\end{align}

From these considerations, it is natural to expect that these new types of 
gauge symmetries can be constructed order by order in $\Psi$, although 
we cannot yet prove it. We give the next-order results as a further
evidence in Appendix. They are also nontrivial in the sense
that the higher-order correction of the constraint is included.

Since all these gauge symmetries, including those provided by imposing
the constraint, must be important to reproduce the unitary amplitudes,
we assume that they have to be respected by the new Feynman rules and propose 
the following alternative prescription:
\begin{itemize}
 \item Use the off-diagonal propagator (\ref{naive R propagator}) for the R string.
 \item Use the vertices (\ref{bare vertices}) as they are without any restriction.
 \item Add two possibilities, $\Xi$ and $\Psi$, for each external fermion, 
and impose the linearized constraint, $Q\Xi=\eta\Psi$, on the on-shell external states.
\end{itemize}
Our claim is that this prescription respecting all the gauge symmetries
is more suitable for the Feynman rules
suggested by the pseudo-action (\ref{R action}).
This is supported by the fact that there is no ambiguity, associated with 
the self-dual$-$anti-self-dual decomposition already mentioned,
in the new Feynman rules. 
The new prescription, in addition, has an advantage that it does not require any special
preparation like the field redefinition (\ref{xi redefinition}).

\section{Amplitudes with external fermions}\label{sec3}

Using the new Feynman rules, we will explicitly calculate in this section the on-shell
four- and five-point amplitudes with external fermions. 
It will be shown that the results agree with the well-known amplitudes obtained in the first
quantized formulation.

\subsection{Four-point amplitudes}

The on-shell four-point amplitudes with external fermions were already
calculated using the self-dual Feynman rules and shown to agree 
with the well-known amplitudes obtained in the first quantized 
formulation.\cite{Kunitomo:2013mqa} 
We first have to confirm that the new Feynman rules also reproduce the same
results.

\begin{figure}[htbp]
\begin{minipage}{0.33\hsize}
 \begin{center}
  \includegraphics[width=4cm]{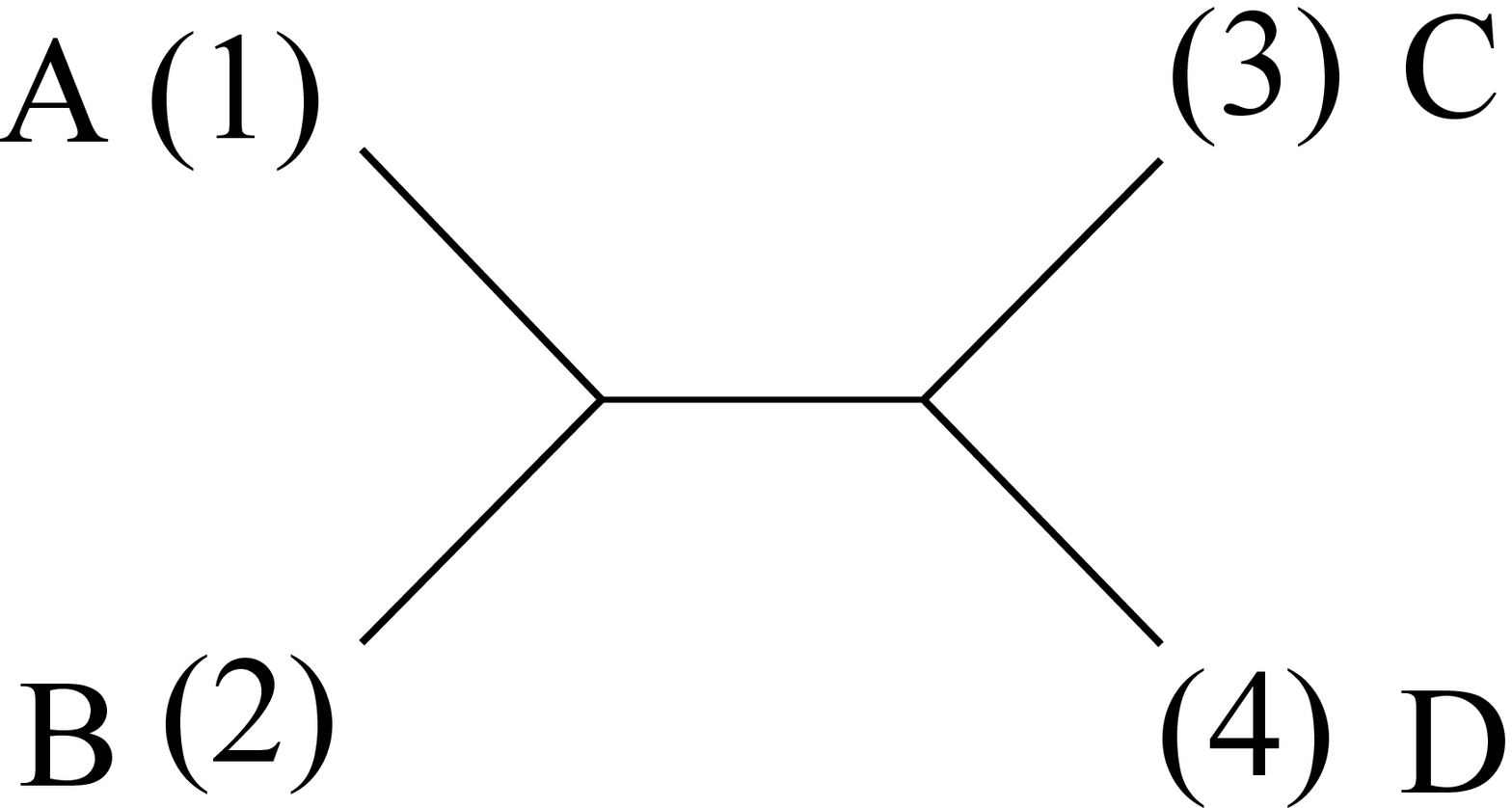}\\
(a)
 \end{center}
\end{minipage}
\begin{minipage}{0.33\hsize}
 \begin{center}
  \includegraphics[width=4cm]{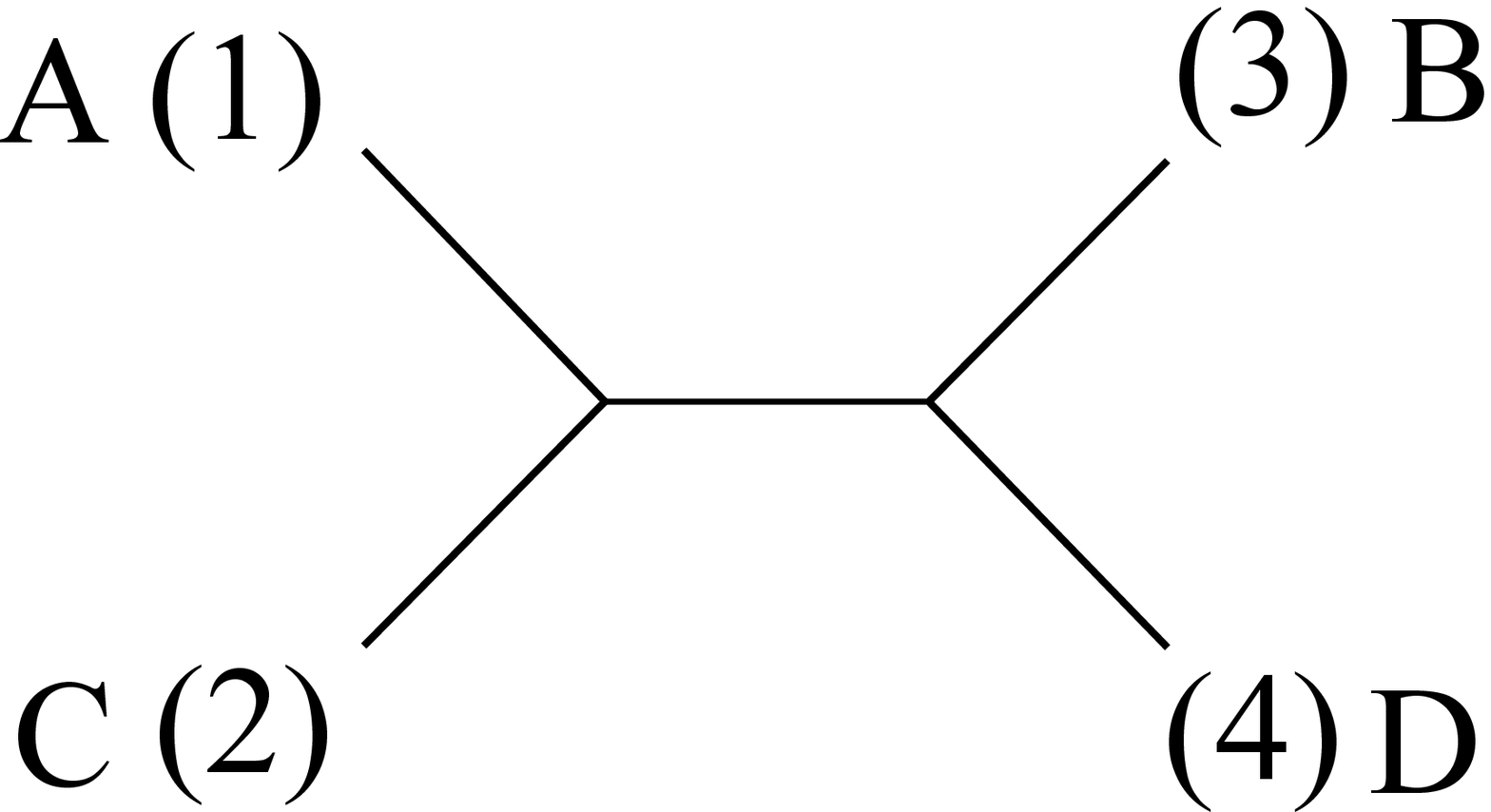}\\
(b)
 \end{center}
\end{minipage}
\begin{minipage}{0.33\hsize}
 \begin{center}
  \includegraphics[width=4cm]{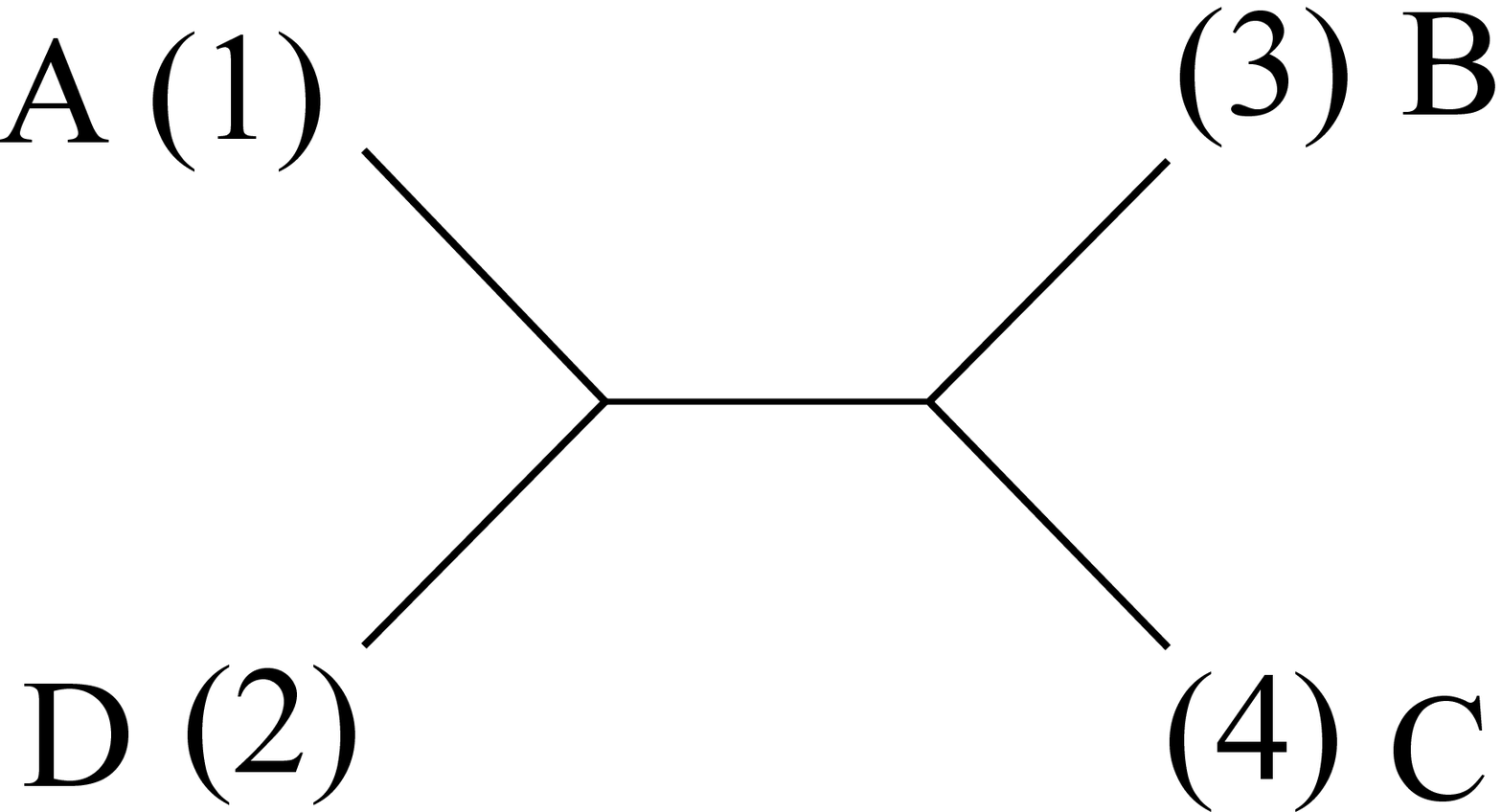}\\
(c)
 \end{center}
\end{minipage}
\\
\caption{Three four-point Feynman diagrams with one propagator:
(a) $s$-channel, (b) $t$-channel and (c) $u$-channel.}
\label{4-1P}
\end{figure}

Let us start from the calculation of the four-fermion amplitude $\mathcal{A}_{F^4}$. 
The contributions come from the $s$-, $t$-, and $u$-channel
diagrams constructed using two three-string vertices, and also a contact-type diagram
containing a four-string vertex.\footnote{
The corresponding string diagrams are depicted in Ref.~\citen{Kunitomo:2013mqa}.} 
In this paper, we denote for example the $s$-channel diagram, 
schematically depicted by Fig.~\ref{4-1P}(a), as $(AB|CD)$, where $A,\ B\ ,C$, and $D$ 
are labels which distinguish external strings. 
Since the order of strings $A$ and $B$,
or $C$ and $D$, has no meaning in the heterotic (closed) string theory,
this has as much information as this type of Feynman diagram. 
The $t$- and $u$-channel diagrams are denoted by $(AC|BD)$ and $(AD|BC)$ 
in this notation, respectively.
Using the new Feynman rules, the $s$-channel contribution is written as
\begin{align}
  \mathcal{A}^{(AB|CD)}_{F^4} =&\ \left(-\frac{\kappa}{2}\right)^2
\int_0^\infty dT\int_0^{2\pi}\frac{d\theta}{2\pi}\
\langle(\eta\Psi_A(1)Q\Xi_B(2)+Q\Xi_A(1)\eta\Psi_B(2))
\nonumber\\
&\hspace{43mm}\times
(\xi_cb_c^-b_c^+ )
(\eta\Psi_C(3)Q\Xi_D(4)+Q\Xi_C(3)\eta\Psi_D(4))\rangle_W,
\label{s-channel}
\end{align}
where the correlation is evaluated as the conformal field theory
on the corresponding string diagram. 
The insertions $\xi_c$, $b_c^-$, and $b_c^+$ are the corresponding fields 
integrated along the contour winding around the propagator. 
The numbers in the parentheses are the labels which distinguish
each leg of the diagram, but they are redundant
if we always arrange the external states in order of the numbers from
the left as in (\ref{s-channel}). 
We omit them hereafter by taking this convention.\footnote{
In Ref.~\citen{Kunitomo:2013mqa}, we have implicitly taken this convention
and distinguished each external string by the numbers $1- 4$
instead of the letters $A - D$.
}
The $t$- and $u$-channel contributions can similarly be
written as
\begin{align}
 \mathcal{A}^{(AC|BD)}_{F^4} =&\ \frac{\kappa^2}{4}
\int d^2T\
\langle(\eta\Psi_A\ Q\Xi_C+Q\Xi_A\ \eta\Psi_C)(\xi_cb_c^-b_c^+)
(\eta\Psi_B\ Q\Xi_D+Q\Xi_B\ \eta\Psi_D)\rangle_W,\\
 \mathcal{A}^{(AD|BC)}_{F^4} =&\ \frac{\kappa^2}{4}
\int d^2T\
\langle(\eta\Psi_A\ Q\Xi_D+Q\Xi_A\ \eta\Psi_D)(\xi_cb_c^-b_c^+)
(\eta\Psi_B\ Q\Xi_C+Q\Xi_B\ \eta\Psi_C)\rangle_W,
\end{align}
where we used the shorthand notation
\begin{equation}
 \int_0^\infty dT\int_0^{2\pi}\frac{d\theta}{2\pi}\ \equiv 
\int d^2T.
\end{equation}
Unlike the open superstring case, a contact-type 
diagram also gives the contribution integrated over a region of the moduli space 
not covered by those from these three diagrams.
It was shown that such a contribution can be realized using 
the four-string interaction represented by the restricted 
tetrahedron,\cite{Saadi:1989tb} or $n$-faced polyhedra for
general $n$-string contact interactions,\cite{Kugo:1989aa} 
parametrized by $\theta_I\ (I=1,\cdots,2(n-3))$ in the notation 
in \citen{Kugo:1989tk}. 
Then the contribution from the contact-type diagram $(ABCD)$ is given by
\begin{align}
 \mathcal{A}^{(ABCD)}_{F^4} =&\ 
\frac{\kappa^2}{12}\int d\theta_1 d\theta_2\
\langle (b_{C_1}b_{C_2})
\Big(
(\eta\Psi_A \Psi_B + \Psi_A \eta\Psi_B) Q\Xi_C Q\Xi_D
\nonumber\\
&
+Q\Xi_A Q\Xi_B (\eta\Psi_C \Psi_D + \Psi_C \eta\Psi_D) 
+\eta\Psi_A Q\Xi_B \Psi_C Q\Xi_D
+\Psi_A Q\Xi_B \eta\Psi_C Q\Xi_D
\nonumber\\
&
+Q\Xi_A \eta\Psi_B Q\Xi_C \Psi_D 
+Q\Xi_A \Psi_B Q\Xi_C \eta\Psi_D 
+Q\Xi_A (\eta\Psi_B \Psi_C
+\Psi_B \eta\Psi_C) Q\Xi_D 
\nonumber\\
&
+\eta\Psi_A Q\Xi_B Q\Xi_C \Psi_D 
+\Psi_A Q\Xi_B Q\Xi_C \eta\Psi_D 
\Big)\rangle_W.
\end{align}
Here the definition of the parameters $\theta_1$ and $\theta_2$,
their integration region and the corresponding contours $C_1$ and $C_2$, 
along which the anti-ghost insertions are integrated, are given 
in Ref.~\citen{Kugo:1989tk}; their explicit forms are not necessary here.
Adding all these contributions and imposing the linearized constraint 
$Q\Xi=\eta\Psi$ on each external state, the on-shell four-fermion amplitude
eventually becomes 
\begin{align}
 \mathcal{A}_{F^4} =&\ \mathcal{A}_{F^4}^{(AB|CD)} + \mathcal{A}_{F^4}^{(AC|BD)} 
+ \mathcal{A}_{F^4}^{(AD|BC)} + \mathcal{A}_{F^4}^{(ABCD)}
\nonumber\\
=&\ \kappa^2 \int d^2T\ \Big(
\llangle (\eta\Psi_A\ \eta\Psi_B\ (b_c^-b_c^+)\ \eta\Psi_C\ \eta\Psi_D)\rrangle_W
+\llangle (\eta\Psi_A\ \eta\Psi_C\ (b_c^-b_c^+)\ \eta\Psi_B\ \eta\Psi_D)\rrangle_W
\nonumber\\
&\hspace{17mm}
+\llangle (\eta\Psi_A\ \eta\Psi_D\ (b_c^-b_c^+)\ \eta\Psi_B\ \eta\Psi_C)\rrangle_W
\Big)
\nonumber\\
& 
+\kappa^2 \int d^2\theta\ 
\llangle (b_{C_1} b_{C_2})\ \eta\Psi_A\ \eta\Psi_B\ \eta\Psi_C\ \eta\Psi_D\rrangle_W,
\label{FFFF amp}
\end{align}
where $\llangle\cdots\rrangle_W$ represents the correlation in the small Hilbert space:
\begin{equation}
 \llangle \mathcal{O}_1\cdots \mathcal{O}_n\rrangle =\
 \langle \xi\ \mathcal{O}_1\cdots \mathcal{O}_n\rangle,
\end{equation}
where $\mathcal{O}_1,\cdots,\mathcal{O}_n$ are the operators in the small Hilbert space.
The $\xi$ on the right-hand side can either be local or integrated.
The correlation is independent of its position or contour since only the zero mode 
gives the non-vanishing contribution.
Although we can, in principle, map this expression (\ref{FFFF amp}) to the well-known form 
in the first quantized formulation evaluated on the complex 
plane,\cite{LeClair:1988sp,LeClair:1988sj,Moeller:2004yy} 
it is not necessary if we notice that each term has the same form 
as that in the bosonic closed string field theory with
the identification of $\eta\Psi$ and the bosonic string fields,
both of which have the same ghost number, $G=2$. 
Using the fact that the bosonic closed string field theory
reproduces the correct perturbative amplitudes, we can conclude that the amplitude 
(\ref{FFFF amp}) agrees with that obtained in the first 
quantized formulation. 

We can similarly calculate the two-boson-two-fermion amplitude.
After a little manipulation, the contributions from the $s$-, $t$- and $u$-channel 
diagrams become
\begin{align}
 \mathcal{A}^{(AB|CD)}_{F^2B^2} =&
-\frac{\kappa^2}{4}\int d^2T\
\langle(\eta\Psi_A Q\Xi_B + Q\Xi_A \eta\Psi_B) (\xi_c b_c^- b_c^+)
(QV_C \eta V_D + \eta V_C QV_D)\rangle_W,\\ 
 \mathcal{A}^{(AC|BD)}_{F^2B^2} =&
-\frac{\kappa^2}{2}\int d^2T
\Big(\langle\eta\Psi_A QV_C ( b_c^- b_c^+) \Xi_B QV_D\rangle_W
+\langle \Xi_A QV_C (b_c^- b_c^+ ) \eta\Psi_B QV_D\rangle_W\Big),\\
 \mathcal{A}^{(AD|BC)}_{F^2B^2} =&
-\frac{\kappa^2}{2}\int d^2T
\Big(\langle\eta\Psi_A QV_D (b_c^- b_c^+ ) \Xi_B QV_C\rangle_W
+\langle \Xi_A QV_D (b_c^- b_c^+ ) \eta\Psi_B QV_C\rangle_W\Big),
\end{align}
respectively. 
The contribution from the contact-type diagram consists 
of two parts coming from the two vertices in (\ref{two-two}):
\begin{align}
 \mathcal{A}^{(ABCD)}_{F^2B^2} =&\ 
-\frac{\kappa^2}{2}\int d^2\theta\
\langle (b_{C_1} b_{C_2}) (\eta\Psi_A \Xi_B+\Xi_A \eta\Psi_B)QV_C QV_D\rangle_W
\nonumber\\
&
-\frac{\kappa^2}{4} \oint d\theta\ 
\langle (\eta\Psi_A \Xi_B+\Xi_A \eta\Psi_B)\ b^-_\theta (QV_C V_D+V_C QV_D)\rangle_W,
\label{4contact}
\end{align}
where 
\begin{equation}
\oint d\theta\ \equiv\ \int_0^{2\pi}\frac{d\theta}{2\pi}
\end{equation}
is the integration over the twist angle of the collapsed propagator, 
and $b^-_\theta$ is the corresponding anti-ghost insertion. 
Although these four contributions other than the second term of (\ref{4contact})
cover the whole moduli space, they are not smoothly connected at each boundary 
since the external states in each contribution appear in different forms (pictures).
This gap is canceled by the remaining contribution, the second term in 
(\ref{4contact}).\footnote{
These discrepancies can be interpreted as coming from the difference of 
the positions of the picture-changing operators.\cite{Friedan:1985ge} 
The second term in (\ref{4contact})
corresponds to the contribution from the vertical integration 
introduced in Ref.~\citen{Sen:2014pia}.}
We can show this by aligning the external bosons 
in the four contributions to the same form, say 
$(QV_C, \eta V_D)$.
This is possible by integrating by parts with respect to $\eta$ and $Q$,
but the latter produces extra boundary contributions
appearing through the relation
\begin{equation}
 \int_0^\infty dT e^{-L_0^+T} \{b_0^+, Q\} =\ 
-\int _0^\infty dT\frac{\partial}{\partial T}e^{-L_0^+T}
\end{equation}
and the similar relation for the anti-ghost insertions in the tetrahedron vertex,
which can be read from the algebraic relation (\ref{fundamental}) satisfied
by the corresponding string products.
After such an alignment, each contribution becomes
\begin{align}
 \mathcal{A}^{(AB|CD)}_{F^2B^2} =&
-\frac{\kappa^2}{2}\int d^2T\
\langle(\eta\Psi_A Q\Xi_B + Q\Xi_A \eta\Psi_B) (\xi_c b_c^- b_c^+)
QV_C \eta V_D\rangle_W\nonumber\\
&+\frac{\kappa^2}{4}\oint d\theta\
\langle(\eta\Psi_A Q\Xi_B + Q\Xi_A \eta\Psi_B)\ b_\theta^- V_C V_D\rangle_W,\\
 \mathcal{A}^{(AC|BD)}_{F^2B^2} =&
-\frac{\kappa^2}{2}\int d^2T\
\Big(
\langle\eta\Psi_A QV_C (\xi_c b_c^- b_c^+) Q\Xi_B \eta V_D\rangle_W
+\langle Q\Xi_A QV_C (\xi_c b_c^- b_c^+) \eta\Psi_B \eta V_D\rangle_W
\Big)\nonumber\\
&+\frac{\kappa^2}{2}\oint d\theta\
\Big(\langle \eta\Psi_A QV_C\ b_\theta^- \Xi_B V_D\rangle_W
+\langle \eta\Psi_B V_D\ b_\theta^- \Xi_A QV_C\rangle_W\Big),\\
 \mathcal{A}^{(AD|BC)}_{F^2B^2} =&
-\frac{\kappa^2}{2}\int d^2T\
\Big(
\langle\eta\Psi_A \eta V_D (\xi_c b_c^- b_c^+) Q\Xi_B QV_C\rangle_W
+\langle Q\Xi_A \eta V_D (\xi_c b_c^- b_c^+) \eta\Psi_B QV_C\rangle_W
\Big)\nonumber\\
&+\frac{\kappa^2}{2}\oint d\theta\
\Big(
\langle \eta\Psi_A V_D\ b_\theta^-  \Xi_B QV_C\rangle_W
+\langle \eta\Psi_B QV_C\ b_\theta^- \Xi_A V_D\rangle_W\Big),\\
\mathcal{A}^{(ABCD)}_{F^2B^2} =&
-\frac{\kappa^2}{2}\int d^2\theta\
\langle\xi (b_{C_1} b_{C_2}) 
(\eta\Psi_A Q\Xi_B+Q\Xi_A \eta\Psi_B) QV_C \eta V_D\rangle_W
\nonumber\\
&
-\frac{\kappa^2}{4}\oint d\theta\ \Big(
\langle(\eta\Psi_A Q\Xi_B+Q\Xi_A \eta\Psi_B)\ b_\theta^- V_C V_D\rangle_W
\nonumber\\
&\hspace{20mm}
+2\langle\eta\Psi_A QV_C\ b_\theta^- \Xi_B V_D\rangle_W
+2\langle \eta\Psi_A V_D\ b_\theta^- \Xi_B QV_C\rangle_W
\nonumber\\
&\hspace{20mm}
+2\langle\eta\Psi_B QV_C\ b_\theta^- \Xi_A V_D\rangle_W
+2\langle \eta\Psi_B V_D\ b_\theta^- \Xi_A QV_C \rangle_W
\Big).
\end{align}
We can easily see that the boundary contributions are completely canceled,
and the total amplitude becomes
\begin{align}
 \mathcal{A}_{F^2B^2} =&\ \mathcal{A}_{F^2B^2}^{(AB|CD)}+
\mathcal{A}_{F^2B^2}^{(AC|BD)}+\mathcal{A}_{F^2B^2}^{(AD|BC)}
+\mathcal{A}_{F^2B^2}^{(ABCD)}
\nonumber\\
=&
-\frac{\kappa^2}{2}\int d^2T\ \Big(
\langle(\eta\Psi_A Q\Xi_B + Q\Xi_A \eta\Psi_B) (\xi_c b_c^- b_c^+)
QV_C \eta V_D\rangle_W\nonumber\\
&\hspace{20mm}
+\langle\eta\Psi_A QV_C (\xi_c b_c^- b_c^+) Q\Xi_B \eta V_D\rangle_W
+\langle Q\Xi_A QV_C (\xi_c b_c^- b_c^+) \eta\Psi_B \eta V_D\rangle_W
\nonumber\\
&\hspace{20mm}
+\langle\eta\Psi_A \eta V_D (\xi_c b_c^- b_c^+) Q\Xi_B QV_C\rangle_W
+\langle Q\Xi_A \eta V_D (\xi_c b_c^- b_c^+) \eta\Psi_B QV_C\rangle_W
\Big)
\nonumber\\
&
-\frac{\kappa^2}{2}\int d^2\theta\
\langle\xi (b_{C_1} b_{C_2}) 
(\eta\Psi_A Q\Xi_B+Q\Xi_A \eta\Psi_B) QV_C \eta V_D\rangle_W,
\end{align}
and can be rewritten as
\begin{align}
\mathcal{A}_{F^2B^2}
=&
-\kappa^2 \int d^2T\ \Big(
\llangle\eta\Psi_A \eta\Psi_B (b_c^- b_c^+) QV_C \eta V_D\rrangle_W
+\llangle\eta\Psi_A QV_C (b_c^- b_c^+) \eta\Psi_B \eta V_D\rrangle_W
\nonumber\\
&\hspace{20mm}
+\llangle\eta\Psi_A \eta V_D (b_c^- b_c^+) \eta\Psi_B QV_C\rrangle_W
\Big)
\nonumber\\
&
-\kappa^2\int d^2\theta\
\llangle (b_{C_1} b_{C_2})\eta\Psi_A \eta\Psi_B QV_C \eta V_D\rrangle_W,
\label{FFBB final}
\end{align}
after imposing the constraint.
Similarly to the case of the four-fermion amplitude,
this final expression agrees with that in the bosonic closed string field
under the identification of the external bosonic strings and 
the external strings in (\ref{FFBB final}), that is, 
$\eta\Psi$, $QV$ and $\eta V$.\footnote{
The overall minus sign should be corrected if we rewrite it using
the physical vertex operators in $\Psi$.} 
Thus, we can again conclude that the well-known amplitude 
in the first quantized formulation is correctly reproduced.

\subsection{Five-point amplitudes}

\begin{figure}[htbp]
\begin{minipage}{0.5\hsize}
\begin{center}
  \includegraphics[width=4.5cm]{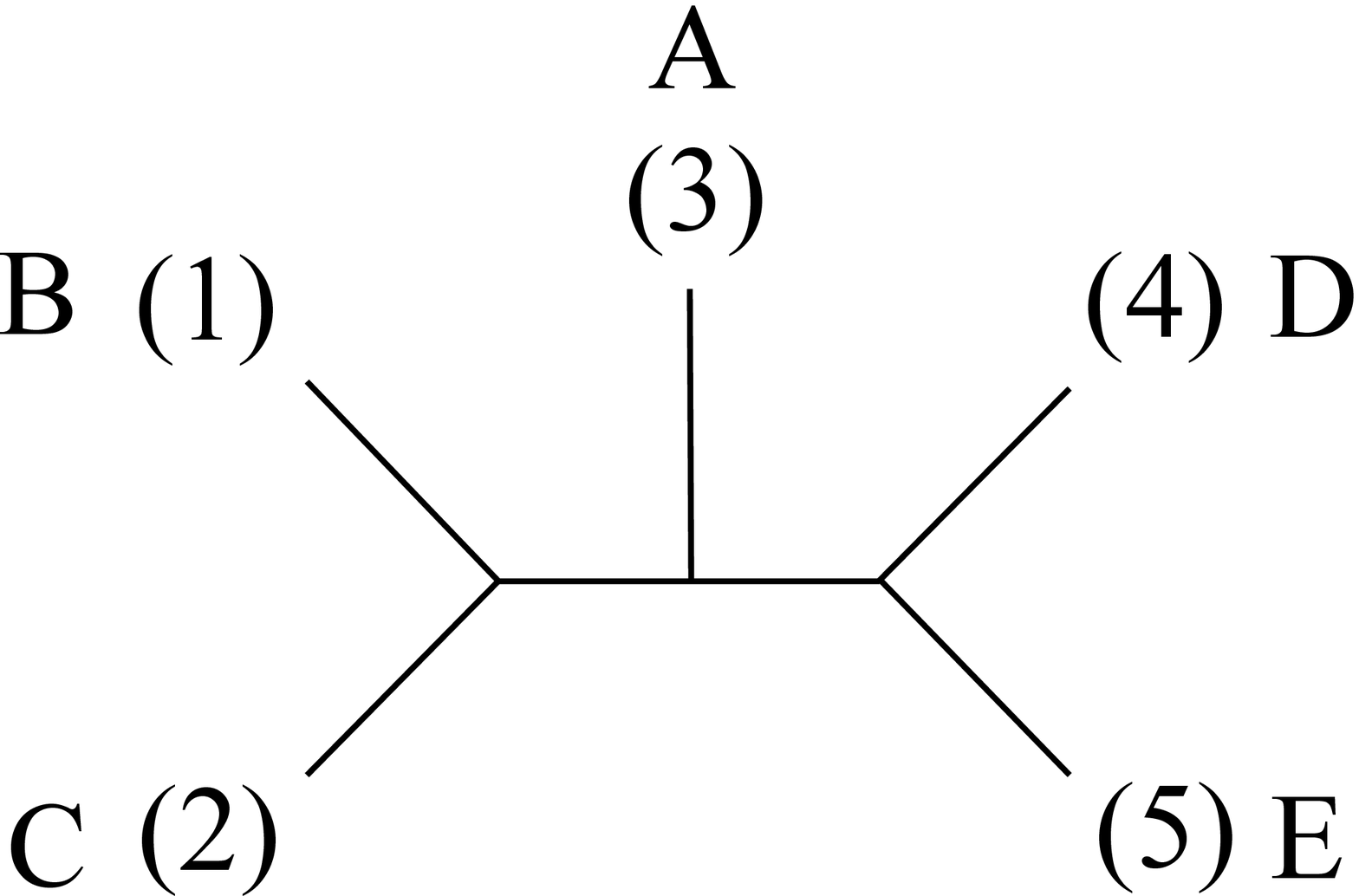}
 \end{center}
\hspace{0cm}
\caption{The topology of the five-point Feynman diagrams with two propagators.}
\label{5-2P}
\end{minipage}
\begin{minipage}{0.5\hsize}
\begin{center}
  \includegraphics[width=4.5cm]{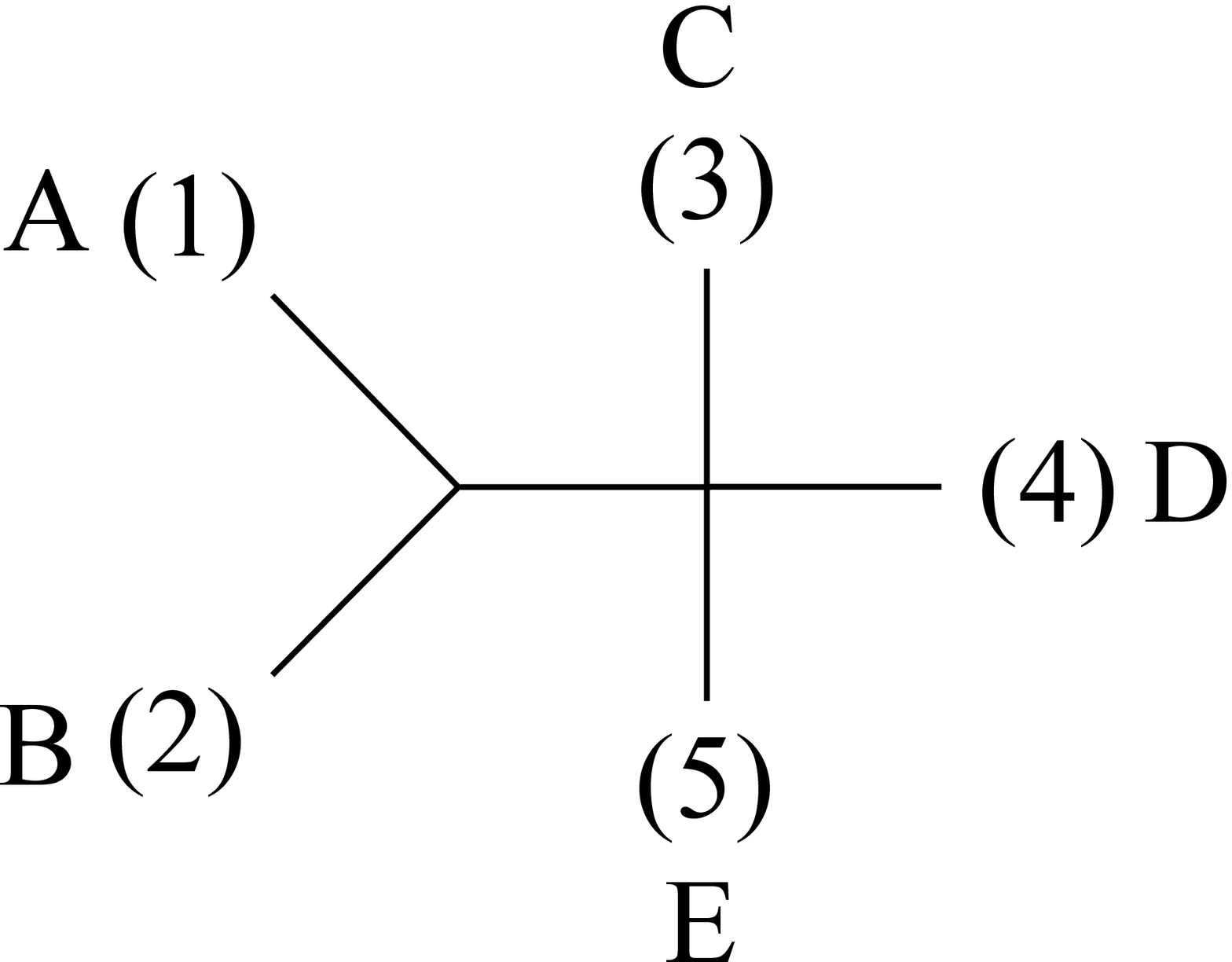}
 \end{center}
\caption{The topology of the five-point Feynman diagrams with one propagator.}
\label{5-1P}
\end{minipage}
\end{figure}

Let us next calculate the on-shell five-point amplitudes with external fermions.
We follow the convention in the previous subsection;
we label the five external strings by $A, B, C, D$, and $E$ arranged in order of
the number assigned to the legs as depicted in Figs.~\ref{5-2P}
and \ref{5-1P}. 
There are three types of diagrams contributing to the five-point amplitudes,
which we refer to as the two-propagator (2P), one-propagator (1P), and no-propagator (NP) 
diagrams corresponding to the number of propagators to be included.
The 2P diagrams contain three three-string vertices and two propagators 
as depicted in Fig.~\ref{5-2P}, which we simply denote as $(BC|A|DE)$.
The 1P diagram contains one three-string vertex,
one four-string vertex, and one propagator as depicted in Fig.~\ref{5-1P}.
We denote this diagram as $(AB|CDE)$.

There are two types of five-point amplitudes including external fermions:
the four-fermion-one-boson ($F^4B$) and two-fermion-three-boson ($F^2B^3$) amplitudes.
Let us first calculate the former, $F^4B$, amplitude. Suppose that the strings 
$A, B, C$, and $D$ are fermions and the string $E$ is a boson. 
We begin with the calculation of the contributions from the fifteen, 
$(_5C_1\times _4C_2)/2$, 2P diagrams.
For example, the contribution of the diagram $(BC|A|DE)$ is calculated as
\begin{align}
\mathcal{A}_{F^4B}^{(BC|A|DE)}
 =&\
\left(-\frac{\kappa}{2}\right)^3(-2)
\int d^2T_1\int d^2T_2\
\Big(\langle (\eta\Psi_B \Xi_C+\Xi_B \eta\Psi_C)
\nonumber\\
&\hspace{60mm}\times
(Q \xi_{c_1} b_{c_1}^- b_{c_1}^+ Q)
\eta\Psi_A (\xi_{c_2} b_{c_2}^- b_{c_2}^+ \eta) \Xi_D QV_E\rangle_W
\nonumber\\
&\hspace{20mm}
+\langle (\eta\Psi_B \Xi_C+\Xi_B \eta\Psi_C)(Q \xi_{c_1} b_{c_1}^- b_{c_1}^+ Q)
\Xi_A (\eta \xi_{c_2} b_{c_2}^- b_{c_2}^+) \eta\Psi_D QV_E\rangle_W
\Big)
\nonumber\\
=&\ \frac{\kappa^3}{4}\int d^2T_1\int d^2T_2\
\Big(
\langle(\eta\Psi_B Q\Xi_C+Q\Xi_B \eta\Psi_C) 
\nonumber\\
&\hspace{60mm}\times
(\xi_{c_1} b_{c_1}^- b_{c_1}^+)
\eta\Psi_A (b_{c_2}^- b_{c_2}^+) Q\Xi_D QV_E\rangle_W
\nonumber\\
&\hspace{20mm}
+\langle(\eta\Psi_B Q\Xi_C+Q\Xi_B \eta\Psi_C) (\xi_{c_1} b_{c_1}^- b_{c_1}^+)
Q\Xi_A (b_{c_2}^- b_{c_2}^+) \eta\Psi_D QV_E\rangle_W
\Big)
\nonumber\\
&-\frac{\kappa^3}{4}\int d^2T \oint d\theta\
\Big(
\langle(\eta\Psi_B Q\Xi_C+Q\Xi_B \eta\Psi_C)
\nonumber\\
&\hspace{45mm}\times
(\xi_{c} b_{c}^- b_{c}^+)
(\eta\Psi_A\ b_\theta^- \Xi_D - \Xi_A\ b_\theta^- \eta\Psi_D) QV_E\rangle_W
\nonumber\\
&\hspace{28mm}
+\langle \Xi_D QV_E (\xi_{c} b_{c}^- b_{c}^+)
\eta\Psi_A\ b_\theta^- (\eta\Psi_B Q\Xi_C+Q\Xi_B \eta\Psi_C)\rangle_W
\Big),
\end{align}
where the inserted operators, $\xi_{c_i}$ or $b_{c_i}^\pm$, are integrated
along the contour winding around the $i$-th propagator.
We moved, by integrating by parts without exchanging the order of $Q$ and $\xi$, 
the operators $Q$ and $\eta$ in a way that acts on
the external states. This produces the boundary contributions, in which
one of the two propagators collapsed. 
Eleven of the remaining fourteen diagrams are obtained by simply relabeling
the external fermions:
\begin{align}
 \mathcal{A}_{F^4B}^{(BD|A|CE)}
 =&\ \frac{\kappa^3}{4}\int d^2T_1\int d^2T_2\
\Big(
\langle(\eta\Psi_B Q\Xi_D+Q\Xi_B \eta\Psi_D) 
\nonumber\\
&\hspace{60mm}\times
(\xi_{c_1} b_{c_1}^- b_{c_1}^+)
\eta\Psi_A (b_{c_2}^- b_{c_2}^+) Q\Xi_C QV_E\rangle_W
\nonumber\\
&\hspace{20mm}
+\langle(\eta\Psi_B Q\Xi_D+Q\Xi_B \eta\Psi_D) (\xi_{c_1} b_{c_1}^- b_{c_1}^+)
Q\Xi_A (b_{c_2}^- b_{c_2}^+) \eta\Psi_C QV_E\rangle_W
\Big)
\nonumber\\
&-\frac{\kappa^3}{4}\int d^2T \oint d\theta\
\Big(
\langle(\eta\Psi_B Q\Xi_D+Q\Xi_B \eta\Psi_D)
\nonumber\\
&\hspace{43mm}\times
(\xi_{c} b_{c}^- b_{c}^+)
(\eta\Psi_A\ b_\theta^- \Xi_C - \Xi_A\ b_\theta^- \eta\Psi_C) QV_E\rangle_W
\nonumber\\
&\hspace{28mm}
+\langle \Xi_C QV_E (\xi_{c} b_{c}^- b_{c}^+)
\eta\Psi_A\ b_{\theta}^- (\eta\Psi_B Q\Xi_D+Q\Xi_B \eta\Psi_D)\rangle_W\Big),\\
\mathcal{A}_{F^4B}^{(CD|A|BE)}
 =&\
\frac{\kappa^3}{4}\int d^2T_1\int d^2T_2\
\Big(
\langle(\eta\Psi_C Q\Xi_D+Q\Xi_C \eta\Psi_D) 
\nonumber\\
&\hspace{60mm}\times
(\xi_{c_1} b_{c_1}^- b_{c_1}^+)
\eta\Psi_A (b_{c_2}^- b_{c_2}^+) Q\Xi_B QV_E\rangle_W
\nonumber\\
&\hspace{20mm}
+\langle(\eta\Psi_C Q\Xi_D+Q\Xi_C \eta\Psi_D) (\xi_{c_1} b_{c_1}^- b_{c_1}^+)
Q\Xi_A (b_{c_2}^- b_{c_2}^+) \eta\Psi_B QV_E\rangle_W
\Big)
\nonumber
\end{align}
\begin{equation}
 \nonumber
\end{equation}
\begin{align}
 &-\frac{\kappa^3}{4}\int d^2T \oint d\theta\
\Big(
\langle(\eta\Psi_C Q\Xi_D+Q\Xi_C \eta\Psi_D)
\nonumber\\
&\hspace{50mm}\times
(\xi_{c} b_{c}^- b_{c}^+)
(\eta\Psi_A\ b_{\theta}^- \Xi_B - \Xi_A\ b_{\theta}^- \eta\Psi_B) QV_E\rangle_W
\nonumber\\
&\hspace{28mm}
+\langle \Xi_B QV_E (\xi_{c} b_{c}^- b_{c}^+)
\eta\Psi_A\ b_{\theta}^- (\eta\Psi_C Q\Xi_D+Q\Xi_C \eta\Psi_D)\rangle_W\Big),\\
\mathcal{A}_{F^4B}^{(AC|B|DE)}
 =&\
\frac{\kappa^3}{4}\int d^2T_1\int d^2T_2\
\Big(
\langle(\eta\Psi_A Q\Xi_C+Q\Xi_A \eta\Psi_C) 
\nonumber\\
&\hspace{60mm}\times
(\xi_{c_1} b_{c_1}^- b_{c_1}^+)
\eta\Psi_B (b_{c_2}^- b_{c_2}^+) Q\Xi_D QV_E\rangle_W
\nonumber\\
&\hspace{20mm}
+\langle(\eta\Psi_A Q\Xi_C+Q\Xi_A \eta\Psi_C) (\xi_{c_1} b_{c_1}^- b_{c_1}^+)
Q\Xi_B (b_{c_2}^- b_{c_2}^+) \eta\Psi_D QV_E\rangle_W
\Big)
\nonumber\\
&-\frac{\kappa^3}{4}\int d^2T \oint d\theta\
\Big(
\langle(\eta\Psi_A Q\Xi_C+Q\Xi_A \eta\Psi_C)
\nonumber\\
&\hspace{50mm}\times
(\xi_{c} b_{c}^- b_{c}^+)
(\eta\Psi_B\ b_{\theta}^- \Xi_D - \Xi_B\ b_{\theta}^- \eta\Psi_D) QV_E\rangle_W
\nonumber\\
&\hspace{28mm}
+\langle \Xi_D QV_E (\xi_{c} b_{c}^- b_{c}^+)
\eta\Psi_B\ b_{\theta}^- (\eta\Psi_A Q\Xi_C+Q\Xi_A \eta\Psi_C)\rangle_W\Big),\\
\mathcal{A}_{F^4B}^{(AD|B|CE)}
 =&\
\frac{\kappa^3}{4}\int d^2T_1\int d^2T_2\
\Big(
\langle(\eta\Psi_A Q\Xi_D+Q\Xi_A \eta\Psi_D) 
\nonumber\\
&\hspace{60mm}\times
(\xi_{c_1} b_{c_1}^- b_{c_1}^+)
\eta\Psi_B (b_{c_2}^- b_{c_2}^+) Q\Xi_C QV_E\rangle_W
\nonumber\\
&\hspace{20mm}
+\langle(\eta\Psi_A Q\Xi_D+Q\Xi_A \eta\Psi_D) (\xi_{c_1} b_{c_1}^- b_{c_1}^+)
Q\Xi_B (b_{c_2}^- b_{c_2}^+) \eta\Psi_C QV_E\rangle_W
\Big)
\nonumber\\
&-\frac{\kappa^3}{4}\int d^2T \oint d\theta\
\Big(
\langle(\eta\Psi_A Q\Xi_D+Q\Xi_A \eta\Psi_D)
\nonumber\\
&\hspace{50mm}\times
(\xi_{c} b_{c}^- b_{c}^+)
(\eta\Psi_B\ b_{\theta}^- \Xi_C - \Xi_B\ b_{\theta}^- \eta\Psi_C) QV_E\rangle_W
\nonumber\\
&\hspace{28mm}
+\langle \Xi_C QV_E (\xi_{c} b_{c}^- b_{c}^+)
\eta\Psi_B\ b_{\theta}^- (\eta\Psi_A Q\Xi_D+Q\Xi_A \eta\Psi_D)\rangle_W\Big),\\
 \mathcal{A}_{F^4B}^{(CD|B|AE)}
 =&\
\frac{\kappa^3}{4}\int d^2T_1\int d^2T_2\
\Big(
\langle(\eta\Psi_C Q\Xi_D+Q\Xi_C \eta\Psi_D) 
\nonumber\\
&\hspace{60mm}\times
(\xi_{c_1} b_{c_1}^- b_{c_1}^+)
\eta\Psi_B (b_{c_2}^- b_{c_2}^+) Q\Xi_A QV_E\rangle_W
\nonumber\\
&\hspace{20mm}
+\langle(\eta\Psi_C Q\Xi_D+Q\Xi_C \eta\Psi_D) (\xi_{c_1} b_{c_1}^- b_{c_1}^+)
Q\Xi_B (b_{c_2}^- b_{c_2}^+) \eta\Psi_A QV_E\rangle_W
\Big)
\nonumber\\
&-\frac{\kappa^3}{4}\int d^2T \oint d\theta\
\Big(
\langle(\eta\Psi_C Q\Xi_D+Q\Xi_C \eta\Psi_D)
\nonumber\\
&\hspace{50mm}
(\xi_{c} b_{c}^- b_{c}^+)
(\eta\Psi_B\ b_{\theta}^- \Xi_A - \Xi_B\ b_{\theta}^- \eta\Psi_A) QV_E\rangle_W
\nonumber\\
&\hspace{28mm}
+\langle \Xi_A QV_E (\xi_{c} b_{c}^- b_{c}^+)
\eta\Psi_B\ b_{\theta}^- (\eta\Psi_C Q\Xi_D+Q\Xi_C \eta\Psi_D)\rangle_W\Big),\\
 \mathcal{A}_{F^4B}^{(AB|C|DE)}
 =&\
\frac{\kappa^3}{4}\int d^2T_1\int d^2T_2\
\Big(
\langle(\eta\Psi_A Q\Xi_B+Q\Xi_A \eta\Psi_B) 
\nonumber\\
&\hspace{60mm}\times
(\xi_{c_1} b_{c_1}^- b_{c_1}^+)
\eta\Psi_C (b_{c_2}^- b_{c_2}^+) Q\Xi_D QV_E\rangle_W
\nonumber\\
&\hspace{20mm}
+\langle(\eta\Psi_A Q\Xi_B+Q\Xi_A \eta\Psi_B) (\xi_{c_1} b_{c_1}^- b_{c_1}^+)
Q\Xi_C (b_{c_2}^- b_{c_2}^+) \eta\Psi_D QV_E\rangle_W
\Big)
\nonumber\\
&-\frac{\kappa^3}{4}\int d^2T \oint d\theta\
\Big(
\langle(\eta\Psi_A Q\Xi_B+Q\Xi_A \eta\Psi_B)
\nonumber\\
&\hspace{50mm}\times
(\xi_{c} b_{c}^- b_{c}^+)
(\eta\Psi_C\ b_{\theta}^- \Xi_D - \Xi_C\ b_{\theta}^- \eta\Psi_D) QV_E\rangle_W
\nonumber\\
&\hspace{28mm}
+\langle \Xi_D QV_E (\xi_{c} b_{c}^- b_{c}^+)
\eta\Psi_C\ b_{\theta}^- (\eta\Psi_A Q\Xi_B+Q\Xi_A \eta\Psi_B)\rangle_W\Big),\\
 \mathcal{A}_{F^4B}^{(AD|C|BE)}
 =&\
\frac{\kappa^3}{4}\int d^2T_1\int d^2T_2\
\Big(
\langle(\eta\Psi_A Q\Xi_D+Q\Xi_A \eta\Psi_D) 
\nonumber\\
&\hspace{60mm}\times
(\xi_{c_1} b_{c_1}^- b_{c_1}^+)
\eta\Psi_C (b_{c_2}^- b_{c_2}^+) Q\Xi_B QV_E\rangle_W
\nonumber\\
&\hspace{20mm}
+\langle(\eta\Psi_A Q\Xi_D+Q\Xi_A \eta\Psi_D) (\xi_{c_1} b_{c_1}^- b_{c_1}^+)
Q\Xi_C (b_{c_2}^- b_{c_2}^+) \eta\Psi_B QV_E\rangle_W
\Big)
\nonumber\\
&-\frac{\kappa^3}{4}\int d^2T \oint d\theta\
\Big(
\langle(\eta\Psi_A Q\Xi_D+Q\Xi_A \eta\Psi_D)
\nonumber\\
&\hspace{50mm}\times
(\xi_{c} b_{c}^- b_{c}^+)
(\eta\Psi_C\ b_{\theta}^- \Xi_B - \Xi_C\ b_{\theta}^- \eta\Psi_B) QV_E\rangle_W
\nonumber\\
&\hspace{28mm}
+\langle \Xi_B QV_E (\xi_{c} b_{c}^- b_{c}^+)
\eta\Psi_C\ b_{\theta}^- (\eta\Psi_A Q\Xi_D+Q\Xi_A \eta\Psi_D)\rangle_W\Big),\\
 \mathcal{A}_{F^4B}^{(BD|C|AE)}
 =&\
\frac{\kappa^3}{4}\int d^2T_1\int d^2T_2\
\Big(
\langle(\eta\Psi_B Q\Xi_D+Q\Xi_B \eta\Psi_D) 
\nonumber\\
&\hspace{60mm}\times
(\xi_{c_1} b_{c_1}^- b_{c_1}^+)
\eta\Psi_C (b_{c_2}^- b_{c_2}^+) Q\Xi_A QV_E\rangle_W
\nonumber\\
&\hspace{20mm}
+\langle(\eta\Psi_B Q\Xi_D+Q\Xi_B \eta\Psi_D) (\xi_{c_1} b_{c_1}^- b_{c_1}^+)
Q\Xi_C (b_{c_2}^- b_{c_2}^+) \eta\Psi_A QV_E\rangle_W
\Big)
\nonumber\\
&-\frac{\kappa^3}{4}\int d^2T \oint d\theta\
\Big(
\langle(\eta\Psi_B Q\Xi_D+Q\Xi_B \eta\Psi_D)
\nonumber\\
&\hspace{50mm}\times
(\xi_{c} b_{c}^- b_{c}^+)
(\eta\Psi_C\ b_{\theta}^- \Xi_A - \Xi_C\ b_{\theta}^- \eta\Psi_A) QV_E\rangle_W
\nonumber\\
&\hspace{28mm}
+\langle \Xi_A QV_E (\xi_{c} b_{c}^- b_{c}^+)
\eta\Psi_C\ b_{\theta}^- (\eta\Psi_B Q\Xi_D+Q\Xi_B \eta\Psi_D)\rangle_W\Big),\\
 \mathcal{A}_{F^4B}^{(AB|D|CE)}
 =&\
\frac{\kappa^3}{4}\int d^2T_1\int d^2T_2\
\Big(
\langle(\eta\Psi_A Q\Xi_B+Q\Xi_A \eta\Psi_B) 
\nonumber\\
&\hspace{60mm}\times
(\xi_{c_1} b_{c_1}^- b_{c_1}^+)
\eta\Psi_D (b_{c_2}^- b_{c_2}^+) Q\Xi_C QV_E\rangle_W
\nonumber\\
&\hspace{20mm}
+\langle(\eta\Psi_A Q\Xi_B+Q\Xi_A \eta\Psi_B) (\xi_{c_1} b_{c_1}^- b_{c_1}^+)
Q\Xi_D (b_{c_2}^- b_{c_2}^+) \eta\Psi_C QV_E\rangle_W
\Big)
\nonumber\\
&-\frac{\kappa^3}{4}\int d^2T \oint d\theta\
\Big(
\langle(\eta\Psi_A Q\Xi_B+Q\Xi_A \eta\Psi_B)
\nonumber\\
&\hspace{50mm}\times
(\xi_{c} b_{c}^- b_{c}^+)
(\eta\Psi_D\ b_{\theta}^- \Xi_C - \Xi_D\ b_{\theta}^- \eta\Psi_C) QV_E\rangle_W
\nonumber\\
&\hspace{28mm}
+\langle \Xi_C QV_E (\xi_{c} b_{c}^- b_{c}^+)
\eta\Psi_D\ b_{\theta}^- (\eta\Psi_A Q\Xi_B+Q\Xi_A \eta\Psi_B)\rangle_W\Big),\\
 \mathcal{A}_{F^4B}^{(AC|D|BE)}
 =&\
\frac{\kappa^3}{4}\int d^2T_1\int d^2T_2\
\Big(
\langle(\eta\Psi_A Q\Xi_C+Q\Xi_A \eta\Psi_C) 
\nonumber\\
&\hspace{60mm}\times
(\xi_{c_1} b_{c_1}^- b_{c_1}^+)
\eta\Psi_D (b_{c_2}^- b_{c_2}^+) Q\Xi_B QV_E\rangle_W
\nonumber\\
&\hspace{20mm}
+\langle(\eta\Psi_A Q\Xi_C+Q\Xi_A \eta\Psi_C) (\xi_{c_1} b_{c_1}^- b_{c_1}^+)
Q\Xi_D (b_{c_2}^- b_{c_2}^+) \eta\Psi_B QV_E\rangle_W
\Big)
\nonumber\\
&-\frac{\kappa^3}{4}\int d^2T \oint d\theta\
\Big(
\langle(\eta\Psi_A Q\Xi_C+Q\Xi_A \eta\Psi_C)
\nonumber\\
&\hspace{50mm}\times
(\xi_{c} b_{c}^- b_{c}^+)
(\eta\Psi_D\ b_{\theta}^- \Xi_B - \Xi_D\ b_{\theta}^- \eta\Psi_B) QV_E\rangle_W
\nonumber\\
&\hspace{28mm}
+\langle \Xi_B QV_E (\xi_{c} b_{c}^- b_{c}^+)
\eta\Psi_D\ b_{\theta}^- (\eta\Psi_A Q\Xi_C+Q\Xi_A \eta\Psi_C)\rangle_W\Big),\\
 \mathcal{A}_{F^4B}^{(BC|D|AE)}
 =&\
\frac{\kappa^3}{4}\int d^2T_1\int d^2T_2\
\Big(
\langle(\eta\Psi_B Q\Xi_C+Q\Xi_B \eta\Psi_C) 
\nonumber\\
&\hspace{60mm}\times
(\xi_{c_1} b_{c_1}^- b_{c_1}^+)
\eta\Psi_D (b_{c_2}^- b_{c_2}^+) Q\Xi_A QV_E\rangle_W
\nonumber\\
&\hspace{20mm}
+\langle(\eta\Psi_B Q\Xi_C+Q\Xi_B \eta\Psi_C) (\xi_{c_1} b_{c_1}^- b_{c_1}^+)
Q\Xi_D (b_{c_2}^- b_{c_2}^+) \eta\Psi_A QV_E\rangle_W
\Big)
\nonumber\\
&-\frac{\kappa^3}{4}\int d^2T \oint d\theta\
\Big(
\langle(\eta\Psi_B Q\Xi_C+Q\Xi_B \eta\Psi_C)
\nonumber\\
&\hspace{50mm}\times
(\xi_{c} b_{c}^- b_{c}^+)
(\eta\Psi_D\ b_{\theta}^- \Xi_A - \Xi_D\ b_{\theta}^- \eta\Psi_A) QV_E\rangle_W
\nonumber\\
&\hspace{28mm}
+\langle \Xi_A QV_E (\xi_{c} b_{c}^- b_{c}^+)
\eta\Psi_D\ b_{\theta}^- (\eta\Psi_B Q\Xi_C+Q\Xi_B \eta\Psi_C)\rangle_W\Big).
\end{align}
The last three contributions, coming from the diagrams including the boson
in the center, are obtained by calculating 
one of them, for example,
\begin{align}
 \mathcal{A}_{F^4B}^{(AB|E|CD)}
 =&\
\left(-\frac{\kappa}{2}\right)^2\frac{\kappa}{3!}\int d^2T_1\int d^2T_2\
\Big(
\langle (\eta\Psi_A Q\Xi_B + Q\Xi_A \eta\Psi_B) 
(\xi_{c_1} b_{c_1}^- b_{c_1}^+ Q) 
\nonumber\\
&\hspace{55mm}\times
V_E (\eta \xi_{c_2} b_{c_2}^- b_{c_2}^+)(\eta\Psi_C Q\Xi_D + Q\Xi_C \eta\Psi_D)\rangle_W
\nonumber\\
&\hspace{45mm}
+\langle (\eta\Psi_A Q\Xi_B + Q\Xi_A \eta\Psi_B)(\xi_{c_1} b_{c_1}^- b_{c_1}^+ \eta)
\nonumber\\
&\hspace{55mm}\times
V_E (Q \xi_{c_2} b_{c_2}^- b_{c_2}^+)(\eta\Psi_C Q\Xi_D + Q\Xi_C \eta\Psi_D)\rangle_W 
\nonumber\\
&\hspace{45mm}
+\langle (\eta\Psi_A Q\Xi_B + Q\Xi_A \eta\Psi_B) (\xi_{c_1} b_{c_1}^- b_{c_1}^+)
\nonumber\\
&\hspace{55mm}\times
QV_E (\eta \xi_{c_2} b_{c_2}^- b_{c_2}^+)(\eta\Psi_C Q\Xi_D + Q\Xi_C \eta\Psi_D)\rangle_W
\nonumber\\
&\hspace{45mm}
+\langle (\eta\Psi_A Q\Xi_B + Q\Xi_A \eta\Psi_B) (\xi_{c_1} b_{c_1}^- b_{c_1}^+ \eta)
\nonumber\\
&\hspace{55mm}\times
QV_E (\xi_{c_2} b_{c_2}^- b_{c_2}^+)(\eta\Psi_C Q\Xi_D + Q\Xi_C \eta\Psi_D)\rangle_W
\nonumber\\
&\hspace{45mm}
+\langle (\eta\Psi_A Q\Xi_B + Q\Xi_A \eta\Psi_B) (\xi_{c_1} b_{c_1}^- b_{c_1}^+ Q)
\nonumber\\
&\hspace{55mm}\times
\eta V_E (\xi_{c_2} b_{c_2}^- b_{c_2}^+)(\eta\Psi_C Q\Xi_D + Q\Xi_C \eta\Psi_D)\rangle_W
\nonumber\\
&\hspace{45mm}
+\langle (\eta\Psi_A Q\Xi_B + Q\Xi_A \eta\Psi_B) (\xi_{c_1} b_{c_1}^- b_{c_1}^+)
\nonumber\\
&\hspace{55mm}\times
\eta V_E (Q \xi_{c_2} b_{c_2}^- b_{c_2}^+)(\eta\Psi_C Q\Xi_D + Q\Xi_C \eta\Psi_D)\rangle_W
\Big) 
\nonumber\\
 =&\
\frac{\kappa^3}{4}\int d^2T_1\int d^2T_2\
\langle(\eta\Psi_A Q\Xi_B+Q\Xi_A \eta\Psi_B) (\xi_{c_1} b_{c_1}^- b_{c_1}^+)
\nonumber\\
&\hspace{60mm}\times
QV_E (b_{c_2}^- b_{c_2}^+) (\eta\Psi_C Q\Xi_D+Q\Xi_C \eta\Psi_D)\rangle_W
\nonumber
\end{align}
\begin{equation}
 \nonumber
\end{equation}
\begin{align}
&
+\frac{\kappa^3}{8}
\int d^2T \oint d\theta\
\Big(
\langle (\eta\Psi_A Q\Xi_B+Q\Xi_A \eta\Psi_B) (\xi_{c} b_{c}^- b_{c}^+) 
\nonumber\\
&\hspace{60mm}\times
V_E\ b_{\theta}^-(\eta\Psi_C Q\Xi_D+Q\Xi_C \eta\Psi_D)\rangle_W
\nonumber\\
&\hspace{30mm}
+\langle (\eta\Psi_C Q\Xi_D+Q\Xi_C \eta\Psi_D) (\xi_{c} b_{c}^- b_{c}^+) 
\nonumber\\
&\hspace{60mm}\times
V_E\ b_{\theta}^- (\eta\Psi_A Q\Xi_B+Q\Xi_A \eta\Psi_B)\rangle_W
\Big),
\end{align}
and relabeling its external fermions as
\begin{align}
 \mathcal{A}_{F^4B}^{(AC|E|BD)}
=&\
\frac{\kappa^3}{4}\int d^2T_1\int d^2T_2\
\langle(\eta\Psi_A Q\Xi_C+Q\Xi_A \eta\Psi_C) (\xi_{c_1} b_{c_1}^- b_{c_1}^+)
\nonumber\\
&\hspace{50mm}\times
QV_E (b_{c_2}^- b_{c_2}^+) (\eta\Psi_B Q\Xi_D+Q\Xi_B \eta\Psi_D)\rangle_W
\nonumber\\
&
+\frac{\kappa^3}{8}\int d^2T \oint d\theta\
\Big(
\langle (\eta\Psi_A Q\Xi_C+Q\Xi_A \eta\Psi_C)
(\xi_{c} b_{c}^- b_{c}^+) 
\nonumber\\
&\hspace{50mm}\times
V_E\ b_{\theta}^-
(\eta\Psi_B Q\Xi_D+Q\Xi_B \eta\Psi_D)\rangle_W
\nonumber\\
&\hspace{30mm}
+\langle (\eta\Psi_B Q\Xi_D+Q\Xi_B \eta\Psi_D) (\xi_{c} b_{c}^- b_{c}^+) 
\nonumber\\
&\hspace{50mm}\times
V_E\ b_{\theta}^- (\eta\Psi_A Q\Xi_C+Q\Xi_A \eta\Psi_C)\rangle_W
\Big),\\
 \mathcal{A}_{F^4B}^{(AD|E|BC)}
 =&\
\frac{\kappa^3}{4}\int d^2T_1\int d^2T_2\
\langle(\eta\Psi_A Q\Xi_D+Q\Xi_A \eta\Psi_D) (\xi_{c_1} b_{c_1}^- b_{c_1}^+)
\nonumber\\
&\hspace{50mm}\times
QV_E (b_{c_2}^- b_{c_2}^+) (\eta\Psi_B Q\Xi_C+Q\Xi_B \eta\Psi_C)\rangle_W
\nonumber\\
&
+\frac{\kappa^3}{8}
\int d^2T \oint d\theta\
\Big(
\langle (\eta\Psi_A Q\Xi_D+Q\Xi_A \eta\Psi_D) (\xi_{c} b_{c}^- b_{c}^+) 
\nonumber\\
&\hspace{50mm}\times
V_E\ b_{\theta}^- (\eta\Psi_B Q\Xi_C+Q\Xi_B \eta\Psi_C)\rangle_W
\nonumber\\
&\hspace{30mm}
+\langle (\eta\Psi_B Q\Xi_C+Q\Xi_B \eta\Psi_C) (\xi_{c} b_{c}^- b_{c}^+) 
\nonumber\\
&\hspace{50mm}\times
V_E\ b_{\theta}^- (\eta\Psi_A Q\Xi_D+Q\Xi_A \eta\Psi_D)\rangle_W
\Big).
\end{align}
Note that the external boson appears in the same form $QV_E$ in all the dominant 
contributions integrated over (a part of) the full moduli space.

There are ten, $_5C_2$, 1P diagrams classified two categories by whether 
the external boson is attached to the three-string vertex or the four-string vertex. 
It is enough to calculate only one of the contributions in each category, 
and the others can be obtained by relabeling the external
fermions. The amplitudes in the first category are given by
\begin{align}
 \mathcal{A}^{(AB|CDE)}_{F^4B}
 =&\
\frac{\kappa^3}{4}\int d^2T \int d^2\theta\
\langle(\eta\Psi_A Q\Xi_B+Q\Xi_A \eta\Psi_B) (\xi_{c} b_{c}^- b_{c}^+)
\nonumber\\
&\hspace{50mm}\times
(b_{C_1} b_{C_2}) (\eta\Psi_C Q\Xi_D+Q\Xi_C \eta\Psi_D) QV_E\rangle_W
\nonumber
\end{align}
\begin{equation}
 \nonumber
\end{equation}
\begin{align}
&
+\frac{\kappa^3}{8}\int d^2 T \oint d\theta\
\Big(
2\langle(\eta\Psi_A Q\Xi_B+Q\Xi_A \eta\Psi_B)
(\xi_{c} b_{c}^- b_{c}^+)
\nonumber\\
&\hspace{50mm}\times
(\eta\Psi_C\ b_{\theta}^- \Xi_D -\Xi_C\ b_{\theta}^- \eta\Psi_D) QV_E\rangle_W
\nonumber\\
&\hspace{30mm}
+2\langle(\eta\Psi_A Q\Xi_B+Q\Xi_A \eta\Psi_B)(\xi_{c} b_{c}^- b_{c}^+)
\nonumber\\
&\hspace{50mm}\times
(\eta\Psi_D\ b_{\theta}^- \Xi_C - \Xi_D\ b_{\theta}^- \eta\Psi_C) QV_E\rangle_W
\nonumber\\
&\hspace{30mm}
-\langle(\eta\Psi_A Q\Xi_B+Q\Xi_A \eta\Psi_B)(\xi_{c} b_{c}^- b_{c}^+)
\nonumber\\
&\hspace{50mm}\times
V_E\ b_{\theta}^- (\eta\Psi_C Q\Xi_D+Q\Xi_C \eta\Psi_D)\rangle_W
\Big),\\
 \mathcal{A}^{(AC|BDE)}_{F^4B}
 =&\
\frac{\kappa^3}{4}\int d^2T \int d^2\theta\
\langle(\eta\Psi_A Q\Xi_C+Q\Xi_A \eta\Psi_C) (\xi_{c} b_{c}^- b_{c}^+)
\nonumber\\
&\hspace{50mm}\times
(b_{C_1} b_{C_2}) (\eta\Psi_B Q\Xi_D+Q\Xi_B \eta\Psi_D) QV_E\rangle_W
\nonumber\\
&
+\frac{\kappa^3}{8}\int d^2 T \oint d\theta\
\Big(
2\langle(\eta\Psi_A Q\Xi_C+Q\Xi_A \eta\Psi_C)
(\xi_{c} b_{c}^- b_{c}^+)
\nonumber\\
&\hspace{50mm}\times
(\eta\Psi_B\ b_{\theta}^- \Xi_D -\Xi_B\ b_{\theta}^- \eta_D) QV_E\rangle_W
\nonumber\\
&\hspace{30mm}
+2\langle(\eta\Psi_A Q\Xi_C+Q\Xi_A \eta\Psi_C)
(\xi_{c} b_{c}^- b_{c}^+)
\nonumber\\
&\hspace{50mm}\times
(\eta\Psi_D\ b_{\theta}^- \Xi_B - \Xi_D\ b_{\theta}^- \eta\Psi_B) QV_E\rangle_W
\nonumber\\
&\hspace{30mm}
-\langle(\eta\Psi_A Q\Xi_C+Q\Xi_A \eta\Psi_C)(\xi_{c} b_{c}^- b_{c}^+)
\nonumber\\
&\hspace{50mm}\times
V_E\ b_{\theta}^- (\eta\Psi_B Q\Xi_D+Q\Xi_B \eta\Psi_D)\rangle_W
\Big),\\
 \mathcal{A}^{(AD|BCE)}_{F^4B}
 =&\
\frac{\kappa^3}{4}\int d^2T \int d^2\theta\
\langle(\eta\Psi_A Q\Xi_D+Q\Xi_A \eta\Psi_D) (\xi_{c} b_{c}^- b_{c}^+)
\nonumber\\
&\hspace{50mm}\times
(b_{C_1} b_{C_2}) (\eta\Psi_B Q\Xi_C+Q\Xi_B \eta\Psi_C) QV_E\rangle_W
\nonumber\\
&
+\frac{\kappa^3}{8}\int d^2 T \oint d\theta\
\Big(
2\langle(\eta\Psi_A Q\Xi_D+Q\Xi_A \eta\Psi_D)
(\xi_{c} b_{c}^- b_{c}^+)
\nonumber\\
&\hspace{50mm}\times
(\eta\Psi_B\ b_{\theta}^- \Xi_C -\Xi_B\ b_{\theta}^- \eta_C) QV_E\rangle_W
\nonumber\\
&\hspace{30mm}
+2\langle(\eta\Psi_A Q\Xi_D+Q\Xi_A \eta\Psi_D)
(\xi_{c} b_{c}^- b_{c}^+)
\nonumber\\
&\hspace{50mm}\times
(\eta\Psi_C\ b_{\theta}^- \Xi_B - \Xi_C\ b_{\theta}^- \eta\Psi_B) QV_E\rangle_W
\nonumber\\
&\hspace{30mm}
-\langle(\eta\Psi_A Q\Xi_D+Q\Xi_A \eta\Psi_D)(\xi_{c} b_{c}^- b_{c}^+)
\nonumber\\
&\hspace{50mm}\times
V_E\ b_{\theta}^- (\eta\Psi_B Q\Xi_C+Q\Xi_B \eta\Psi_C)\rangle_W
\Big),\\
 \mathcal{A}^{(BC|ADE)}_{F^4B}
 =&\
\frac{\kappa^3}{4}\int d^2T \int d^2\theta\
\langle(\eta\Psi_B Q\Xi_C+Q\Xi_B \eta\Psi_C) (\xi_{c} b_{c}^- b_{c}^+)
\nonumber\\
&\hspace{50mm}\times
(b_{C_1} b_{C_2}) (\eta\Psi_A Q\Xi_D+Q\Xi_A \eta\Psi_D) QV_E\rangle_W
\nonumber\\
&
+\frac{\kappa^3}{8}\int d^2 T \oint d\theta\
\Big(
2\langle(\eta\Psi_B Q\Xi_C+Q\Xi_B \eta\Psi_C)
(\xi_{c} b_{c}^- b_{c}^+)
\nonumber\\
&\hspace{50mm}\times
(\eta\Psi_A\ b_{\theta}^- \Xi_D -\Xi_A\ b_{\theta}^- \eta_D) QV_E\rangle_W
\nonumber\\
&\hspace{30mm}
+2\langle(\eta\Psi_B Q\Xi_C+Q\Xi_B \eta\Psi_C)
(\xi_{c} b_{c}^- b_{c}^+)
\nonumber\\
&\hspace{50mm}\times
(\eta\Psi_D\ b_{\theta}^- \Xi_A - \Xi_D\ b_{\theta}^- \eta\Psi_A) QV_E\rangle_W
\nonumber\\
&\hspace{30mm}
-\langle(\eta\Psi_B Q\Xi_C+Q\Xi_B \eta\Psi_C)(\xi_{c} b_{c}^- b_{c}^+)
\nonumber\\
&\hspace{50mm}\times
V_E\ b_{\theta}^- (\eta\Psi_A Q\Xi_D+Q\Xi_A \eta\Psi_D)\rangle_W
\Big),\\
 \mathcal{A}^{(BD|ACE)}_{F^4B}
=&\
\frac{\kappa^3}{4}\int d^2T \int d^2\theta\
\langle(\eta\Psi_B Q\Xi_D+Q\Xi_B \eta\Psi_D) (\xi_{c} b_{c}^- b_{c}^+)
\nonumber\\
&\hspace{50mm}\times
(b_{C_1} b_{C_2}) (\eta\Psi_A Q\Xi_C+Q\Xi_A \eta\Psi_C) QV_E\rangle_W
\nonumber\\
&
+\frac{\kappa^3}{8}\int d^2 T \oint d\theta\
\Big(
2\langle(\eta\Psi_B Q\Xi_D+Q\Xi_B \eta\Psi_D)
(\xi_{c} b_{c}^- b_{c}^+)
\nonumber\\
&\hspace{50mm}\times
(\eta\Psi_A\ b_{\theta}^- \Xi_C -\Xi_A\ b_{\theta}^- \eta_C) QV_E\rangle_W
\nonumber\\
&\hspace{30mm}
+2\langle(\eta\Psi_B Q\Xi_D+Q\Xi_B \eta\Psi_D)
(\xi_{c} b_{c}^- b_{c}^+)
\nonumber\\
&\hspace{50mm}\times
(\eta\Psi_C\ b_{\theta}^- \Xi_A - \Xi_C\ b_{\theta}^- \eta\Psi_A) QV_E\rangle_W
\nonumber\\
&\hspace{30mm}
-\langle(\eta\Psi_B Q\Xi_D+Q\Xi_B \eta\Psi_D)(\xi_{c} b_{c}^- b_{c}^+)
\nonumber\\
&\hspace{50mm}\times
V_E\ b_{\theta}^- (\eta\Psi_A Q\Xi_C+Q\Xi_A \eta\Psi_C)\rangle_W
\Big),\\
 \mathcal{A}^{(CD|ABE)}_{F^4B}
 =&\
\frac{\kappa^3}{4}\int d^2T \int d^2\theta\
\langle(\eta\Psi_C Q\Xi_D+Q\Xi_C \eta\Psi_D) (\xi_{c} b_{c}^- b_{c}^+)
\nonumber\\
&\hspace{50mm}\times
(b_{C_1} b_{C_2}) (\eta\Psi_A Q\Xi_B+Q\Xi_A \eta\Psi_B) QV_E\rangle_W
\nonumber\\
&
+\frac{\kappa^3}{8}\int d^2 T \oint d\theta\
\Big(
2\langle(\eta\Psi_C Q\Xi_D+Q\Xi_C \eta\Psi_D)
(\xi_{c} b_{c}^- b_{c}^+)
\nonumber\\
&\hspace{50mm}\times
(\eta\Psi_A\ b_{\theta}^- \Xi_B -\Xi_A\ b_{\theta}^- \eta_B) QV_E\rangle_W
\nonumber\\
&\hspace{30mm}
+2\langle(\eta\Psi_C Q\Xi_D+Q\Xi_C \eta\Psi_D)
(\xi_{c} b_{c}^- b_{c}^+)
\nonumber\\
&\hspace{50mm}\times
(\eta\Psi_B\ b_{\theta}^- \Xi_A - \Xi_B\ b_{\theta}^- \eta\Psi_A) QV_E\rangle_W
\nonumber\\
&\hspace{30mm}
-\langle(\eta\Psi_C Q\Xi_D+Q\Xi_C \eta\Psi_D)(\xi_{c} b_{c}^- b_{c}^+)
\nonumber\\
&\hspace{50mm}\times
V_E\ b_{\theta}^- (\eta\Psi_A Q\Xi_B+Q\Xi_A \eta\Psi_B)\rangle_W
\Big),
\end{align}
and those in the second category are
\begin{align}
 \mathcal{A}^{(AE|BCD)}_{F^4B}
 =&\
\frac{\kappa^3}{6}\int d^2T \int d^2\theta\
\Big(
\langle\eta\Psi_A QV_E (\xi_{c} b_{c}^- b_{c}^+) (b_{C_1}b_{C_2}) 
\nonumber\\
&\hspace{35mm}\times
\big(\eta\Psi_B Q\Xi_C Q\Xi_D+Q\Xi_B \eta\Psi_C Q\Xi_D
+Q\Xi_B Q \Xi_C \eta\Psi_D\big)\rangle_W
\nonumber\\
&\hspace{28mm}
+\langle Q\Xi_A QV_E (\xi_{c} b_{c}^- b_{c}^+) (b_{C_1}b_{C_2}) 
\nonumber\\
&\hspace{35mm}\times
\big(Q\Xi_B \eta\Psi_C \eta\Psi_D 
+\eta\Psi_B Q\Xi_C \eta\Psi_D 
+\eta\Psi_B \eta\Psi_C Q\Xi_D 
\big)\rangle_W\Big)
\nonumber
\end{align}
\begin{equation}
 \nonumber
\end{equation}
\begin{align}
&
+\frac{\kappa^3}{6}\int d^2T \oint d\theta\
\Big(
\langle \Xi_A QV_E (\xi_{c} b_{c}^- b_{c}^+)
\nonumber\\
&\hspace{33mm}\times
\big(
\eta\Psi_B\ b_{\theta}^- (\eta\Psi_C Q\Xi_D
+Q\Xi_C \eta\Psi_D)  
+Q\Xi_B\ b_{\theta}^- \eta\Psi_C \eta\Psi_D
\nonumber\\
&\hspace{35mm}
+\eta\Psi_C\ b_{\theta}^- (\eta\Psi_B Q\Xi_D
+Q\Xi_B \eta\Psi_D)
+Q\Xi_C\ b_{\theta}^- \eta\Psi_B \eta\Psi_D
\nonumber\\
&\hspace{35mm}
+\eta\Psi_D\ b_{\theta}^- (\eta\Psi_B Q\Xi_C
+Q\Xi_B \eta\Psi_C) 
+Q\Xi_D\ b_{\theta}^- \eta\Psi_B \eta\Psi_C
\big)\rangle_W
\Big)
\nonumber\\
&
-\frac{\kappa^3}{12}\int d^2\theta \oint d\theta\
\langle (b_{C_1} b_{C_2})\big(\Psi_B \eta\Psi_C Q\Xi_D+\Psi_B \eta\Psi_D Q\Xi_C
+\Psi_C \eta\Psi_D Q\Xi_B
\nonumber\\
&\hspace{20mm}
+\Psi_C \eta\Psi_B Q\Xi_D+\Psi_D \eta\Psi_B Q\Xi_C
+\Psi_D \eta\Psi_C Q\Xi_B\big)\ b_{\theta}^- \Xi_A QV_E\rangle_W,\\
 \mathcal{A}^{(BE|ACD)}_{F^4B}
 =&\
\frac{\kappa^3}{6}\int d^2T \int d^2\theta\
\Big(
\langle\eta\Psi_B QV_E (\xi_{c} b_{c}^- b_{c}^+) (b_{C_1}b_{C_2})
\nonumber\\
&\hspace{38mm}\times
\big(\eta\Psi_A Q\Xi_C Q\Xi_D+Q\Xi_A \eta\Psi_C Q\Xi_D
+Q\Xi_A Q\Xi_C \eta\Psi_D\big)\rangle_W
\nonumber\\
&\hspace{28mm}
+\langle Q\Xi_B QV_E (\xi_{c} b_{c}^- b_{c}^+) (b_{C_1}b_{C_2}) 
\nonumber\\
&\hspace{38mm}\times
\big(Q\Xi_A \eta\Psi_C \eta\Psi_D+\eta\Psi_A Q\Xi_C \eta\Psi_D
+\eta\Psi_A \eta\Psi_C Q\Xi_D\big)\rangle_W\Big)
\nonumber\\
&
+\frac{\kappa^3}{6}\int d^2T \oint d\theta\
\Big(
\langle \Xi_B QV_E (\xi_{c} b_{c}^- b_{c}^+)
\nonumber\\
&\hspace{33mm}\times
\big(
\eta\Psi_A\ b_{\theta}^- (\eta\Psi_C Q\Xi_D
+ Q\Xi_C \eta\Psi_D) 
+Q\Xi_A\ b_{\theta}^- \eta\Psi_C \eta\Psi_D
\nonumber\\
&\hspace{35mm}
+\eta\Psi_C\ b_{\theta}^- (\eta\Psi_A Q\Xi_D
+Q\Xi_A \eta\Psi_D)
+Q\Xi_C\ b_{\theta}^- \eta\Psi_A \eta\Psi_D
\nonumber\\
&\hspace{35mm}
+\eta\Psi_D\ b_{\theta}^- (\eta\Psi_A Q\Xi_C
+Q\Xi_A \eta\Psi_C) 
+Q\Xi_D\ b_{\theta}^- \eta\Psi_A \eta\Psi_C
\big)\rangle_W
\Big)
\nonumber\\
&
-\frac{\kappa^3}{12}\int d^2\theta \oint d\theta\
\langle (b_{C_1} b_{C_2})\big(\Psi_A \eta\Psi_C Q\Xi_D+\Psi_A \eta\Psi_D Q\Xi_C
+\Psi_C \eta\Psi_D Q\Xi_A
\nonumber\\
&\hspace{20mm}
+\Psi_C \eta\Psi_A Q\Xi_D+\Psi_D \eta\Psi_A Q\Xi_C
+\Psi_D \eta\Psi_C Q\Xi_A\big)\ b_{\theta}^- \Xi_B QV_E\rangle_W,\\
 \mathcal{A}^{(CE|ABD)}_{F^4B}
 =&\
\frac{\kappa^3}{6}\int d^2T \int d^2\theta\
\Big(
\langle\eta\Psi_C QV_E (\xi_{c} b_{c}^- b_{c}^+) (b_{C_1}b_{C_2})
\nonumber\\
&\hspace{38mm}\times
\big(\eta\Psi_A Q\Xi_B Q\Xi_D+Q\Xi_A \eta\Psi_B Q\Xi_D
+Q\Xi_A Q\Xi_B \eta\Psi_D\big)\rangle_W
\nonumber
\\
&\hspace{28mm}
+\langle Q\Xi_C QV_E (\xi_{c} b_{c}^- b_{c}^+) (b_{C_1}b_{C_2}) 
\nonumber\\
&\hspace{38mm}\times
\big(Q\Xi_A \eta\Psi_B \eta\Psi_D+\eta\Psi_A Q\Xi_B \eta\Psi_D
+\eta\Psi_A \eta\Psi_B Q\Xi_D\big)\rangle_W\Big)
\nonumber\\
&
+\frac{\kappa^3}{6}\int d^2T \oint d\theta\
\langle \Xi_C QV_E (\xi_{c} b_{c}^- b_{c}^+)
\nonumber\\
&\hspace{33mm}\times
\big(
\eta\Psi_A\ b_{\theta}^- (\eta\Psi_B Q\Xi_D
+Q\Xi_B \eta\Psi_D) 
+Q\Xi_A\ b_{\theta}^- \eta\Psi_B \eta\Psi_D
\nonumber\\
&\hspace{35mm}
+\eta\Psi_B\ b_{\theta}^- (\eta\Psi_A Q\Xi_D
+Q\Xi_A \eta\Psi_D)
+Q\Xi_B\ b_{\theta}^- \eta\Psi_A \eta\Psi_D
\nonumber\\
&\hspace{35mm}
+\eta\Psi_D\ b_{\theta}^- (\eta\Psi_A Q\Xi_B
+Q\Xi_A \eta\Psi_B) 
+Q\Xi_D\ b_{\theta}^- \eta\Psi_A \eta\Psi_B
\big)\rangle_W
\Big)
\nonumber
\end{align}
\begin{equation}
 \nonumber
\end{equation}
\begin{align}
&
-\frac{\kappa^3}{12}\int d^2\theta \oint d\theta
\langle (b_{C_1} b_{C_2})\big(\Psi_A \eta\Psi_B Q\Xi_D+\Psi_A \eta\Psi_D Q\Xi_B
+\Psi_B \eta\Psi_D Q\Xi_A
\nonumber\\
&\hspace{20mm}
+\Psi_B \eta\Psi_A Q\Xi_D+\Psi_D \eta\Psi_A Q\Xi_B
+\Psi_D \eta\Psi_B Q\Xi_A\big)\ b_{\theta}^- \Xi_C QV_E\rangle_W,\\
 \mathcal{A}^{(DE|ABC)}_{F^4B}
 =&\
\frac{\kappa^3}{6}\int d^2T \int d^2\theta\
\Big(
\langle\eta\Psi_D QV_E (\xi_{c} b_{c}^- b_{c}^+) (b_{C_1}b_{C_2})
\nonumber\\
&\hspace{38mm}\times
\big(\eta\Psi_A Q\Xi_B Q\Xi_C+Q\Xi_A \eta\Psi_B Q\Xi_C
+Q\Xi_A Q\Xi_B \eta\Psi_C\big)\rangle_W
\nonumber\\
&\hspace{28mm}
+\langle Q\Xi_D QV_E (\xi_{c} b_{c}^- b_{c}^+) (b_{C_1}b_{C_2}) 
\nonumber\\
&\hspace{38mm}\times
\big(Q\Xi_A \eta\Psi_B \eta\Psi_C+\eta\Psi_A Q\Xi_B \eta\Psi_C
+\eta\Psi_A \eta\Psi_B Q\Xi_C\big)\rangle_W\Big)
\nonumber\\
&
+\frac{\kappa^3}{6}\int d^2T \oint d\theta\
\langle \Xi_D QV_E (\xi_{c} b_{c}^- b_{c}^+)
\nonumber\\
&\hspace{33mm}\times
\big(
\eta\Psi_A\ b_{\theta}^- (\eta\Psi_B Q\Xi_C
+Q\Xi_B \eta\Psi_C)  
+Q\Xi_A\ b_{\theta}^- \eta\Psi_B \eta\Psi_C
\nonumber\\
&\hspace{33mm}
+\eta\Psi_B\ b_{\theta}^- (\eta\Psi_A Q\Xi_C
+ Q\Xi_A \eta\Psi_C)
+Q\Xi_B\ b_{\theta}^- \eta\Psi_A \eta\Psi_C
\nonumber\\
&\hspace{33mm}
+\eta\Psi_C\ b_{\theta}^- (\eta\Psi_A Q\Xi_B
+ Q\Xi_A \eta\Psi_B)  
+Q\Xi_C\ b_{\theta}^- \eta\Psi_A \eta\Psi_B
\big)\rangle_W
\Big)
\nonumber\\
&
-\frac{\kappa^3}{12}\int d^2\theta \oint d\theta\
\langle (b_{C_1} b_{C_2})\big(\Psi_A \eta\Psi_B Q\Xi_C+\Psi_A \eta\Psi_C Q\Xi_B
+\Psi_B \eta\Psi_C Q\Xi_A
\nonumber\\
&\hspace{20mm}
+\Psi_B \eta\Psi_A Q\Xi_C+\Psi_C \eta\Psi_A Q\Xi_B
+\Psi_C \eta\Psi_B Q\Xi_A\big)\ b_{\theta}^- \Xi_D QV_E\rangle_W.
\end{align}
The contributions from the last (NP) diagram can also be divided into 
two parts; the dominant part integrated over the whole moduli space and 
the boundary part coming from the first and the second
four-string vertices in (\ref{four-one}), respectively:
\begin{align}
 \mathcal{A}_{F^4B}^{(ABCDE)}
 =&\
\frac{\kappa^3}{6}\int d^4\theta\
\langle \xi (b_{C_1} b_{C_2} b_{C_3} b_{C_4}) 
\big(
\eta\Psi_A \eta\Psi_B Q\Xi_C Q\Xi_D
+\eta\Psi_A Q\Xi_B \eta\Psi_C Q\Xi_D
\nonumber\\
&\hspace{40mm}
+\eta\Psi_A Q\Xi_B Q\Xi_C \eta\Psi_D 
+Q\Xi_A \eta\Psi_B \eta\Psi_C Q\Xi_D 
\nonumber\\
&\hspace{40mm}
+Q\Xi_A \eta\Psi_B Q\Xi_C \eta\Psi_D 
+Q\Xi_A Q\Xi_B \eta\Psi_C \eta\Psi_D 
\big) QV_E\rangle_W
\nonumber\\
&
+\frac{\kappa^3}{12}\int d^2\theta \oint d\theta\
\Big(
\langle(b_{C_1}b_{C_2})
\big(\Psi_A \eta\Psi_B Q\Xi_C+\Psi_A \eta\Psi_C Q\Xi_B 
\nonumber\\
&\hspace{45mm}
+\Psi_B \eta\Psi_C Q\Xi_A
+\Psi_B \eta\Psi_A Q\Xi_C 
\nonumber\\
&\hspace{45mm}
+\Psi_C \eta\Psi_A Q\Xi_B+\Psi_C \eta\Psi_B Q\Xi_A\big)\
b_{\theta}^- \Xi_D QV_E\rangle_W
\nonumber\\
&\hspace{30mm}
+ \langle(b_{C_1}b_{C_2})\big(\Psi_B \eta\Psi_C Q\Xi_D+\Psi_B \eta\Psi_D Q\Xi_C 
\nonumber\\
&\hspace{45mm}
+\Psi_C \eta\Psi_D Q\Xi_B
+\Psi_C \eta\Psi_B Q\Xi_D 
\nonumber\\
&\hspace{45mm}
+\Psi_D \eta\Psi_B Q\Xi_C+\Psi_D \eta\Psi_C Q\Xi_B\big)\
b_{\theta}^- \Xi_A QV_E\rangle_W
\nonumber\\
&\hspace{30mm}
+\langle(b_{C_1}b_{C_2})\big(\Psi_A \eta\Psi_C Q\Xi_D+\Psi_A \eta\Psi_D Q\Xi_C 
\nonumber\\
&\hspace{45mm}
+\Psi_C \eta\Psi_D Q\Xi_A
+\Psi_C \eta\Psi_A Q\Xi_D 
\nonumber\\
&\hspace{45mm}
+\Psi_D \eta\Psi_A Q\Xi_C+\Psi_D \eta\Psi_C Q\Xi_A\big)\
b_{\theta}^- \Xi_B QV_E\rangle_W
\nonumber\\
&\hspace{30mm}
+\langle(b_{C_1}b_{C_2})\big( 
\Psi_A \eta\Psi_B Q\Xi_D
+\Psi_A \eta\Psi_D Q\Xi_B 
\nonumber\\
&\hspace{45mm}
+\Psi_B \eta\Psi_D Q\Xi_A
+\Psi_B \eta\Psi_A Q\Xi_D
\nonumber\\
&\hspace{45mm}
+\Psi_D \eta\Psi_A Q\Xi_B+\Psi_D \eta\Psi_B Q\Xi_A\big)\
b_{\theta}^- \Xi_C QV_E\rangle_W
\Big).
\end{align}
The total amplitude is obtained by summing up all these contributions.
Almost all the boundary contributions
are canceled, except for a small portion given by
\begin{align}
&-\frac{\kappa^3}{12}\int d^2T \oint d\theta\
\times 
\Big(
\langle \Xi_A QV_E (\xi_{c} b_{c}^- b_{c}^+) 
\nonumber\\
&\hspace{40mm}\times
\big(\eta\Psi_B\ b_{\theta}^- 
(\eta\Psi_C Q\Xi_D+Q\Xi_C \eta\Psi_D)-2Q\Xi_B\ b_{\theta}^- \eta\Psi_C \eta\Psi_D
\nonumber\\
&\hspace{42mm}
+\eta\Psi_C\ b_{\theta}^- 
(\eta\Psi_B Q\Xi_D+Q\Xi_B \eta\Psi_D)-2Q\Xi_C\ b_{\theta}^- \eta\Psi_B \eta\Psi_D
\nonumber\\
&\hspace{42mm}
+\eta\Psi_D\ b_{\theta}^- 
(\eta\Psi_B Q\Xi_C+Q\Xi_B \eta\Psi_C)-2Q\Xi_D\ b_{\theta}^- \eta\Psi_B \eta\Psi_C\big)\rangle_W
\nonumber\\
&\hspace{10mm}
+\langle \Xi_B QV_E (\xi_{c} b_{c}^- b_{c}^+) \big(\eta\Psi_A\ b_{\theta}^- 
(\eta\Psi_C Q\Xi_D+Q\Xi_C \eta\Psi_D)-2Q\Xi_A\ b_{\theta}^- \eta\Psi_C \eta\Psi_D
\nonumber\\
&\hspace{42mm}
+\eta\Psi_C\ b_{\theta}^- 
(\eta\Psi_A Q\Xi_D+Q\Xi_A \eta\Psi_D)-2Q\Xi_C\ b_{\theta}^- \eta\Psi_A \eta\Psi_D
\nonumber\\
&\hspace{42mm}
+\eta\Psi_D\ b_{\theta}^- 
(\eta\Psi_A Q\Xi_C+Q\Xi_A \eta\Psi_C)-2Q\Xi_D\ b_{\theta}^- \eta\Psi_A \eta\Psi_C\big)\rangle_W
\nonumber\\
&\hspace{10mm}
+\langle \Xi_C QV_E (\xi_{c} b_{c}^- b_{c}^+) \big(\eta\Psi_A\ b_{\theta}^- 
(\eta\Psi_B Q\Xi_D+Q\Xi_B \eta\Psi_D)-2Q\Xi_A\ b_{\theta}^- \eta\Psi_B \eta\Psi_D
\nonumber\\
&\hspace{42mm}
+\eta\Psi_B\ b_{\theta}^- 
(\eta\Psi_A Q\Xi_D+Q\Xi_A \eta\Psi_D)-2Q\Xi_B\ b_{\theta}^- \eta\Psi_A \eta\Psi_D
\nonumber\\
&\hspace{42mm}
+\eta\Psi_D\ b_{\theta}^- 
(\eta\Psi_A Q\Xi_B+Q\Xi_A \eta\Psi_B)-2Q\Xi_D\ b_{\theta}^- \eta\Psi_A \eta\Psi_B)\rangle_W
\nonumber\\
&\hspace{10mm}
+\langle \Xi_D QV_E (\xi_{c} b_{c}^- b_{c}^+) \big(\eta\Psi_A\ b_{\theta}^- 
(\eta\Psi_B Q\Xi_C+Q\Xi_B \eta\Psi_C)-2Q\Xi_A\ b_{\theta}^- \eta\Psi_B \eta\Psi_C
\nonumber\\
&\hspace{42mm}
+\eta\Psi_B\ b_{\theta}^- 
(\eta\Psi_A Q\Xi_C+Q\Xi_A \eta\Psi_C)-2Q\Xi_B\ b_{\theta}^- \eta\Psi_A \eta\Psi_C
\nonumber\\
&\hspace{42mm}
+
\eta\Psi_C\ b_{\theta}^- 
(\eta\Psi_A Q\Xi_B+Q\Xi_A \eta\Psi_B)-2Q\Xi_C\ b_{\theta}^- \eta\Psi_A \eta\Psi_B)\rangle_W
\Big),
\end{align} %
which vanishes if we impose the constraint $Q\Xi=\eta\Psi$.
In consequence, the total amplitude can be written as
the sum of the dominant contribution of each diagram,
which can be evaluated as the correlations in the small Hilbert space as
\begin{align}
\mathcal{A}_{F^4B}
=&\  \kappa^3 \int d^2T_1 d^2T_2\
\Big(
\llangle \eta\Psi_B \eta\Psi_C (b_{c_1}^- b_{c_1}^+) \eta\Psi_A 
(b_{c_2}^- b_{c_2}^+) \eta\Psi_D QV_E\rrangle 
+ \textrm{14\ terms}
\Big)
\nonumber\\
&
+\kappa^3 \int d^2T d^2\theta\
\Big(
\llangle \eta\Psi_A \eta\Psi_B (b_c^- b_c^+) (b_{C_1}b_{C_2})
\eta\Psi_C \eta\Psi_D QV_E \rrangle
+ \textrm{9\ terms}
\Big)
\nonumber\\
&
+\kappa^3 \int d^4\theta\
\llangle (b_{C_1} b_{C_2} b_{C_3} b_{C_4}) \eta\Psi_A \eta\Psi_B
\eta\Psi_C \eta\Psi_D QV_E \rrangle.
\end{align}
after imposing the constraint. The first, second and third lines
come from the 2P, 1P, and NP diagrams, respectively.
Each of these contributions has the same form as that in the bosonic 
closed string field theory if we identify the bosonic string fields with $\eta\Psi$ 
or $QV$. 
Hence the four-fermion-one-boson amplitude calculated by the new Feynman rules
agrees with the well-known amplitude in the first quantized formulation.

We can similarly calculate the two-fermion-three-boson, $F^2B^3$, amplitude. 
The 2P diagram $(BC|A|DE)$ is, for example, given by 
\begin{align}
 \mathcal{A}^{(BC|A|DE)}_{F^2B^3}
 =&\
\left(-\frac{\kappa}{2}\right)^2\ \frac{\kappa}{2}\
(-2) \int d^2T_1 \int d^2T_2\
\Big(
\langle \eta\Psi_B QV_C\ 
(\xi_{c_1} b_{c_1}^- b_{c_1}^+ \eta) 
\nonumber\\
&\hspace{50mm}\times
\Xi_A 
(Q \xi_{c_2} b_{c_2}^- b_{c_2}^+) (QV_D \eta V_E+\eta V_D QV_E)\rangle_W
\nonumber\\
&\hspace{35mm}
+\langle \Xi_B QV_C\ (\eta \xi_{c_1} b_{c_1}^- b_{c_1}^+) \eta\Psi_A 
\nonumber\\
&\hspace{50mm}\times
(Q \xi_{c_2} b_{c_2}^- b_{c_2}^+) (QV_D \eta V_E+\eta V_D QV_E)\rangle_W
\Big),
\end{align}
using the new Feynman rules.
We can move $Q$, by integrating by parts, so as to act on $\Xi$, 
and align the external bosons as $(QV_C, QV_D, \eta V_E)$,
which are uniquely realized by requiring not to exchange 
the order of $Q$ and $\xi$:
\begin{align}
 \mathcal{A}^{(BC|A|DE)}_{F^2B^3}
 =&\
-\frac{\kappa^3}{2}\int d^2T_1 \int d^2T_2\
\Big(
\langle \eta\Psi_B QV_C (\xi_{c_1} b_{c_1}^- b_{c_1}^+) Q\Xi_A 
(b_{c_2}^- b_{c_2}^+) QV_D \eta V_E\rangle_W
\nonumber\\
&\hspace{35mm}
+\langle Q\Xi_B QV_C (\xi_{c_1} b_{c_1}^- b_{c_1}^+) \eta\Psi_A 
(b_{c_2}^- b_{c_2}^+) QV_D \eta V_E\rangle_W
\Big)
\nonumber\\
&+\frac{\kappa^3}{4}\int d^2T \oint d\theta\ 
\Big(
\langle\big(\eta\Psi_B QV_C (b_{c}^- b_{c}^+) Q\Xi_A
\nonumber\\
&\hspace{60mm}
+ Q\Xi_B QV_C (b_{c}^- b_{c}^+) \eta\Psi_A\big)\ b_{\theta}^- V_D V_E\rangle_W
\nonumber\\
&\hspace{20mm}
+\langle \Xi_B QV_C (\xi_{c} b_{c}^- b_{c}^+) \eta\Psi_A\ b_{\theta}^-
(QV_D \eta V_E+\eta V_D QV_E)\rangle_W
\nonumber\\
&\hspace{20mm}
+\langle (QV_D \eta V_E+\eta V_D QV_E) 
(\xi_{c} b_{c}^- b_{c}^+)
\nonumber\\
&\hspace{50mm}\times
\big(\eta\Psi_A\ b_{\theta}^- \Xi_B-\Xi_A\ b_{\theta}^- \eta\Psi_B\big) QV_C\rangle_W
\nonumber\\
&\hspace{20mm}
+\langle V_D V_E (b_{c}^- b_{c}^+) 
\big(\eta\Psi_A\ b_{\theta}^- Q\Xi_B
+Q\Xi_A\ b_{\theta}^- \eta\Psi_B\big) QV_C\rangle_W
\Big).
\end{align}
According to this recipe,
the contributions from the other fourteen diagrams 
are similarly calculated as
\begin{align}
 \mathcal{A}^{(BD|A|CE)}_{F^2B^3}
 =&\
-\frac{\kappa^3}{2}\int d^2T_1 \int d^2T_2\
\Big(
\langle \eta\Psi_B QV_D (\xi_{c_1} b_{c_1}^- b_{c_1}^+) Q\Xi_A 
(b_{c_2}^- b_{c_2}^+) QV_C \eta V_E\rangle_W
\nonumber\\
&\hspace{35mm}
+\langle Q\Xi_B QV_D (\xi_{c_1} b_{c_1}^- b_{c_1}^+) \eta\Psi_A 
(b_{c_2}^- b_{c_2}^+) QV_C \eta V_E\rangle_W
\Big)
\nonumber\\
&+\frac{\kappa^3}{4}\int d^2T \oint d\theta\ 
\Big(
\langle\big(\eta\Psi_B QV_D (b_{c}^- b_{c}^+) Q\Xi_A
\nonumber\\
&\hspace{60mm}
+ Q\Xi_B QV_D (b_{c}^- b_{c}^+) \eta\Psi_A\big)\ b_{\theta}^- V_C V_E\rangle_W
\nonumber\\
&\hspace{30mm}
+\langle \Xi_B QV_D (\xi_{c} b_{c}^- b_{c}^+) \eta\Psi_A\ b_{\theta}^-
(QV_C \eta V_E+\eta V_C QV_E)\rangle_W
\nonumber\\
&\hspace{30mm}
+\langle (QV_C \eta V_E+\eta V_C QV_E) (\xi_{c} b_{c}^- b_{c}^+)
\nonumber\\
&\hspace{60mm}\times
\big(\eta\Psi_A\ b_{\theta}^- \Xi_B-\Xi_A\ b_{\theta}^- \eta\Psi_B\big) QV_D\rangle_W
\nonumber\\
&\hspace{28mm}
+\langle V_C V_E (b_{c}^- b_{c}^+) \big(\eta\Psi_A\ b_{\theta}^- Q\Xi_B
+Q\Xi_A\ b_{\theta}^- \eta\Psi_B\big) QV_D\rangle_W
\Big),\\
 \mathcal{A}^{(BE|A|CD)}_{F^2B^3}
 =&\
-\frac{\kappa^3}{2}\int d^2T_1 \int d^2T_2\
\Big(
\langle \eta\Psi_B \eta V_E (\xi_{c_1} b_{c_1}^- b_{c_1}^+)
Q\Xi_A\ (b_{c_2}^- b_{c_2}^+) QV_C QV_D\rangle_W
\nonumber\\
&\hspace{35mm} 
+\langle Q\Xi_B \eta V_E (\xi_{c_1} b_{c_1}^- b_{c_1}^+)
\eta\Psi_A\ (b_{c_2}^- b_{c_2}^+) QV_C QV_D\rangle_W
\Big)
\nonumber\\
&+\frac{\kappa^3}{4}\int d^2T \oint d\theta\ 
\Big(
\langle\big(\eta\Psi_B V_E (b_{c}^- b_{c}^+) Q\Xi_A
+ Q\Xi_B V_E (b_{c}^- b_{c}^+) \eta\Psi_A\big)\ 
\nonumber\\
&\hspace{80mm}\times
b_{\theta}^- 
(QV_C V_D+V_C QV_D)\rangle_W
\nonumber\\
&\hspace{30mm}
+\langle \Xi_B QV_E (\xi_{c} b_{c}^- b_{c}^+) \eta\Psi_A\ b_{\theta}^-
(QV_C \eta V_D+\eta V_C QV_D)\rangle_W
\nonumber\\
&\hspace{30mm}
+\langle (QV_C \eta V_D+\eta V_C QV_D) (\xi_{c} b_{c}^- b_{c}^+)
\nonumber\\
&\hspace{55mm}\times
\big(\eta\Psi_A\ b_{\theta}^- \Xi_B-\Xi_A\ b_{\theta}^- \eta\Psi_B\big) QV_E\rangle_W
\nonumber\\
&\hspace{30mm}
-\langle (QV_C V_D + V_C QV_D) (b_{c}^- b_{c}^+) 
\nonumber\\
&\hspace{55mm}\times
\big(\eta\Psi_A\ b_{\theta}^- Q\Xi_B
+Q\Xi_A\ b_{\theta}^- \eta\Psi_B\big) V_E\rangle_W
\Big),\\
 \mathcal{A}^{(AC|B|DE)}_{F^2B^3}
 =&\
-\frac{\kappa^3}{2}\int d^2T_1 \int d^2T_2\
\Big(
\langle \eta\Psi_A QV_C (\xi_{c_1} b_{c_1}^- b_{c_1}^+) Q\Xi_B 
(b_{c_2}^- b_{c_2}^+) QV_D \eta V_E\rangle_W
\nonumber\\
&\hspace{35mm}
+\langle Q\Xi_A QV_C (\xi_{c_1} b_{c_1}^- b_{c_1}^+) \eta\Psi_B 
(b_{c_2}^- b_{c_2}^+) QV_D \eta V_E\rangle_W
\Big)
\nonumber\\
&+\frac{\kappa^3}{4}\int d^2T \oint d\theta\ 
\Big(
\langle\big(\eta\Psi_A QV_C (b_{c}^- b_{c}^+) Q\Xi_B
\nonumber\\
&\hspace{60mm}
+ Q\Xi_A QV_C (b_{c}^- b_{c}^+) \eta\Psi_B\big)\ b_{\theta}^- V_D V_E\rangle_W
\nonumber\\
&\hspace{20mm}
+\langle \Xi_A QV_C (\xi_{c} b_{c}^- b_{c}^+) \eta\Psi_B\ b_{\theta}^-
(QV_D \eta V_E+\eta V_D QV_E)\rangle_W
\nonumber\\
&\hspace{20mm}
+\langle (QV_D \eta V_E+\eta V_D QV_E) (\xi_{c} b_{c}^- b_{c}^+)
\nonumber\\
&\hspace{50mm}\times
\big(\eta\Psi_B\ b_{\theta}^- \Xi_A-\Xi_B\ b_{\theta}^- \eta\Psi_A\big) QV_C\rangle_W
\nonumber\\
&\hspace{20mm}
+\langle V_D V_E (b_{c}^- b_{c}^+) 
\big(\eta\Psi_B\ b_{\theta}^- Q\Xi_A
+Q\Xi_B\ b_{\theta}^- \eta\Psi_A\big) QV_C\rangle_W
\Big),\\
 \mathcal{A}^{(AD|B|CE)}_{F^2B^3}
 =&\
-\frac{\kappa^3}{2}\int d^2T_1 \int d^2T_2\
\Big(
\langle \eta\Psi_A QV_D (\xi_{c_1} b_{c_1}^- b_{c_1}^+) Q\Xi_B 
(b_{c_2}^- b_{c_2}^+) QV_C \eta V_E\rangle_W
\nonumber\\
&\hspace{35mm}
+\langle Q\Xi_A QV_D (\xi_{c_1} b_{c_1}^- b_{c_1}^+) \eta\Psi_B 
(b_{c_2}^- b_{c_2}^+) QV_C \eta V_E\rangle_W
\Big)
\nonumber\\
&+\frac{\kappa^3}{4}\int d^2T \oint d\theta\ 
\Big(
\langle\big(\eta\Psi_A QV_D (b_{c}^- b_{c}^+) Q\Xi_B
\nonumber\\
&\hspace{60mm}
+ Q\Xi_A QV_D (b_{c}^- b_{c}^+) \eta\Psi_B\big)\ b_{\theta}^- V_C V_E\rangle_W
\nonumber\\
&\hspace{20mm}
+\langle \Xi_A QV_D (\xi_{c} b_{c}^- b_{c}^+) \eta\Psi_B\ b_{\theta}^-
(QV_C \eta V_E+\eta V_C QV_E)\rangle_W
\nonumber\\
&\hspace{20mm}
+\langle (QV_C \eta V_E+\eta V_C QV_E) 
\nonumber\\
&\hspace{50mm}\times
(\xi_{c} b_{c}^- b_{c}^+)
\big(\eta\Psi_B\ b_{\theta}^- \Xi_A-\Xi_B\ b_{\theta}^- \eta\Psi_A\big) QV_D\rangle_W
\nonumber\\
&\hspace{20mm}
+\langle V_C V_E (b_{c}^- b_{c}^+) 
\big(\eta\Psi_B\ b_{\theta}^- Q\Xi_A
+Q\Xi_B\ b_{\theta}^- \eta\Psi_A\big) QV_D\rangle_W
\Big),\\
 \mathcal{A}^{(AE|B|CD)}_{F^2B^3}
 =&\
-\frac{\kappa^3}{2}\int d^2T_1 \int d^2T_2\
\Big(
\langle \eta\Psi_A \eta V_E (\xi_{c_1} b_{c_1}^- b_{c_1}^+)
Q\Xi_B\ (b_{c_2}^- b_{c_2}^+) QV_C QV_D\rangle_W
\nonumber\\
&\hspace{35mm} 
+\langle Q\Xi_A \eta V_E (\xi_{c_1} b_{c_1}^- b_{c_1}^+)
\eta\Psi_B\ (b_{c_2}^- b_{c_2}^+) QV_C QV_D\rangle_W
\Big)
\nonumber\\
&+\frac{\kappa^3}{4}\int d^2T \oint d\theta\ 
\Big(
\langle\big(\eta\Psi_A V_E (b_{c}^- b_{c}^+) Q\Xi_B
+ Q\Xi_A V_E (b_{c}^- b_{c}^+) \eta\Psi_B\big)\ 
\nonumber\\
&\hspace{80mm}\times
b_{\theta}^- 
(QV_C V_D+V_C QV_D)\rangle_W
\nonumber\\
&\hspace{30mm}
+\langle \Xi_A QV_E (\xi_{c} b_{c}^- b_{c}^+) \eta\Psi_B\ b_{\theta}^-
(QV_C \eta V_D+\eta V_C QV_D)\rangle_W
\nonumber\\
&\hspace{30mm}
+\langle (QV_C \eta V_D+\eta V_C QV_D) 
(\xi_{c} b_{c}^- b_{c}^+)
\nonumber\\
&\hspace{50mm}\times
\big(\eta\Psi_B\ b_{\theta}^- \Xi_A-\Xi_B\ b_{\theta}^- \eta\Psi_A\big) QV_E\rangle_W
\nonumber\\
&\hspace{30mm}
-\langle (QV_C V_D + V_C QV_D) (b_{c}^- b_{c}^+) 
\nonumber\\
&\hspace{50mm}\times
\big(\eta\Psi_B\ b_{\theta}^- Q\Xi_A
+Q\Xi_B\ b_{\theta}^- \eta\Psi_A\big) V_E\rangle_W
\Big),\\
 \mathcal{A}^{(AB|C|DE)}_{F^2B^3}
 =&\
-\frac{\kappa^3}{2}\int d^2T_1 \int d^2T_2\
\langle(\eta\Psi_A Q\Xi_B+Q\Xi_A \eta\Psi_B) 
\nonumber\\
&\hspace{45mm}\times
(\xi_{c_1} b_{c_1}^- b_{c_1}^+) 
QV_C (b_{c_2}^- b_{c_2}^+) QV_D \eta V_E\rangle_W
\nonumber\\
&+\frac{\kappa^3}{8}\int d^2T \oint d\theta\
\Big(
2\langle (\eta\Psi_A Q\Xi_B+Q\Xi_A \eta\Psi_B) (b_{c}^- b_{c}^+)
QV_C\ b_{\theta}^- V_D V_E\rangle_W
\nonumber\\
&\hspace{30mm}
-\langle (\eta\Psi_A Q\Xi_B+Q\Xi_A \eta\Psi_B) (\xi_{c} b_{c}^- b_{c}^+)
\nonumber\\
&\hspace{55mm}\times
V_C\ b_{\theta}^- (QV_D \eta V_E+\eta V_D QV_E)\rangle_W
\nonumber\\
&\hspace{30mm}
-\langle (QV_D \eta V_E+\eta V_D QV_E) (\xi_{c} b_{c}^- b_{c}^+)
\nonumber\\
&\hspace{55mm}\times
V_C\ b_{\theta}^- (\eta\Psi_A Q\Xi_B+Q\Xi_A \eta\Psi_B)\rangle_W
\nonumber\\
&\hspace{30mm}
+2\langle V_D V_E (b_{c}^- b_{c}^+) QV_C\ b_{\theta}^-
(\eta\Psi_A Q\Xi_B+Q\Xi_A \eta\Psi_B)\rangle_W
 \Big),\\
 \mathcal{A}^{(AD|C|BE)}_{F^2B^3}
 =&\
-\frac{\kappa^3}{2}\int d^2T_1 \int d^2T_2\
\Big(
\langle \eta\Psi_A QV_D (\xi_{c_1} b_{c_1}^- b_{c_1}^+) QV_C 
(b_{c_2}^- b_{c_2}^+) Q\Xi_B \eta V_E\rangle_W
\nonumber\\
&\hspace{35mm}
+\langle Q\Xi_A QV_D (\xi_{c_1} b_{c_1}^- b_{c_1}^+) QV_C 
(b_{c_2}^- b_{c_2}^+) \eta\Psi_B \eta V_E\rangle_W
\Big)
\nonumber\\
&+\frac{\kappa^3}{2}\int d^2T \oint d\theta\
\Big(
\langle \eta\Psi_A QV_D (b_{c}^- b_{c}^+) QV_C\ b_{\theta}^-
\Xi_B V_E\rangle_W 
\nonumber\\
&\hspace{35mm}
-\langle \Xi_A QV_D (b_{c}^- b_{c}^+) QV_C\ b_{\theta}^-
\eta\Psi_B V_E\rangle_W 
\nonumber\\
&\hspace{35mm}
+\langle \eta\Psi_B V_E (b_{c}^- b_{c}^+) QV_C\ b_{\theta}^-
\Xi_A QV_D\rangle_W 
\nonumber\\
&\hspace{35mm}
+\langle \Xi_B V_E (b_{c}^- b_{c}^+) QV_C\ b_{\theta}^-
\eta\Psi_A QV_D\rangle_W 
\Big),\\
 \mathcal{A}^{(AE|C|BD)}_{F^2B^3}
 =&\
-\frac{\kappa^3}{2}\int d^2T_1 \int d^2T_2\
\Big(
\langle \eta\Psi_A \eta V_E (\xi_{c_1} b_{c_1}^- b_{c_1}^+) QV_C 
(b_{c_2}^- b_{c_2}^+) Q\Xi_B QV_D\rangle_W
\nonumber\\
&\hspace{35mm}
+\langle Q\Xi_A \eta V_E (\xi_{c_1} b_{c_1}^- b_{c_1}^+) QV_C 
(b_{c_2}^- b_{c_2}^+) \eta\Psi_B QV_D\rangle_W
\Big)
\nonumber\\
&+\frac{\kappa^3}{2}\int d^2T \oint d\theta\
\Big(
\langle \eta\Psi_A V_E (b_{c}^- b_{c}^+) QV_C\ b_{\theta}^-
\Xi_B QV_D\rangle_W 
\nonumber\\
&\hspace{35mm}
+\langle \Xi_A V_E (b_{c}^- b_{c}^+) QV_C\ b_{\theta}^-
\eta\Psi_B QV_D\rangle_W 
\nonumber\\
&\hspace{35mm}
+\langle \eta\Psi_B QV_D (b_{c}^- b_{c}^+) QV_C\ b_{\theta}^-
\Xi_A V_E\rangle_W 
\nonumber\\
&\hspace{35mm}
-\langle \Xi_B QV_D (b_{c}^- b_{c}^+) QV_C\ b_{\theta}^-
\eta\Psi_A V_E\rangle_W 
\Big),\\
 \mathcal{A}^{(AB|D|CE)}_{F^2B^3}
 =&\
-\frac{\kappa^3}{2}\int d^2T_1 \int d^2T_2\
\langle(\eta\Psi_A Q\Xi_B+Q\Xi_A \eta\Psi_B) 
\nonumber\\
&\hspace{45mm}\times
(\xi_{c_1} b_{c_1}^- b_{c_1}^+) 
QV_D (b_{c_2}^- b_{c_2}^+) QV_C \eta V_E\rangle_W
\nonumber\\
&+\frac{\kappa^3}{8}\int d^2T \oint d\theta\
\Big(
2\langle (\eta\Psi_A Q\Xi_B+Q\Xi_A \eta\Psi_B) (b_{c}^- b_{c}^+)
QV_D\ b_{\theta}^- V_C V_E\rangle_W
\nonumber\\
&\hspace{30mm}
-\langle (\eta\Psi_A Q\Xi_B+Q\Xi_A \eta\Psi_B) (\xi_{c} b_{c}^- b_{c}^+)
\nonumber\\
&\hspace{50mm}\times 
V_D\ b_{\theta}^- (QV_C \eta V_E+\eta V_C QV_E)\rangle_W
\nonumber\\
&\hspace{30mm}
-\langle (QV_C \eta V_E+\eta V_C QV_E) (\xi_{c} b_{c}^- b_{c}^+)
\nonumber\\
&\hspace{50mm}\times
V_D\ b_{\theta}^- (\eta\Psi_A Q\Xi_B+Q\Xi_A \eta\Psi_B)\rangle_W
\nonumber\\
&\hspace{30mm}
+2\langle V_C V_E (b_{c}^- b_{c}^+) QV_D\ b_{\theta}^-
(\eta\Psi_A Q\Xi_B+Q\Xi_A \eta\Psi_B)\rangle_W
 \Big),\\
 \mathcal{A}^{(AC|D|BE)}_{F^2B^3}
 =&\
-\frac{\kappa^3}{2}\int d^2T_1 \int d^2T_2\ \Big(
\langle \eta\Psi_A QV_C (\xi_{c_1} b_{c_1}^- b_{c_1}^+) QV_D 
(b_{c_2}^- b_{c_2}^+) Q\Xi_B \eta V_E\rangle_W
\nonumber\\
&\hspace{35mm}
+\langle Q\Xi_A QV_C (\xi_{c_1} b_{c_1}^- b_{c_1}^+) QV_D 
(b_{c_2}^- b_{c_2}^+) \eta\Psi_B \eta V_E\rangle_W
\Big)
\nonumber\\
&+\frac{\kappa^3}{2}\int d^2T \oint d\theta\
\Big(
\langle \eta\Psi_A QV_C (b_{c}^- b_{c}^+) QV_D\ b_{\theta}^-
\Xi_B V_E\rangle_W 
\nonumber\\
&\hspace{35mm}
-\langle \Xi_A QV_C (b_{c}^- b_{c}^+) QV_D\ b_{\theta}^-
\eta\Psi_B V_E\rangle_W 
\nonumber\\
&\hspace{35mm}
+\langle \eta\Psi_B V_E (b_{c}^- b_{c}^+) QV_D\ b_{\theta}^-
\Xi_A QV_C\rangle_W 
\nonumber\\
&\hspace{35mm}
+\langle \Xi_B V_E (b_{c}^- b_{c}^+) QV_D\ b_{\theta}^-
\eta\Psi_A QV_C\rangle_W 
\Big),\\
 \mathcal{A}^{(AE|D|BC)}_{F^2B^3}
 =&\
-\frac{\kappa^3}{2}\int d^2T_1 \int d^2T_2\ \Big(
\langle \eta\Psi_A \eta V_E (\xi_{c_1} b_{c_1}^- b_{c_1}^+) QV_D 
(b_{c_2}^- b_{c_2}^+) Q\Xi_B QV_C\rangle_W
\nonumber\\
&\hspace{35mm}
+\langle Q\Xi_A \eta V_E (\xi_{c_1} b_{c_1}^- b_{c_1}^+) QV_D 
(b_{c_2}^- b_{c_2}^+) \eta\Psi_B QV_C\rangle_W
\Big)
\nonumber\\
&+\frac{\kappa^3}{2}\int d^2T \oint d\theta\
\Big(
\langle \eta\Psi_A V_E (b_{c}^- b_{c}^+) QV_D\ b_{\theta}^-
\Xi_B QV_C\rangle_W 
\nonumber\\
&\hspace{35mm}
+\langle \Xi_A V_E (b_{c}^- b_{c}^+) QV_D\ b_{\theta}^-
\eta\Psi_B QV_C\rangle_W 
\nonumber\\
&\hspace{35mm}
+\langle \eta\Psi_B QV_C (b_{c}^- b_{c}^+) QV_D\ b_{\theta}^-
\Xi_A V_E\rangle_W 
\nonumber\\
&\hspace{35mm}
-\langle \Xi_B QV_C (b_{c}^- b_{c}^+) QV_D\ b_{\theta}^-
\eta\Psi_A V_E\rangle_W 
\Big),
\end{align}
\begin{equation}
 \nonumber
\end{equation}
\begin{align}
 \mathcal{A}^{(AB|E|CD)}_{F^2B^3}
 =&\
-\frac{\kappa^3}{2}\int d^2T_1 \int d^2T_2\
\langle(\eta\Psi_A Q\Xi_B+Q\Xi_A \eta\Psi_B) 
\nonumber\\
&\hspace{60mm}\times
(\xi_{c_1} b_{c_1}^- b_{c_1}^+) 
\eta V_E (b_{c_2}^- b_{c_2}^+) QV_C QV_D\rangle_W
\nonumber\\
&
+\frac{\kappa^3}{8}\int d^2T \oint d\theta\
\Big(
2\langle (\eta\Psi_A Q\Xi_B+Q\Xi_A \eta\Psi_B) (b_{c}^- b_{c}^+)
\nonumber\\
&\hspace{60mm}\times
V_E\ b_{\theta}^- (QV_C V_D+V_C QV_D)\rangle_W
\nonumber\\
&\hspace{30mm}
-\langle (\eta\Psi_A Q\Xi_B+Q\Xi_A \eta\Psi_B) (\xi_{c} b_{c}^- b_{c}^+)
\nonumber\\
&\hspace{60mm}\times
V_E\ b_{\theta}^- (QV_C \eta V_D+\eta V_C QV_D)\rangle_W
\nonumber\\
&\hspace{30mm}
-\langle (QV_C \eta V_D+\eta V_C QV_D) (\xi_{c} b_{c}^- b_{c}^+)
\nonumber\\
&\hspace{60mm}\times
V_E\ b_{\theta}^- (\eta\Psi_A Q\Xi_B+Q\Xi_A \eta\Psi_B)\rangle_W
\nonumber\\
&\hspace{30mm}
+2\langle (QV_C V_D+V_C QV_D) (b_{c}^- b_{c}^+) 
\nonumber\\
&\hspace{60mm}\times
V_E\ 
b_{\theta}^- (\eta\Psi_A Q\Xi_B+Q\Xi_A \eta\Psi_B)\rangle_W
 \Big),\\
 \mathcal{A}^{(AC|E|BD)}_{F^2B^3}
 =&\
-\frac{\kappa^3}{2}\int d^2T_1 \int d^2T_2\ \Big(
\langle \eta\Psi_A QV_C (\xi_{c_1} b_{c_1}^- b_{c_1}^+) \eta V_E 
(b_{c_2}^- b_{c_2}^+) Q\Xi_B QV_D\rangle_W
\nonumber\\
&\hspace{35mm}
+\langle Q\Xi_A QV_C (\xi_{c_1} b_{c_1}^- b_{c_1}^+) \eta V_E 
(b_{c_2}^- b_{c_2}^+) \eta\Psi_B QV_D\rangle_W
\Big)
\nonumber\\
&+\frac{\kappa^3}{2}\int d^2T \oint d\theta\ \Big(
\langle\eta\Psi_A QV_C (b_{c}^- b_{c}^+) 
V_E\ b_{\theta}^- \Xi_B QV_D\rangle_W
\nonumber\\
&\hspace{35mm}
+\langle \Xi_A QV_C (b_{c}^- b_{c}^+)
V_E\ b_{\theta}^- \eta\Psi_B QV_D\rangle_W
\nonumber\\
&\hspace{35mm}
+\langle \eta\Psi_B QV_D\ (b_{c}^- b_{c}^+) V_E\ 
b_{\theta}^- \Xi_A QV_C\rangle_W
\nonumber\\
&\hspace{35mm}
+\langle \Xi_B QV_D (b_{c}^- b_{c}^+)
V_E\ b_{\theta}^- \eta\Psi_A QV_C\rangle_W
\Big),\\
 \mathcal{A}^{(BC|E|AD)}_{F^2B^3}
 =&\
-\frac{\kappa^3}{2}\int d^2T_1 \int d^2T_2\ \Big(
\langle \eta\Psi_B QV_C (\xi_{c_1} b_{c_1}^- b_{c_1}^+) \eta V_E 
(b_{c_2}^- b_{c_2}^+) Q\Xi_A QV_D \rangle_W
\nonumber\\
&\hspace{35mm}
+\langle Q\Xi_B QV_C (\xi_{c_1} b_{c_1}^- b_{c_1}^+) \eta V_E 
(b_{c_2}^- b_{c_2}^+) \eta\Psi_A QV_D \rangle_W
\Big)
\nonumber\\
&+\frac{\kappa^3}{2}\int d^2T \oint d\theta\ \Big(
\langle\eta\Psi_B QV_C (b_{c}^- b_{c}^+) 
V_E\ b_{\theta}^- \Xi_A QV_D\rangle_W
\nonumber\\
&\hspace{35mm}
+\langle \Xi_B QV_C (b_{c}^- b_{c}^+)
V_E\ b_{\theta}^- \eta\Psi_A QV_D\rangle_W
\nonumber\\
&\hspace{35mm}
+\langle \eta\Psi_A QV_D\ (b_{c}^- b_{c}^+) V_E\ 
b_{\theta}^- \Xi_B QV_C\rangle_W
\nonumber\\
&\hspace{35mm}
+\langle \Xi_A QV_D (b_{c}^- b_{c}^+)
V_E\ b_{\theta}^- \eta\Psi_B QV_C\rangle_W
\Big).
\end{align}
The contributions from the 1P diagrams are also calculated in the same manner, 
for example:
\begin{align}
 \mathcal{A}_{F^2B^3}^{(AB|CDE)}
 =&\
-\frac{\kappa^3}{2}\int d^2T \int d^2\theta\
\langle(\eta\Psi_A Q\Xi_B+Q\Xi_A \eta\Psi_B) 
\nonumber\\
&\hspace{45mm}\times
(\xi_{c} b_{c}^- b_{c}^+) 
(b_{C_1} b_{C_2}) QV_C QV_D \eta V_E\rangle_W
\nonumber\\
&
-\frac{\kappa^3}{8}\int d^2T \oint d\theta\
\Big(
2\langle (\eta\Psi_A Q\Xi_B+Q\Xi_A \eta\Psi_B) (b_{c}^- b_{c}^+)
QV_C\ b_{\theta}^- V_D V_E\rangle_W
\nonumber\\
&\hspace{30mm}
-\langle (\eta\Psi_A Q\Xi_B+Q\Xi_A \eta\Psi_B) (\xi_{c} b_{c}^- b_{c}^+)
\nonumber\\
&\hspace{60mm}\times
V_C\ b_{\theta}^- (QV_D \eta V_E+\eta V_D QV_E)\rangle_W
\nonumber\\
&\hspace{30mm}
+2\langle (\eta\Psi_A Q\Xi_B+Q\Xi_A \eta\Psi_B) (b_{c}^- b_{c}^+)
QV_D\ b_{\theta}^- V_C V_E\rangle_W
\nonumber\\
&\hspace{30mm}
-\langle (\eta\Psi_A Q\Xi_B+Q\Xi_A \eta\Psi_B) (\xi_{c} b_{c}^- b_{c}^+)
\nonumber\\
&\hspace{60mm}\times
V_D\ b_{\theta}^- (QV_C \eta V_E+\eta V_C QV_E)\big)\rangle_W
\nonumber\\
&\hspace{30mm}
+2\langle (\eta\Psi_A Q\Xi_B+Q\Xi_A \eta\Psi_B) (b_{c}^- b_{c}^+)
\nonumber\\
&\hspace{60mm}\times
V_E\ b_{\theta}^- (QV_C V_D+V_C QV_D))\rangle_W
\nonumber\\
&\hspace{30mm}
-\langle (\eta\Psi_A Q\Xi_B+Q\Xi_A \eta\Psi_B) (\xi_{c} b_{c}^- b_{c}^+)
\nonumber\\
&\hspace{60mm}\times
V_E\ b_{\theta}^- (QV_C \eta V_D+\eta V_C QV_D)\rangle_W\Big)
\nonumber\\
&
+\frac{\kappa^3}{6}\int d^2\theta \oint d\theta\
\langle(b_{C_1} b_{C_2}) (QV_C V_D+V_C QV_D) 
\nonumber\\
&\hspace{60mm}\times
V_E\ b_{\theta}^-
(\eta\Psi_A Q\Xi_B+Q\Xi_A \eta\Psi_B)\rangle_W,
\nonumber\\
&
+\frac{\kappa^3}{12}\oint d\theta \oint d\theta'\
\langle (\eta\Psi_A Q\Xi_B+Q\Xi_A \eta\Psi_B)
\nonumber\\
&\hspace{60mm}\times 
b_{\theta}^- 
\big(V_C\ b_{\theta'}^- V_D +V_D\ b_{\theta'}^- V_C\big) V_E\rangle_W.
\end{align}
The external bosons in the dominant contribution, 
the first term, are again aligned as $(QV_C,QV_D,\eta V_E)$. 
The contributions from the other nine 1P diagrams are also 
calculated as
\begin{align}
 \mathcal{A}_{F^2B^3}^{(AC|BDE)}
 =&\
-\frac{\kappa^3}{2}\int d^2T \int d^2\theta\
\Big(
\langle \eta\Psi_A QV_C (\xi_{c} b_{c}^- b_{c}^+) 
(b_{C_1} b_{C_2}) Q\Xi_B QV_D \eta V_E\rangle_W
\nonumber\\
&\hspace{35mm}
+\langle Q\Xi_A QV_C (\xi_{c} b_{c}^- b_{c}^+) 
(b_{C_1} b_{C_2}) \eta\Psi_B QV_D \eta V_E\rangle_W 
\Big)
\nonumber\\
&
-\frac{\kappa^3}{4}\int d^2T \oint d\theta\
\Big(
\langle 
\big(\eta\Psi_A QV_C (b_{c}^- b_{c}^+) Q\Xi_B
+Q\Xi_A QV_C (b_{c}^- b_{c}^+) \eta\Psi_B)\ b_{\theta}^- V_D V_E\rangle_W
\nonumber\\
&\hspace{35mm}
+\langle \Xi_A QV_C (\xi_{c} b_{c}^- b_{c}^+) \eta\Psi_B\ b_{\theta}^-
(QV_D \eta V_E+\eta V_D QV_E)\rangle_W
\nonumber\\
&\hspace{35mm}
+2\langle \eta\Psi_A QV_C (b_{c}^- b_{c}^+)
QV_D\ b_{\theta}^- \Xi_B V_E\rangle_W
\nonumber\\
&\hspace{35mm}
-2\langle \Xi_A QV_C (b_{c}^- b_{c}^+)
QV_D\ b_{\theta}^- \eta\Psi_B V_E\rangle_W
\nonumber\\
&\hspace{35mm}
+2\langle \eta\Psi_A QV_C (b_{c}^- b_{c}^+) V_E\ b_{\theta}^- \Xi_B QV_D\rangle_W
\nonumber\\
&\hspace{35mm}
+2\langle \Xi_A QV_C (b_{c}^- b_{c}^+)  V_E\ b_{\theta}^- \eta\Psi_B QV_D\rangle_W
\Big)
\nonumber\\
&
+\frac{\kappa^3}{2}\int d^2\theta \oint d\theta\
\Big(
\langle(b_{C_1} b_{C_2}) \eta\Psi_B QV_D V_E\ b_{\theta}^-
\Xi_A QV_C\rangle_W 
\nonumber\\
&\hspace{35mm}
+\langle(b_{C_1} b_{C_2}) \Xi_B QV_D V_E\ b_{\theta}^-
\eta\Psi_A QV_C\rangle_W 
\Big)
\nonumber\\
&
+\frac{\kappa^3}{4}\oint d\theta \oint d\theta'\
\langle \big(\eta\Psi_A QV_C\ b_{\theta}^- \Xi_B
- \Xi_A QV_C\ b_{\theta}^- \eta\Psi_B\big)\ b_{\theta'}^- V_D V_E\rangle_W,\\
 \mathcal{A}_{F^2B^3}^{(AD|BCE)}
 =&\
-\frac{\kappa^3}{2}\int d^2T \int d^2\theta\
\Big(
\langle \eta\Psi_A QV_D (\xi_{c} b_{c}^- b_{c}^+) 
(b_{C_1} b_{C_2}) Q\Xi_B QV_C \eta V_E\rangle_W
\nonumber\\
&\hspace{35mm}
+\langle Q\Xi_A QV_D (\xi_{c} b_{c}^- b_{c}^+) 
(b_{C_1} b_{C_2}) \eta\Psi_B QV_C \eta V_E\rangle_W 
\Big)
\nonumber\\
&
-\frac{\kappa^3}{4}\int d^2T \oint d\theta\
\Big(
\langle \big(\eta\Psi_A QV_D (b_{c}^- b_{c}^+) Q\Xi_B
+\ Q\Xi_A QV_D (b_{c}^- b_{c}^+) \eta\Psi_B)\ b_{\theta}^- V_C V_E\rangle_W
\nonumber\\
&\hspace{35mm}
+\langle \Xi_A QV_D (\xi_{c} b_{c}^- b_{c}^+) \eta\Psi_B\ b_{\theta}^-
(QV_C \eta V_E+\eta V_C QV_E)\rangle_W
\nonumber\\
&\hspace{35mm}
+2\langle \eta\Psi_A QV_D (b_{c}^- b_{c}^+)
QV_C\ b_{\theta}^- \Xi_B V_E\rangle_W
\nonumber\\
&\hspace{35mm}
-2\langle \Xi_A QV_D (b_{c}^- b_{c}^+)
QV_C\ b_{\theta}^- \eta\Psi_B V_E\rangle_W
\nonumber\\
&\hspace{35mm}
+2\langle \eta\Psi_A QV_D (b_{c}^- b_{c}^+) V_E\ b_{\theta}^- \Xi_B QV_C\rangle_W
\nonumber\\
&\hspace{35mm}
+2\langle \Xi_A QV_D (b_{c}^- b_{c}^+)  V_E\ b_{\theta}^- \eta\Psi_B QV_C\rangle_W
\Big)
\nonumber\\
&
+\frac{\kappa^3}{2}\int d^2\theta \oint d\theta\
\Big(
\langle(b_{C_1} b_{C_2}) \eta\Psi_B QV_C V_E\ b_{\theta}^-
\Xi_A QV_D\rangle_W 
\nonumber\\
&\hspace{35mm}
+\langle(b_{C_1} b_{C_2}) \Xi_B QV_C V_E\ b_{\theta}^-
\eta\Psi_A QV_D\rangle_W 
\Big)
\nonumber\\
&
+\frac{\kappa^3}{4}\oint d\theta \oint d\theta'\
\langle \big(\eta\Psi_A QV_D\ b_{\theta}^- \Xi_B
- \Xi_A QV_D\ b_{\theta}^- \eta\Psi_B\big)\ b_{\theta'}^- V_C V_E\rangle_W,\\
 \mathcal{A}_{F^2B^3}^{(AE|BCD)}
 =&\
-\frac{\kappa^3}{2}\int d^2T \int d^2\theta\
\Big(
\langle \eta\Psi_A \eta V_E (\xi_{c} b_{c}^- b_{c}^+) 
(b_{C_1} b_{C_2}) Q\Xi_B QV_C QV_D\rangle
\nonumber\\
&\hspace{35mm}
+\langle Q\Xi_A \eta V_E (\xi_{c} b_{c}^- b_{c}^+) 
(b_{C_1} b_{C_2}) \eta\Psi_B QV_C QV_D\rangle_W 
\Big)
\nonumber\\
&
-\frac{\kappa^3}{4}\int d^2T \oint d\theta\
\Big(
\big(\langle \eta\Psi_A V_E (b_{c}^- b_{c}^+) Q\Xi_B
+Q\Xi_A V_E (b_{c}^- b_{c}^+) \eta\Psi_B\big)
\nonumber\\ 
&\hspace{75mm} \times
b_{\theta}^- (QV_C V_D+V_C QV_D)\rangle_W
\nonumber\\
&\hspace{35mm}
+\langle \Xi_A QV_E (\xi_{c} b_{c}^- b_{c}^+) \eta\Psi_B\ b_{\theta}^- 
(QV_C \eta V_D+\eta V_C QV_D)\rangle_W
\nonumber\\
&\hspace{35mm}
+2\langle \eta\Psi_A V_E (b_{c}^- b_{c}^+)
QV_C\ b_{\theta}^- \Xi_B QV_D\rangle_W
\nonumber\\
&\hspace{35mm}
+2\langle \Xi_A V_E (b_{c}^- b_{c}^+)
QV_C\ b_{\theta}^- \eta\Psi_B QV_D\rangle_W
\nonumber\\
&\hspace{35mm}
+2\langle \eta\Psi_A V_E (b_{c}^- b_{c}^+)
QV_D\ b_{\theta}^- \Xi_B QV_C\rangle_W 
\nonumber\\
&\hspace{35mm}
+2\langle \Xi_A V_E (b_{c}^- b_{c}^+)
QV_D\ b_{\theta}^- \eta\Psi_B QV_C\rangle_W
\Big)
\nonumber\\
&
+\frac{\kappa^3}{2}\int d^2\theta \oint d\theta\
\Big(
\langle(b_{C_1} b_{C_2}) \eta\Psi_B QV_C QV_D\ 
b_{\theta}^- \Xi_A V_E\rangle_W 
\nonumber\\
&\hspace{35mm}
-\langle(b_{C_1} b_{C_2}) \Xi_B QV_C QV_D\ 
b_{\theta}^- \eta\Psi_A V_E\rangle_W 
\Big)
\nonumber
\end{align} 
\begin{equation}
 \nonumber
\end{equation}\begin{align}
&
+\frac{\kappa^3}{4}\oint d\theta \oint d\theta'\
\langle \big(\eta\Psi_A V_E\ b_{\theta}^- \Xi_B
+\Xi_A V_E\ b_{\theta}^- \eta\Psi_B\big)\ 
\nonumber\\
&\hspace{70mm}\times
b_{\theta'}^- (QV_C V_D + V_C QV_D)\rangle_W,\\
 \mathcal{A}_{F^2B^3}^{(BC|ADE)}
 =&\
-\frac{\kappa^3}{2}\int d^2T \int d^2\theta\
\Big(
\langle \eta\Psi_B QV_C (\xi_{c} b_{c}^- b_{c}^+) 
(b_{C_1} b_{C_2}) Q\Xi_A QV_D \eta V_E\rangle_W
\nonumber\\
&\hspace{35mm}
+\langle Q\Xi_B QV_C (\xi_{c} b_{c}^- b_{c}^+) 
(b_{C_1} b_{C_2}) \eta\Psi_A QV_D \eta V_E\rangle_W 
\Big)
\nonumber\\
&
-\frac{\kappa^3}{4}\int d^2T \oint d\theta\
\Big(
\langle \big(\eta\Psi_B QV_C (b_{c}^- b_{c}^+) Q\Xi_A
+ Q\Xi_B QV_C (b_{c}^- b_{c}^+) \eta\Psi_A\big)\ b_{\theta}^- V_D V_E\rangle_W
\nonumber\\
&\hspace{35mm}
+\langle \Xi_B QV_C (\xi_{c} b_{c}^- b_{c}^+) \eta\Psi_A\ b_{\theta}^-
(QV_D \eta V_E+\eta V_D QV_E)\rangle_W
\nonumber\\
&\hspace{35mm}
+2\langle \eta\Psi_B QV_C (b_{c}^- b_{c}^+)
QV_D\ b_{\theta}^- \Xi_A V_E\rangle_W
\nonumber\\
&\hspace{35mm}
-2\langle \Xi_B QV_C (b_{c}^- b_{c}^+)
QV_D\ b_{\theta}^- \eta\Psi_A V_E\rangle_W
\nonumber\\
&\hspace{35mm}
+2\langle \eta\Psi_B QV_C (b_{c}^- b_{c}^+) V_E\ b_{\theta}^- \Xi_A QV_D\rangle_W
\nonumber\\
&\hspace{35mm}
+2\langle \Xi_B QV_C (b_{c}^- b_{c}^+)  V_E\ b_{\theta}^- \eta\Psi_A QV_D\rangle_W
\Big)
\nonumber\\
&
+\frac{\kappa^3}{2}\int d^2\theta \oint d\theta\ \Big(
\langle(b_{C_1} b_{C_2}) \eta\Psi_A QV_D V_E\ b_{\theta}^-
\Xi_B QV_C\rangle_W 
\nonumber\\
&\hspace{35mm}
+\langle(b_{C_1} b_{C_2}) \Xi_A QV_D V_E\ b_{\theta}^-
\eta\Psi_B QV_C\rangle_W 
\Big)
\nonumber\\
&
+\frac{\kappa^3}{4}\oint d\theta \oint d\theta'\
\langle \big(\eta\Psi_B QV_C\ b_{\theta}^- \Xi_A
- \Xi_B QV_C\ b_{\theta}^- \eta\Psi_A\big)\ b_{\theta'}^- V_D V_E\rangle_W,\\
 \mathcal{A}_{F^2B^3}^{(BD|ACE)}
 =&\
-\frac{\kappa^3}{2}\int d^2T \int d^2\theta\
\Big(
\langle \eta\Psi_B QV_D (\xi_{c} b_{c}^- b_{c}^+) 
(b_{C_1} b_{C_2}) Q\Xi_A QV_C \eta V_E\rangle_W
\nonumber\\
&\hspace{35mm}
+\langle Q\Xi_B QV_D (\xi_{c} b_{c}^- b_{c}^+) 
(b_{C_1} b_{C_2}) \eta\Psi_A QV_C \eta V_E\rangle_W 
\Big)
\nonumber\\
&
-\frac{\kappa^3}{4}\int d^2T \oint d\theta\
\Big(
\langle \big(\eta\Psi_B QV_D (b_{c}^- b_{c}^+) Q\Xi_A
+Q\Xi_B QV_D (b_{c}^- b_{c}^+) \eta\Psi_A\big)\ b_{\theta}^- V_C V_E\rangle_W
\nonumber\\
&\hspace{35mm}
+\langle \Xi_B QV_D (\xi_{c} b_{c}^- b_{c}^+) \eta\Psi_A\ b_{\theta}^-
(QV_C \eta V_E+\eta V_C QV_E)\rangle_W
\nonumber\\
&\hspace{35mm}
+2\langle \eta\Psi_B QV_D (b_{c}^- b_{c}^+)
QV_C\ b_{\theta}^- \Xi_A V_E\rangle_W
\nonumber\\
&\hspace{35mm}
-2\langle \Xi_B QV_D (b_{c}^- b_{c}^+)
QV_C\ b_{\theta}^- \eta\Psi_A V_E\rangle_W
\nonumber\\
&\hspace{35mm}
+2\langle \eta\Psi_B QV_D (b_{c}^- b_{c}^+) V_E\ b_{\theta}^- \Xi_A QV_C\rangle_W
\nonumber\\
&\hspace{35mm}
+2\langle \Xi_B QV_D (b_{c}^- b_{c}^+)  V_E\ b_{\theta}^- \eta\Psi_A QV_C\rangle_W
\Big)
\nonumber\\
&
+\frac{\kappa^3}{2}\int d^2\theta \oint d\theta\ \Big(
\langle(b_{C_1} b_{C_2}) \eta\Psi_A QV_C V_E\ b_{\theta}^-
\Xi_B QV_D\rangle_W 
\nonumber\\
&\hspace{35mm}
+\langle(b_{C_1} b_{C_2}) \Xi_A QV_C V_E\ b_{\theta}^-
\eta\Psi_B QV_D\rangle_W 
\Big)
\nonumber\\
&
+\frac{\kappa^3}{4}\oint d\theta \oint d\theta'\
\langle \big(\eta\Psi_B QV_D\ b_{\theta}^- \Xi_A
- \Xi_B QV_D\ b_{\theta}^- \eta\Psi_A\big)\ b_{\theta'}^- V_C V_E\rangle_W,\\
 \mathcal{A}_{F^2B^3}^{(BE|ACD)}
 =&\
-\frac{\kappa^3}{2}\int d^2T \int d^2\theta\
\Big(
\langle \eta\Psi_B \eta V_E (\xi_{c} b_{c}^- b_{c}^+) 
(b_{C_1} b_{C_2}) Q\Xi_A QV_C QV_D\rangle
\nonumber\\
&\hspace{35mm}
+\langle Q\Xi_B \eta V_E (\xi_{c} b_{c}^- b_{c}^+) 
(b_{C_1} b_{C_2}) \eta\Psi_A QV_C QV_D\rangle_W 
\Big)
\nonumber\\
&
-\frac{\kappa^3}{4}\int d^2T \oint d\theta\
\Big(
\langle \big(\eta\Psi_B V_E (b_{c}^- b_{c}^+) Q\Xi_A
+ Q\Xi_B V_E (b_{c}^- b_{c}^+) \eta\Psi_A\big)
\nonumber\\ 
&\hspace{75mm}\times
b_{\theta}^- (QV_C V_D+V_C QV_D)\rangle_W
\nonumber\\
&\hspace{35mm}
+\langle \Xi_B QV_E (\xi_{c} b_{c}^- b_{c}^+) \eta\Psi_A\ b_{\theta}^- 
(QV_C \eta V_D+\eta V_C QV_D)\rangle_W
\nonumber\\
&\hspace{35mm}
+2\langle \eta\Psi_B V_E (b_{c}^- b_{c}^+)
QV_C\ b_{\theta}^- \Xi_A QV_D\rangle_W
\nonumber\\
&\hspace{35mm}
+2\langle \Xi_B V_E (b_{c}^- b_{c}^+)
QV_C\ b_{\theta}^- \eta\Psi_A QV_D\rangle_W
\nonumber\\
&\hspace{35mm}
+2\langle \eta\Psi_B V_E (b_{c}^- b_{c}^+)
QV_D\ b_{\theta}^- \Xi_A QV_C\rangle_W 
\nonumber\\
&\hspace{35mm}
+2\langle \Xi_B V_E (b_{c}^- b_{c}^+)
QV_D\ b_{\theta}^- \eta\Psi_A QV_C\rangle_W
\Big)
\nonumber\\
&
+\frac{\kappa^3}{2}\int d^2\theta \oint d\theta\
\Big(
\langle(b_{C_1} b_{C_2}) \eta\Psi_A QV_C QV_D\ 
b_{\theta}^- \Xi_B V_E\rangle_W 
\nonumber\\
&\hspace{35mm}
-\langle(b_{C_1} b_{C_2}) \Xi_A QV_C QV_D\ 
b_{\theta}^- \eta\Psi_B V_E\rangle_W 
\Big)
\nonumber\\
&
+\frac{\kappa^3}{4}\oint d\theta \oint d\theta'\
\langle \big(\eta\Psi_B V_E\ b_{\theta}^- \Xi_A
+ \Xi_B V_E\ b_{\theta}^- \eta\Psi_A\big)\ 
\nonumber\\
&\hspace{70mm}\times
b_{\theta'}^- (QV_C V_D + V_C QV_D)\rangle_W,\\
\mathcal{A}_{F^2B^3}^{(CD|ABE)}
 =&\
-\frac{\kappa^3}{2}\int d^2T \int d^2\theta\
\langle QV_C QV_D (\xi_{c} b_{c}^- b_{c}^+) (b_{C_1} b_{C_2})
(\eta\Psi_A Q\Xi_B+Q\Xi_A \eta\Psi_B) \eta V_E\rangle_W
\nonumber\\
&
-\frac{\kappa^3}{8}\int d^2T \oint d\theta\
\Big(
\langle (QV_C \eta V_D+\eta V_C QV_D) (\xi_{c} b_{c}^- b_{c}^+)
\nonumber\\
&\hspace{25mm}\times
\Big(2\big(\eta\Psi_A\ b_{\theta}^- \Xi_B-\Xi_A\ b_{\theta}^- \eta\Psi_B
+\eta\Psi_B\ b_{\theta}^- \Xi_A-\Xi_B\ b_{\theta}^- \eta\Psi_A\big) QV_E
\nonumber\\
&\hspace{35mm}
-V_E\ b_{\theta}^- (\eta\Psi_A Q\Xi_B + Q\Xi_A \eta\Psi_B)\Big)\rangle_W
\nonumber\\
&\hspace{28mm}
-2\langle (QV_C V_D+V_C QV_D) (b_{c}^- b_{c}^+)
\nonumber\\
&\hspace{25mm}\times
\Big(
\big(\eta\Psi_A\ b_{\theta}^-  Q\Xi_B+Q\Xi_A\ b_{\theta}^-  \eta\Psi_B
+\eta\Psi_B\ b_{\theta}^-  Q\Xi_A+Q\Xi_B\ b_{\theta}^-  \eta\Psi_A\big) V_E
\nonumber\\
&\hspace{35mm}
-V_E\ b_{\theta}^- (\eta\Psi_A Q\Xi_B+Q\Xi_A \eta\Psi_B)\big)\rangle_W
\Big)
\nonumber\\
&
+\frac{\kappa^3}{4}\int d^2\theta \oint d\theta\
\langle(b_{C_1}b_{C_2})(\eta\Psi_A Q\Xi_B+Q\Xi_A \eta\Psi_B) V_E
\nonumber\\
&\hspace{70mm}\times
b_{\theta}^-
(QV_C V_D+V_C QV_D)\rangle_W,\\
\mathcal{A}_{F^2B^3}^{(CE|ABD)}
 =&\
-\frac{\kappa^3}{2}\int d^2T \int d^2\theta\
\langle QV_C \eta V_E (\xi_{c} b_{c}^- b_{c}^+) (b_{C_1} b_{C_2})
(\eta\Psi_A Q\Xi_B+Q\Xi_A \eta\Psi_B) QV_D\rangle_W
\nonumber
\end{align}
\begin{equation}
\nonumber 
\end{equation}
\begin{align}
&
-\frac{\kappa^3}{8}\int d^2T \oint d\theta\
\langle(QV_C \eta V_E+\eta V_C QV_E)(\xi_{c} b_{c}^- b_{c}^+) 
\nonumber\\
&\hspace{25mm}\times
\Big(2\big(\eta\Psi_A\ b_{\theta}^- \Xi_B -\Xi_A\ b_{\theta}^- \eta\Psi_B
+\eta\Psi_B\ b_{\theta}^- \Xi_A -\Xi_B\ b_{\theta}^- \eta\Psi_A\big) QV_D
\nonumber\\
&\hspace{35mm}
-V_D\ b_{\theta}^- (\eta\Psi_A Q\Xi_B+Q\Xi_A \eta\Psi_B)
\Big)\rangle_W
\nonumber\\
&\hspace{28mm}
+2\langle V_C V_E (b_{c}^- b_{c}^+)
\nonumber\\
&\hspace{25mm}\times
\Big(
\big(\eta\Psi_A\ b_{\theta}^- Q\Xi_B+Q\Xi_A\ b_{\theta}^- \eta\Psi_B
+\eta\Psi_B b_{\theta}^- Q\Xi_A +Q\Xi_B b_{\theta}^- \eta\Psi_A\big)QV_D
\nonumber\\
&\hspace{35mm}
+QV_D b_{\theta}^- (\eta\Psi_A Q\Xi_B+Q\Xi_A \eta\Psi_B)
\big)\rangle_W 
\Big)
\nonumber\\
&
+\frac{\kappa^3}{4}\int d^2\theta \oint d\theta\
\langle(b_{C_1}b_{C_2}) (\eta\Psi_A Q\Xi_B+Q\Xi_A \eta\Psi_B) QV_D\ 
b_{\theta}^- V_C V_E\rangle_W,\\
\mathcal{A}_{F^2B^3}^{(DE|ABC)}
 =&\
-\frac{\kappa^3}{2}\int d^2T \int d^2\theta\
\langle QV_D \eta V_E (\xi_{c} b_{c}^- b_{c}^+) 
\nonumber\\
&\hspace{45mm}\times
(b_{C_1} b_{C_2})
(\eta\Psi_A Q\Xi_B+Q\Xi_A \eta\Psi_B) QV_C\rangle_W
\nonumber\\
&
-\frac{\kappa^3}{8}\int d^2T \oint d\theta\
\Big(
\langle(QV_D \eta V_E+\eta V_D QV_E)(\xi_{c} b_{c}^- b_{c}^+) 
\nonumber\\
&\hspace{25mm}\times
\Big(
2\big(\eta\Psi_A\ b_{\theta}^- \Xi_B -\Xi_A\ b_{\theta}^- \eta\Psi_B
+\eta\Psi_B\ b_{\theta}^- \Xi_A -\Xi_B\ b_{\theta}^- \eta\Psi_A\big) QV_C
\nonumber\\
&\hspace{35mm}
-V_C\ b_{\theta}^- (\eta\Psi_A Q\Xi_B+Q\Xi_A \eta\Psi_B)
\Big)\rangle_W
\nonumber\\
&\hspace{28mm}
+2\langle V_D V_E (b_{c}^- b_{c}^+)
\nonumber\\
&\hspace{25mm}\times
\Big(
\big(\eta\Psi_A\ b_{\theta}^- Q\Xi_B+Q\Xi_A\ b_{\theta}^- \eta\Psi_B
+\eta\Psi_B b_{\theta}^- Q\Xi_A +Q\Xi_B b_{\theta}^- \eta\Psi_A\big)QV_C
\nonumber\\
&\hspace{35mm}
+QV_C b_{\theta}^- (\eta\Psi_A Q\Xi_B+Q\Xi_A \eta\Psi_B)
\Big)\rangle_W 
\Big)
\nonumber\\
&
+\frac{\kappa^3}{4}\int d^2\theta \oint d\theta\
\langle(b_{C_1}b_{C_2}) (\eta\Psi_A Q\Xi_B+Q\Xi_A \eta\Psi_B) QV_C\ 
b_{\theta}^- V_D V_E\rangle_W.
\end{align}
The last contribution from the NP diagram can be divided into three parts: 
those integrated by four, three, and two moduli parameters, respectively. 
After a little calculation to align the bosons in the first part,
the dominant contribution, we obtain:
\begin{align}
 \mathcal{A}_{F^2B^3}^{(ABCDE)}
 =&\
-\frac{\kappa^3}{2}\int d^4\theta\ 
\langle \xi (b_{C_1} b_{C_2} b_{C_3} b_{C_4})
(\eta\Psi_A Q\Xi_B+Q\Xi_A \eta\Psi_B) QV_C QV_D \eta V_E\rangle_W
\nonumber\\
&
-\frac{\kappa^3}{12}\int d^2\theta \oint d\theta\
 \Big(
6\langle (b_{C_1} b_{C_2}) \eta\Psi_A QV_C QV_D\ b_{\theta}^- \Xi_B V_E\rangle_W
\nonumber\\
&\hspace{35mm}
-6\langle (b_{C_1} b_{C_2}) \Xi_A QV_C QV_D\ b_{\theta}^- \eta\Psi_B V_E\rangle_W
\nonumber\\
&\hspace{35mm}
+6\langle (b_{C_1} b_{C_2}) \eta\Psi_B QV_C QV_D\ b_{\theta}^- \Xi_A V_E\rangle_W
\nonumber\\
&\hspace{35mm}
-6\langle (b_{C_1} b_{C_2}) \Xi_B QV_C QV_D\ b_{\theta}^- \eta\Psi_A V_E\rangle_W
\nonumber\\
&\hspace{35mm}
+6\langle(b_{C_1} b_{C_2}) \eta\Psi_A QV_C V_E\ b_{\theta}^- \Xi_B QV_D\rangle_W
\nonumber\\
&\hspace{35mm}
+6\langle(b_{C_1} b_{C_2}) \Xi_A QV_C V_E\ b_{\theta}^- \eta\Psi_B QV_D\rangle_W
\nonumber\\
&\hspace{35mm}
+6\langle(b_{C_1} b_{C_2}) \eta\Psi_A QV_D V_E\ b_{\theta}^- \Xi_B QV_C\rangle_W
\nonumber\\
&\hspace{35mm}
+6\langle(b_{C_1} b_{C_2}) \Xi_A QV_D V_E\ b_{\theta}^- \eta\Psi_B QV_C\rangle_W
\nonumber\\
&\hspace{35mm}
+6\langle(b_{C_1} b_{C_2}) \eta\Psi_B QV_C V_E\ b_{\theta}^- \Xi_A QV_D\rangle_W
\nonumber\\
&\hspace{35mm}
+6\langle(b_{C_1} b_{C_2}) \Xi_B QV_C V_E\ b_{\theta}^- \eta\Psi_A QV_D\rangle_W
\nonumber\\
&\hspace{35mm}
+6\langle(b_{C_1} b_{C_2}) \eta\Psi_B QV_D V_E\ b_{\theta}^- \Xi_A QV_C\rangle_W
\nonumber\\
&\hspace{35mm}
+6\langle(b_{C_1} b_{C_2}) \Xi_B QV_D V_E\ b_{\theta}^- \eta\Psi_A QV_C\rangle_W
\nonumber\\
&\hspace{35mm}
+3\langle(b_{C_1} b_{C_2}) (\eta\Psi_A Q\Xi_B+Q\Xi_A \eta\Psi_B) 
QV_D\ b_{\theta}^- V_C V_E\rangle_W
\nonumber\\
&\hspace{35mm}
+3\langle(b_{C_1} b_{C_2}) (\eta\Psi_A Q\Xi_B+Q\Xi_A \eta\Psi_B) 
QV_C\ b_{\theta}^- V_D V_E\rangle_W
\nonumber\\
&\hspace{35mm}
+3\langle(b_{C_1} b_{C_2})
 (\eta\Psi_A Q\Xi_B+Q\Xi_A \eta\Psi_B) V_E
\nonumber\\ 
&\hspace{75mm}\times
b_{\theta}^- (QV_C V_D+V_C QV_D)\rangle_W
\nonumber\\
&\hspace{35mm}
+2\langle (b_{C_1} b_{C_2}) (QV_C V_D+V_C QV_D) V_E
\nonumber\\ 
&\hspace{75mm}\times
b_{\theta}^- (\eta\Psi_A Q\Xi_B+Q\Xi_A \eta\Psi_B)\rangle_W
\Big)
\nonumber\\
&
-\frac{\kappa^3}{12}\oint d\theta \oint d\theta'\
\Big(
\langle (\eta\Psi_A Q\Xi_B+Q\Xi_A \eta\Psi_B)\ b_{\theta}^- 
\big(V_C\ b_{\theta'}^- V_D + V_D\ b_{\theta'}^- V_C\big) V_E\rangle_W 
\nonumber\\
&\hspace{35mm}
+3\langle \big(
\eta\Psi_A QV_C\ b_{\theta}^- \Xi_B-\Xi_A QV_C\ b_{\theta}^- \eta\Psi_B
\nonumber\\
&\hspace{45mm}
+\eta\Psi_B QV_C\ b_{\theta}^- \Xi_A-\Xi_B QV_C\ b_{\theta}^- \eta\Psi_A
\big)\ b_{\theta'}^- V_D V_E\rangle_W
\nonumber\\
&\hspace{35mm}
+3\langle \big(
\eta\Psi_A QV_D\ b_{\theta}^- \Xi_B-\Xi_A QV_D\ b_{\theta}^- \eta\Psi_B
\nonumber\\
&\hspace{45mm}
+\eta\Psi_B QV_D\ b_{\theta}^- \Xi_A-\Xi_B QV_D\ b_{\theta}^- \eta\Psi_A
\big)\ b_{\theta'}^- V_C V_E\rangle_W
\nonumber\\
&\hspace{35mm}
+3\langle \big(
\eta\Psi_A V_E\ b_{\theta}^- \Xi_B+\Xi_A V_E\ b_{\theta}^- \eta\Psi_B
\nonumber\\
&\hspace{45mm}
+\eta\Psi_B V_E\ b_{\theta}^- \Xi_A+\Xi_B V_E\ b_{\theta}^- \eta\Psi_A
\big)
\nonumber\\ 
&\hspace{75mm}\times
b_{\theta'}^- (QV_C V_D+V_C QV_D)\rangle_W
\Big).
\end{align}
The total amplitude is given by summing all these contributions. 
One can show that the boundary contributions integrated over less
(two or three) moduli parameters are canceled, 
and consequently the total amplitude becomes the sum of the dominant
contribution of each diagram:
\begin{align}
\mathcal{A}_{F^2B^3}
=&\  \kappa^3 \int d^2T_1 d^2T_2\
\Big(
\llangle \eta\Psi_B QV_C (b_{c_1}^- b_{c_1}^+) \eta\Psi_A 
(b_{c_2}^- b_{c_2}^+) QV_D \eta V_E\rrangle 
+ \textrm{14\ terms}
\Big)
\nonumber\\
&
+\kappa^3 \int d^2T d^2\theta\
\Big(
\llangle \eta\Psi_A \eta\Psi_B (b_c^- b_c^+) (b_{C_1}b_{C_2})
QV_C QV_D \eta V_E \rrangle
+ \textrm{9\ terms}
\Big)
\nonumber\\
&
+\kappa^3 \int d^4\theta\
\llangle (b_{C_1} b_{C_2} b_{C_3} b_{C_4}) \eta\Psi_A \eta\Psi_B
QV_C QV_D \eta V_E \rrangle.
\end{align}
Each contribution again has the same form as that in the bosonic closed 
string field theory after imposing the constraint 
if we identify the external bosonic strings and $\eta\Psi$, $QV$ or $\eta V$. 
Hence the two-fermion-three-boson amplitude is also reproduced by the new Feynman rules.

\section{Conclusion and discussion}\label{sec4}

In this paper we have reconsidered the symmetries of the pseudo-action
of the heterotic string field theory. It has been found, at some lower order
in the fermion expansion, that the missing gauge symmetries,
 which were considered to be present only in the equations of motion,
are realized as the symmetries provided we impose the constraint
after the transformation. Respecting also this type of gauge symmetry, 
we have proposed a prescription for the new Feynman rules and shown that 
they actually reproduce the correct tree-level amplitudes in the case 
of the four- and five-external strings including fermions. 

An important remaining task is to prove that the new Feynman rules 
actually reproduce an arbitrary on-shell amplitude at the tree level.
For this purpose, it is necessary to complete the pseudo-action,
which has only been obtained at some lower order in the number
of fermions or string products.\cite{Kunitomo:2013mqa} 
The new kind of symmetries must play an important role
in this construction and proof. 
The Feynman rules should also be extended to be able
to calculate general loop amplitudes, for which we need to 
introduce an infinite sequence of ghosts for ghosts and 
construct the quantum action satisfying the Batalin-Vilkovisky 
master equation.\cite{Zwiebach:1992ie,Kroyter:2012ni}
It is still unclear what role the pseudo-action can play.
It is important to clarify whether the apparent 
difficulty coming from the duplicated off-shell fermions actually 
causes an inconsistency.
It is also worthwhile studying the off-shell amplitudes obtained 
by the new Feynman rules and comparing the results with those obtained by 
the rules proposed recently.\cite{Sen:2014pia}

\section*{Acknowledgments}

This work was initiated at the workshop on 
\lq\lq String Field Theory and Related Aspects VI\rq\rq\ 
held, from 28 July to 1 August 2014, at SISSA in Trieste, Italy .
The author would like to thank the organizers, particularly Loriano Bonora, 
for their hospitality and providing a stimulating atmosphere.

\newpage

\appendix

\section{}

The gauge symmetries provided by the constraint given in \S\S\ref{2-3} have
only been shown to exist at some lower order in the fermion expansion. 
Up to the order discussed in the text, however, 
the transformation of the pseudo-action is proportional to the constraint in the lowest 
order of the fermion expansion: $Q_G\Xi=\eta\Psi$. 
It is therefore worthwhile to show that the transformation including the next-order
corrections properly transforms the pseudo-action to the form proportional to
the constraint correctly including the next-order corrections.
Including the next-order pseudo-action,\cite{Kunitomo:2013mqa}
\begin{align}
 S_{R[6]}=&-\frac{\kappa^4}{6!}\langle\epsi,\mspd{\Psi,\qpsi,(\qxi)^3}
+\frac{2}{6!}\kappa^4\langle\epsi,\mspd{\Psi,\mspd{\Psi,(\qxi)^3}}\rangle\nonumber\\
&-\frac{2}{6!}\kappa^4\langle\epsi,\mspd{\Psi,\qxi,\mspd{\Psi,(\qxi)^2}}\rangle
-\frac{3}{6!}\kappa^4\langle\epsi,\mspd{\Psi,(\qxi)^2,\mspd{\Psi,\qxi}}\rangle,
\end{align}
we can find that the next-order $\Lambda_1$-transformation has to be
\begin{align}
\delta^{[4]}_{\Lambda_1}\Psi =&\
-\frac{\kappa^5}{5!}\mspd{\Psi,(\qpsi)^2,(\qxi)^2,\eLambda}
+\frac{3}{5!}\kappa^5\mspd{\Psi,\mspd{\Psi,\qpsi,(\qxi)^2,\eLambda}}
\nonumber\\
& 
+\frac{8\kappa^5}{6!}\mspd{\Psi,\qpsi,\mspd{\Psi,(\qxi)^2,\eLambda}}
+\frac{4\kappa^5}{6!}\mspd{\Psi,\qxi,\mspd{\Psi,\qpsi,\qxi,\eLambda}}
\nonumber\\
& 
+\frac{4\kappa^5}{5!}\mspd{\Psi,\qpsi,\qxi,\mspd{\Psi,\qxi,\eLambda}}
-\frac{\kappa^5}{5!}\mspd{\Psi,(\qxi)^2,\mspd{\Psi,\qpsi,\eLambda}}
\nonumber\\
& 
-\frac{\kappa^5}{5!}\mspd{\mspd{\Psi,\qpsi},\Psi,(\qxi)^2,\eLambda}
-\frac{6\kappa^5}{5!}\mspd{\mspd{\Psi,\qxi},\Psi,\qpsi,\qxi,\eLambda}
\nonumber\\
& 
-\frac{16\kappa^5}{6!}\mspd{\mspd{\Psi,\qpsi,\qxi},\Psi,\qxi,\eLambda}
-\frac{2\kappa^5}{5!}\mspd{\mspd{\Psi,(\qxi)^2},\Psi,\qpsi,\eLambda}
\nonumber\\
& 
-\frac{2\kappa^5}{5!}\mspd{\mspd{\Psi,\qpsi,(\qxi)^2},\Psi,\eLambda}
-\frac{10\kappa^5}{6!}\kappa^5\mspd{\Psi,\mspd{\Psi,\mspd{\Psi,(\qxi)^2,\eLambda}}}
\nonumber\\
& 
-\frac{\kappa^5}{4!}\mspd{\Psi,\mspd{\Psi,\qxi,\mspd{\Psi,\qxi,\eLambda}}}
+\frac{10\kappa^5}{6!}\mspd{\Psi,\qxi,\mspd{\Psi,\mspd{\Psi,\qxi,\eLambda}}}
\nonumber\\
& 
+\frac{\kappa^5}{4!}\mspd{\Psi,\mspd{\Psi,\qxi},\mspd{\Psi,\qxi,\eLambda}}
+\frac{2\kappa^5}{4!}\mspd{\Psi,\mspd{\mspd{\Psi,\qxi},\Psi,\qxi,\eLambda}}
\nonumber\\
& 
+\frac{20\kappa^5}{6!}\mspd{\Psi,\mspd{\mspd{\Psi,(\qxi)^2},\Psi,\eLambda}}
+\frac{10\kappa^5}{6!}\mspd{\Psi,\qxi,\mspd{\mspd{\Psi,\qxi},\Psi,\eLambda}}
\nonumber\\
& 
-\frac{20\kappa^5}{6!}\kappa^5\mspd{\mspd{\Psi,\mspd{\Psi,\qxi}},\Psi,\qxi,\eLambda}
-\frac{10\kappa^5}{6!}\mspd{\mspd{\Psi,\mspd{\Psi,(\qxi)^2}},\Psi,\eLambda}
\nonumber\\
& 
-\frac{\kappa^5}{4!}\mspd{\mspd{\Psi,\qxi,\mspd{\Psi,\qxi}},\Psi,\eLambda}
-\frac{\kappa^5}{4!}\mspd{\mspd{\Psi,\qxi},\mspd{\Psi,\qxi},\Psi,\eLambda},\\
\delta^{[4]}_{\Lambda_1}\Xi =&\
-\frac{\kappa^5}{5!}\mspd{\Psi,\qpsi,(\qxi)^3,\eLambda}
+\frac{\kappa^5}{5!}\mspd{\Psi,\mspd{\Psi,(\qxi)^3,\eLambda}}
\nonumber\\
& 
+\frac{4}{5!}\kappa^5\mspd{\Psi,\qxi,\mspd{\Psi,(\qxi)^2,\eLambda}}
+\frac{6}{5!}\kappa^5\mspd{\Psi,(\qxi)^2,\mspd{\Psi,\qxi,\eLambda}}
\nonumber\\
& 
-\frac{4}{5!}\kappa^5\mspd{\mspd{\Psi,\qxi},\Psi,(\qxi)^2,\eLambda}
-\frac{16\kappa^5}{6!}\mspd{\mspd{\Psi,(\qxi)^2},\Psi,\qxi,\eLambda}
\nonumber\\
& 
+\frac{\kappa^5}{5!}\mspd{\mspd{\Psi,(\qxi)^3},\Psi,\eLambda}
+\frac{\kappa^5}{4!}\mspd{\Xi,\mspd{\Psi,\qpsi,(\qxi)^2,\eLambda}}
\nonumber\\
& 
-\frac{\kappa^5}{4!}\mspd{\Xi,\mspd{\Psi,\mspd{\Psi,(\qxi)^2,\eLambda}}}
-\frac{\kappa^5}{8}\mspd{\Xi,\mspd{\Psi,\qxi,\mspd{\Psi,\qxi,\eLambda}}}
\nonumber\\
& 
+\frac{\kappa^5}{8}\mspd{\Xi,\mspd{\mspd{\Psi,\qxi},\Psi,\qxi,\eLambda}}
+\frac{\kappa^5}{4!}\mspd{\Xi,\mspd{\mspd{\Psi,(\qxi)^2},\Psi,\eLambda}},\\
 B_{\delta_{\Lambda_1}}^{[6]} =&\
\frac{\kappa^6}{6!}\mspd{\Psi,(\qpsi)^2,(\qxi)^3,\eLambda}
-\frac{2\kappa^6}{6!}\mspd{\Psi,\mspd{\Psi,\qpsi,(\qxi)^3,\eLambda}}
\nonumber\\
& 
-\frac{\kappa^6}{6!}\mspd{\Psi,\qpsi,\mspd{\Psi,(\qxi)^3,\eLambda}}
-\frac{9\kappa^6}{6!}\mspd{\Psi,\qxi,\mspd{\Psi,\qpsi,(\qxi)^2,\eLambda}}
\nonumber\\
& 
-\frac{4\kappa^6}{6!}\mspd{\Psi,\qpsi,\qxi,\mspd{\Psi,(\qxi)^2,\eLambda}}
-\frac{\kappa^6}{6!}\mspd{\Psi,(\qxi)^2,\mspd{\Psi,\qpsi,\qxi,\eLambda}}
\nonumber\\
& 
-\frac{\kappa^6}{5!}\mspd{\Psi,\qpsi,(\qxi)^2,\mspd{\Psi,\qxi,\eLambda}}
+\frac{\kappa^6}{6!}\mspd{\Psi,(\qxi)^3,\mspd{\Psi,\qpsi,\eLambda}}
\nonumber\\
& 
+\frac{\kappa^6}{6!}\mspd{\mspd{\Psi,\qpsi},\Psi,(\qxi)^3,\eLambda}
+\frac{9\kappa^6}{6!}\mspd{\mspd{\Psi,\qxi},\Psi,\qpsi,(\qxi)^2,\eLambda}
\nonumber\\
& 
+\frac{4\kappa^6}{6!}\mspd{\mspd{\Psi,\qpsi,\qxi},\Psi,(\qxi)^2,\eLambda}
+\frac{\kappa^6}{5!}\mspd{\mspd{\Psi,(\qxi)^2},\Psi,\qpsi,\qxi,\eLambda}
\nonumber\\
& 
+\frac{\kappa^6}{5!}\mspd{\mspd{\Psi,\qpsi,(\qxi)^2},\Psi,\qxi,\eLambda}
-\frac{\kappa^6}{6!}\mspd{\mspd{\Psi,(\qxi)^3},\Psi,\qpsi,\eLambda}
\nonumber\\
& 
+\frac{\kappa^6}{6!}\mspd{\mspd{\Psi,\qpsi,(\qxi)^3},\Psi,\eLambda}
+\frac{\kappa^6}{6!}\mspd{\Psi,\mspd{\Psi,\mspd{\Psi,(\qxi)^3,\eLambda}}}
\nonumber\\
& 
+\frac{4\kappa^6}{6!}\mspd{\Psi,\mspd{\Psi,\qxi,\mspd{\Psi,(\qxi)^2,\eLambda}}}
+\frac{\kappa^6}{5!}\mspd{\Psi,\mspd{\Psi,(\qxi)^2,\mspd{\Psi,\qxi,\eLambda}}}
\nonumber\\
& 
+\frac{5\kappa^6}{6!}\mspd{\Psi,\qxi,\mspd{\Psi,\mspd{\Psi,(\qxi)^2,\eLambda}}}
+\frac{15\kappa^6}{6!}\mspd{\Psi,\qxi,\mspd{\Psi,\qxi,\mspd{\Psi,\qxi,\eLambda}}}
\nonumber\\
& 
-\frac{5\kappa^6}{2\cdot 6!}\mspd{\Psi,(\qxi)^2,\mspd{\Psi,\mspd{\Psi,\qxi,\eLambda}}}
-\frac{9\kappa^6}{6!}\mspd{\Psi,\mspd{\mspd{\Psi,\qxi},\Psi,(\qxi)^2,\eLambda}}
\nonumber\\
& 
-\frac{\kappa^6}{5!}\mspd{\Psi,\mspd{\mspd{\Psi,(\qxi)^2},\Psi,\qxi,\eLambda}}
+\frac{\kappa^6}{6!}\mspd{\Psi,\mspd{\mspd{\Psi,(\qxi)^3},\Psi,\eLambda}}
\nonumber\\
& 
-\frac{\kappa^6}{4!}\mspd{\Psi,\qxi,\mspd{\mspd{\Psi,\qxi},\Psi,\qxi,\eLambda}}
-\frac{10\kappa^6}{6!}\mspd{\Psi,\qxi,\mspd{\mspd{\Psi,(\qxi)^2},\Psi,\eLambda}}
\nonumber\\
& 
-\frac{5\kappa^6}{2\cdot 6!}\mspd{\Psi,(\qxi)^2,\mspd{\mspd{\Psi,\qxi},\Psi,\eLambda}}
-\frac{5\kappa^6}{6!}\mspd{\Psi,\mspd{\Psi,\qxi},\mspd{\Psi,(\qxi)^2,\eLambda}}
\nonumber\\
& 
-\frac{5\kappa^6}{6!}\mspd{\Psi,\mspd{\Psi,(\qxi)^2},\mspd{\Psi,\qxi,\eLambda}}
-\frac{15\kappa^6}{6!}\mspd{\Psi,\qxi,\mspd{\Psi,\qxi},\mspd{\Psi,\qxi,\eLambda}}
\nonumber\\
& 
+\frac{15\kappa^6}{6!}\mspd{\mspd{\Psi,\qxi},\mspd{\Psi,\qxi},\Psi,\qxi,\eLambda}
+\frac{10\kappa^6}{6!}\mspd{\mspd{\Psi,\qxi},\mspd{\Psi,(\qxi)^2},\Psi,\eLambda}
\nonumber\\
& 
+\frac{5\kappa^6}{6!}\mspd{\mspd{\Psi,\mspd{\Psi,\qxi}},\Psi,(\qxi)^2,\eLambda}
+\frac{5\kappa^6}{6!}\mspd{\mspd{\Psi,\mspd{\Psi,(\qxi)^2}},\Psi,\qxi,\eLambda}
\nonumber\\
& 
-\frac{2\kappa^6}{6!}\mspd{\mspd{\Psi,\mspd{\Psi,(\qxi)^3}},\Psi,\eLambda}
+\frac{15\kappa^6}{6!}\mspd{\mspd{\Psi,\qxi,\mspd{\Psi,\qxi}},\Psi,\qxi,\eLambda}
\nonumber\\
& 
+\frac{2\kappa^6}{6!}\mspd{\mspd{\Psi,\qxi,\mspd{\Psi,(\qxi)^2}},\Psi,\eLambda}
\nonumber\\
&
+\frac{3\kappa^6}{6!}\mspd{\mspd{\Psi,(\qxi)^2,\mspd{\Psi,\qxi}},\Psi,\eLambda}.
\end{align}
Then the transformation of the pseudo-action at this order is given by
\begin{align}
 &\delta^{[6]}_{\Lambda_1}S_{NS}+\delta^{[4]}_{\Lambda_1}S_{R[2]}
+\delta^{[2]}_{\Lambda_1}S_{R[4]}+\delta^{[0]}S_{R[6]}
\nonumber\\
=&\
\frac{\kappa^3}{4!}\langle\eLambda,\mspd{\Psi,\qxi,\mspd{
\left(\frac{\kappa^2}{3!}\mspd{\Psi,\epsi,\qxi}\right),\qxi}}\rangle
\nonumber\\
&\hspace{60mm}
-\frac{\kappa^3}{4!}\mspd{\Psi,
\left(\frac{\kappa^2}{3!}\mspd{\Psi,\epsi,\qxi}\right),\mspd{(\qxi)^2}}\rangle
\nonumber\\
&
+\frac{\kappa^3}{4!}\langle\eLambda,\mspd{(\qxi)^2,\mspd{\Psi,
\left(\frac{\kappa^2}{3!}\mspd{\Psi,\epsi,\qxi}\right)}}\rangle
\nonumber\\
&\hspace{60mm}
-\frac{\kappa^3}{4!}\langle\eLambda,\mspd{
\left(\frac{\kappa^2}{3!}\mspd{\Psi,\epsi,\qxi}\right),\qxi,\mspd{\Psi,\qxi}}\rangle
\nonumber\\
&
-\frac{3\kappa^5}{6!}\langle\eLambda,\mspd{\Psi,\qxi,\mspd{\qpsi,\epsi,(\qxi)^2}}\rangle
+\frac{3\kappa^5}{6!}\langle\eLambda,\mspd{\Psi,\epsi,\mspd{\qpsi,(\qxi)^3}}\rangle
\nonumber\\
&
-\frac{3\kappa^5}{6!}\langle\eLambda,\mspd{(\qxi)^2,\mspd{\Psi,\qpsi,\epsi,\qxi}}\rangle
+\frac{3\kappa^5}{6!}\langle\eLambda,\mspd{\epsi,\qxi,\mspd{\Psi,\qpsi,(\qxi)^2}}\rangle
\nonumber\\
&
-\frac{\kappa^5}{6!}\langle\eLambda,\mspd{\Psi,\qpsi,\qxi,\mspd{\epsi,(\qxi)^2}}\rangle
+\frac{\kappa^5}{6!}\langle\eLambda,\mspd{\Psi,\qpsi,\epsi,\mspd{(\qxi)^3}}\rangle
\nonumber\\
&
-\frac{\kappa^5}{6!}\langle\eLambda,\mspd{\Psi,(\qxi)^2,\mspd{\qpsi,\epsi,\qxi}}\rangle
+\frac{\kappa^5}{6!}\langle\eLambda,\mspd{\Psi,\epsi,\qxi,\mspd{\qpsi,(\qxi)^2}}\rangle
\nonumber\\
&
-\frac{\kappa^5}{6!}\langle\eLambda,\mspd{\qpsi,(\qxi)^2,\mspd{\Psi,\epsi,\qxi}}\rangle
+\frac{\kappa^5}{6!}\langle\eLambda,\mspd{\qpsi,\epsi,\qxi,\mspd{\Psi,(\qxi)^2}}\rangle
\nonumber\\
&
-\frac{\kappa^5}{6!}\langle\eLambda,\mspd{(\qxi)^3,\mspd{\Psi,\qpsi,\epsi}}\rangle
+\frac{\kappa^5}{6!}\langle\eLambda,\mspd{\epsi,(\qxi)^2,\mspd{\Psi,\qpsi,\qxi}}\rangle
\nonumber\\
&
-\frac{3\kappa^5}{6!}\langle\eLambda,\mspd{\Psi,\qpsi,(\qxi)^2,\mspd{\epsi,\qxi}}\rangle
+\frac{3\kappa^5}{6!}\langle\eLambda,\mspd{\Psi,\qpsi,\epsi,\qxi,\mspd{(\qxi)^2}}\rangle
\nonumber\\
&
-\frac{3\kappa^5}{6!}\langle\eLambda,\mspd{\qpsi,(\qxi)^3,\mspd{\Psi,\epsi}}\rangle
+\frac{3\kappa^5}{6!}\langle\eLambda,\mspd{\qpsi,\epsi,(\qxi)^2,\mspd{\Psi,\qxi}}\rangle
\nonumber\\
&
-\frac{3\kappa^5}{6!}\langle\eLambda,\mspd{(\qxi)^2,\mspd{\Psi,\mspd{\Psi,\epsi,\qxi}}}\rangle
+\frac{3\kappa^5}{6!}\langle\eLambda,\mspd{\epsi,\qxi,\mspd{\Psi,\mspd{\Psi,(\qxi)^2}}}\rangle
\nonumber\\
&
-\frac{3\kappa^5}{6!}\langle\eLambda,\mspd{\Psi,\qxi,\mspd{\qxi,\mspd{\Psi,\epsi,\qxi}}}\rangle
-\frac{5\kappa^5}{6!}\langle\eLambda,\mspd{\Psi,\qxi,\mspd{\epsi,\mspd{\Psi,(\qxi)^2}}}\rangle
\nonumber\\
&
+\frac{8\kappa^5}{6!}\langle\eLambda,\mspd{\Psi,\epsi,\mspd{\qxi,\mspd{\Psi,(\qxi)^2}}}\rangle
\nonumber\\
&
+\frac{2\kappa^5}{6!}\langle\eLambda,\mspd{\Psi,\qxi,\mspd{\Psi,\mspd{\epsi,(\qxi)^2}}}\rangle
-\frac{2\kappa^5}{6!}\langle\eLambda,\mspd{\Psi,\epsi,\mspd{\Psi,\mspd{(\qxi)^3}}}\rangle
\nonumber\\
&
-\frac{\kappa^5}{5!}\langle\eLambda,\mspd{(\qxi)^2,\mspd{\Psi,\qxi,\mspd{\Psi,\epsi}}}\rangle
-\frac{3\kappa^5}{6!}\langle\eLambda,\mspd{(\qxi)^2,\mspd{\Psi,\epsi,\mspd{\Psi,\qxi}}}\rangle
\nonumber\\
&
+\frac{9\kappa^5}{6!}\langle\eLambda,\mspd{\epsi,\qxi,\mspd{\Psi,\qxi,\mspd{\Psi,\qxi}}}\rangle
\nonumber\\
&
-\frac{\kappa^5}{5!}\langle\eLambda,\mspd{\Psi,\qxi,\mspd{(\qxi)^2,\mspd{\Psi,\epsi}}}\rangle
-\frac{3\kappa^5}{6!}\langle\eLambda,\mspd{\Psi,\qxi,\mspd{\epsi,\qxi,\mspd{\Psi,\qxi}}}\rangle
\nonumber\\
&
+\frac{9\kappa^5}{6!}\langle\eLambda,\mspd{\Psi,\epsi,\mspd{(\qxi)^2,\mspd{\Psi,\qxi}}}
\nonumber\\
&
-\frac{\kappa^5}{5!}\langle\eLambda,\mspd{\Psi,\qxi,\mspd{\Psi,\qxi,\mspd{\epsi,\qxi}}}\rangle
+\frac{2}{5!}\kappa^5\langle\eLambda,\mspd{\Psi,\qxi,\mspd{\Psi,\epsi,\mspd{(\qxi)^2}}}\rangle
\nonumber\\
&
-\frac{\kappa^5}{5!}\langle\eLambda,\mspd{\Psi,\epsi,\mspd{\Psi,\qxi,\mspd{(\qxi)^2}}}\rangle
\nonumber\\
&
-\frac{2\kappa^5}{6!}\langle\eLambda,\mspd{(\qxi)^3,\mspd{\Psi,\mspd{\Psi,\epsi}}}\rangle
+\frac{2\kappa^5}{6!}\langle\eLambda,\mspd{\epsi,(\qxi)^2,\mspd{\Psi,\mspd{\Psi,\qxi}}}\rangle
\nonumber\\
&
-\frac{2\kappa^5}{6!}\langle\eLambda,\mspd{\Psi,(\qxi)^2,\mspd{\qxi,\mspd{\Psi,\epsi}}}\rangle
+\frac{2\kappa^5}{6!}\langle\eLambda,\mspd{\Psi,\epsi,\qxi,\mspd{\qxi,\mspd{\Psi,\qxi}}}\rangle
\nonumber\\
&
-\frac{2\kappa^5}{6!}\langle\eLambda,\mspd{\Psi,(\qxi)^2,\mspd{\Psi,\mspd{\epsi,\qxi}}}\rangle
+\frac{2\kappa^5}{6!}\langle\eLambda,\mspd{\Psi,\epsi,\qxi,\mspd{\Psi,\mspd{(\qxi)^2}}}\rangle
\nonumber\\
&
+\frac{3\kappa^5}{6!}\langle\eLambda,\mspd{\qxi,\mspd{\Psi,\qxi},\mspd{\Psi,\epsi,\qxi}}\rangle
-\frac{8\kappa^5}{6!}\langle\eLambda,\mspd{\qxi,\mspd{\Psi,\epsi},\mspd{\Psi,(\qxi)^2}}\rangle
\nonumber\\
&
+\frac{5\kappa}{6!}\langle\eLambda,\mspd{\epsi,\mspd{\Psi,\qxi},\mspd{\Psi,(\qxi)^2}}\rangle
\nonumber\\
&
+\frac{3\kappa^5}{6!}\langle\eLambda,\mspd{\Psi,\mspd{(\qxi)^2},\mspd{\Psi,\epsi,\qxi}}\rangle
-\frac{3\kappa^5}{6!}\langle\eLambda,\mspd{\Psi,\mspd{\epsi,\qxi},\mspd{\Psi,(\qxi)^2}}\rangle
\nonumber\\
&
-\frac{2\kappa^5}{6!}\kappa^5\langle\eLambda,\mspd{\Psi,\mspd{\Psi,\qxi},\mspd{\epsi,(\qxi)^2}}\rangle
+\frac{2\kappa^5}{6!}\langle\eLambda,\mspd{\Psi,\mspd{\Psi,\epsi},\mspd{(\qxi)^3}}\rangle
\nonumber\\
&
-\frac{9\kappa^5}{6!}\langle\eLambda,\mspd{(\qxi)^2,\mspd{\Psi,\epsi},\mspd{\Psi,\qxi}}\rangle
+\frac{9\kappa^5}{6!}\langle\eLambda,\mspd{\epsi,\qxi,\mspd{\Psi,\qxi},\mspd{\Psi,\qxi}}\rangle
\nonumber\\
&
-\frac{9\kappa^5}{6!}\langle\eLambda,\mspd{\Psi,\qxi,\mspd{\Psi,\qxi},\mspd{\epsi,\qxi}}\rangle
+\frac{\kappa^5}{5!}\langle\eLambda,\mspd{\Psi,\qxi,\mspd{\Psi,\epsi},\mspd{(\qxi)^2}}\rangle
\nonumber\\
&
+\frac{3\kappa^5}{6!}\langle\eLambda,\mspd{\Psi,\epsi,\mspd{\Psi,\qxi},\mspd{(\qxi)^2}}\rangle,
\end{align}
where the first four terms give the $\mathcal{O}(\Psi^3)$ corrections to the constraint in
the previous order result (\ref{tf lambda1}).

The $\Lambda_{1/2}$-transformation at the next order is similarly obtained as
\begin{align}
\delta^{[4]}_{\Lambda_{1/2}}\Psi =&\
\frac{\kappa^4}{5!}\mspd{\Psi,\qpsi,(\qxi)^2,\qLambda}
-\frac{\kappa^4}{5!}\mspd{\Psi,\mspd{\Psi,(\qxi)^2,\qLambda}}
\nonumber\\
&
+\frac{2\kappa^4}{5!}\mspd{\Psi,\qxi,\mspd{\Psi,\qxi,\qLambda}}
+\frac{2\kappa^4}{5!}\mspd{\Psi,(\qxi)^2,\mspd{\Psi,\qLambda}}
\nonumber\\
&
+\frac{2\kappa^4}{5!}\mspd{\mspd{\Psi,\qxi},\Psi,\qxi,\qLambda}
+\frac{4\kappa^4}{6!}\mspd{\mspd{\Psi,(\qxi)^2},\Psi,\qLambda},\\
\delta^{[4]}_{\Lambda_{1/2}}\Xi =&\
\frac{\kappa^4}{5!}\mspd{\Psi,(\qxi)^3,\qLambda}
-\frac{\kappa^4}{4!}\mspd{\Xi,\mspd{\Psi,(\qxi)^2,\qLambda}},\\
 B_{\delta_{\Lambda_{1/2}}}^{[6]} =&\
-\frac{\kappa^5}{6!}\mspd{\Psi,\qpsi,(\qxi)^3,\qLambda}
+\frac{\kappa^5}{6!}\mspd{\Psi,\mspd{\Psi,(\qxi)^3,\qLambda}}
\nonumber\\
&
+\frac{3\kappa^5}{6!}\mspd{\Psi,\qxi,\mspd{\Psi,(\qxi)^2,\qLambda}}
-\frac{3\kappa^5}{6!}\mspd{\Psi,(\qxi)^2,\mspd{\Psi,\qxi,\qLambda}}
\nonumber\\
&
-\frac{2\kappa^5}{6!}\mspd{\Psi,(\qxi)^3,\mspd{\Psi,\qLambda}}
-\frac{3\kappa^5}{6!}\mspd{\mspd{\Psi,\qxi},\Psi,(\qxi)^2,\qLambda}
\nonumber\\
&
-\frac{2\kappa^5}{6!}\mspd{\mspd{\Psi,(\qxi)^2},\Psi,\qxi,\qLambda}
+\frac{2\kappa^5}{6!}\mspd{\mspd{\Psi,(\qxi)^3},\Psi,\qLambda},
\end{align}
which transform the pseudo-action to
\begin{align}
   &\delta^{[6]}_{\Lambda_{1/2}}S_{NS}+\delta^{[4]}_{\Lambda_{1/2}}S_{R[2]}
+\delta^{[2]}_{\Lambda_{1/2}}S_{R[4]}+\delta^{[0]}_{\Lambda_{1/2}}S_{R[6]}
\nonumber\\
=&\
\frac{\kappa^2}{12}
\langle\qLambda,\mspd{\qxi,\mspd{\Xi,
\left(\frac{\kappa^2}{3!}\mspd{\Psi,\epsi,\qxi}\right)}}\rangle
\nonumber\\
&\hspace{60mm}
-\frac{\kappa^2}{12}\langle\qLambda,\mspd{
\left(\frac{\kappa^2}{3!}\mspd{\Psi,\epsi,\qxi}\right),\mspd{\Xi,\qxi}}\rangle
\nonumber\\
&
-\frac{\kappa^4}{5!}\langle\qLambda,\mspd{\qxi,\mspd{\Psi,\epsi,(\qxi)^2}}\rangle
+\frac{\kappa^4}{5!}\langle\qLambda,\mspd{\epsi,\mspd{\Psi,(\qxi)^3}}\rangle
\nonumber\\
&
+\frac{2\kappa^4}{6!}\langle\qLambda,\mspd{\Psi,\qxi,\mspd{\epsi,(\qxi)^2}}\rangle
-\frac{2\kappa^4}{6!}\langle\qLambda,\mspd{\Psi,\epsi,\mspd{(\qxi)^3}}\rangle
\nonumber\\
&
-\frac{2\kappa^4}{6!}\langle\qLambda,\mspd{(\qxi)^2,\mspd{\Psi,\epsi,\qxi}}\rangle
+\frac{2\kappa^4}{6!}\langle\qLambda,\mspd{\epsi,\qxi,\mspd{\Psi,(\qxi)^2}}\rangle
\nonumber\\
&
+\frac{\kappa^4}{5!}\langle\qLambda,\mspd{\Psi,(\qxi)^2,\mspd{\epsi,\qxi}}\rangle
-\frac{\kappa^4}{5!}\langle\qLambda,\mspd{\Psi,\epsi,\qxi,\mspd{(\qxi)^2}}\rangle.
\end{align}
The first two terms give the correction to the constraint in (\ref{tf lambda1half}).

The pseudo-action is invariant under the $\Lambda_{3/2}$-transformation
up to the order discussed in the text. 
If we improve the transformation by adding the next-order transformation,
\begin{align}
\delta^{[4]}_{\Lambda_{3/2}}\Psi =&\
-\frac{2\kappa^4}{5!}\mspd{\Psi,\qpsi,(\qxi)^2,\eLambdaf}
+\frac{3\kappa^4}{5!}\mspd{\Psi,\mspd{\Psi,(\qxi)^2,\eLambdaf}}
\nonumber\\
&
+\frac{4\kappa^4}{6!}\mspd{\Psi,\qxi,\mspd{\Psi,\qxi,\eLambdaf}}
-\frac{\kappa^4}{5!}\mspd{\Psi,(\qxi)^2,\mspd{\Psi,\eLambdaf}}
\nonumber\\
&
-\frac{6\kappa^4}{5!}\mspd{\mspd{\Psi,\qxi},\Psi,\qxi,\eLambdaf}
-\frac{2\kappa^4}{5!}\mspd{\mspd{\Psi,(\qxi)^2},\Psi,\eLambdaf}\\
\delta^{[4]}_{\Lambda_{3/2}}\Xi =&\
-\frac{\kappa^4}{5!}\mspd{\Psi,(\qxi)^3,\eLambdaf}
+\frac{\kappa^4}{4!}\mspd{\Xi,\mspd{\Psi,(\qxi)^2,\eLambdaf}},\\
 B_{\delta_{\Lambda_{3/2}}}^{[6]} =&\
\frac{2\kappa^5}{6!}\kappa^5\mspd{\Psi,\qpsi,(\qxi)^3,\eLambdaf}
-\frac{2\kappa^5}{6!}\mspd{\Psi,\mspd{\Psi,(\qxi)^3,\eLambdaf}}
\nonumber\\
&
-\frac{9\kappa^5}{6!}\mspd{\Psi,\qxi,\mspd{\Psi,(\qxi)^2,\eLambdaf}}
-\frac{\kappa^5}{6!}\mspd{\Psi,(\qxi)^2,\mspd{\Psi,\qxi,\eLambdaf}}
\nonumber\\
&
+\frac{\kappa^5}{6!}\mspd{\Psi,(\qxi)^3,\mspd{\Psi,\eLambdaf}}
+\frac{9\kappa^5}{6!}\mspd{\mspd{\Psi,\qxi},\Psi,(\qxi)^2,\eLambdaf}
\nonumber\\
&
+\frac{\kappa^5}{5!}\mspd{\mspd{\Psi,(\qxi)^2},\Psi,\qxi,\eLambdaf}
-\frac{\kappa^5}{6!}\mspd{\mspd{\Psi,(\qxi)^3},\Psi,\eLambdaf},
\end{align}
it transforms the pseudo-action nontrivially as:
\begin{align}
  &\delta^{[6]}_{\Lambda_{3/2}}S_{NS}+\delta^{[4]}_{\Lambda_{3/2}}S_{R[2]}
+\delta^{[2]}_{\Lambda_{3/2}}S_{R[4]}+\delta^{[0]}_{\Lambda_{3/2}}S_{R[6]}
\nonumber\\
=&\
-\frac{\kappa^4}{6!}\langle\eLambdaf,\mspd{\Psi,\qxi,\mspd{\epsi,(\qxi)^2}}\rangle
+\frac{\kappa^4}{6!}\langle\eLambdaf,\mspd{\Psi,\epsi,\mspd{(\qxi)^3}}\rangle
\nonumber\\
&
-\frac{\kappa^4}{6!}\langle\eLambdaf,\mspd{(\qxi)^2,\mspd{\Psi,\epsi,\qxi}}\rangle
+\frac{\kappa^4}{6!}\langle\eLambdaf,\mspd{\epsi,\qxi,\mspd{\Psi,(\qxi)^2}}\rangle
\nonumber\\
&
-\frac{3\kappa^4}{6!}\langle\eLambdaf,\mspd{\Psi,(\qxi)^2,\mspd{\epsi,\qxi}}\rangle
+\frac{3\kappa^4}{6!}\langle\eLambdaf,\mspd{\Psi,\epsi,\qxi,\mspd{(\qxi)^2}}\rangle
\nonumber\\
&
-\frac{3\kappa^4}{6!}\langle\eLambdaf,\mspd{(\qxi)^3,\mspd{\Psi,\epsi}}\rangle
+\frac{3\kappa^4}{6!}\langle\eLambdaf,\mspd{\epsi,(\qxi)^2,\mspd{\Psi,\qxi}}\rangle.
\end{align}
The right-hand side vanishes under the constraint.

Last of all, the $\tilde{\Lambda}_{1/2}$-transformation can be found as:
\begin{align}
\delta^{[4]}_{\tilde{\Lambda}_{1/2}}\Psi =&\
-\frac{2\kappa^4}{5!}\mspd{\Psi,(\qpsi)^2,\qxi,\eLambdax}
+\frac{7\kappa^4}{5!}\mspd{\Psi,\mspd{\Psi,\qpsi,\qxi,\eLambdax}}
\nonumber\\
&
+\frac{3\kappa^4}{5!}\mspd{\Psi,\qpsi,\mspd{\Psi,\qxi,\eLambdax}}
+\frac{28\kappa^4}{6!}\mspd{\Psi,\qxi,\mspd{\Psi,\qpsi,\eLambdax}}
\nonumber\\
&
+\frac{6\kappa^4}{5!}\mspd{\Psi,\qpsi,\qxi,\mspd{\Psi,\eLambdax}}
-\frac{2\kappa^4}{5!}\mspd{\mspd{\Psi,\qpsi},\Psi,\qxi,\eLambdax}
\nonumber\\
&
-\frac{7\kappa^4}{5!}\mspd{\mspd{\Psi,\qxi},\Psi,\qpsi,\eLambdax}
-\frac{3\kappa^4}{5!}\mspd{\mspd{\Psi,\qpsi,\qxi},\Psi,\eLambdax}
\nonumber\\
&
-\frac{4\kappa^4}{5!}\mspd{\Psi,\mspd{\Psi,\mspd{\Psi,\qxi,\eLambdax}}}
-\frac{8\kappa^4}{5!}\mspd{\Psi,\mspd{\Psi,\qxi,\mspd{\Psi,\eLambdax}}}
\nonumber\\
&
-\frac{32\kappa^4}{6!}\mspd{\Psi,\qxi,\mspd{\Psi,\mspd{\Psi,\eLambdax}}}
+\frac{16\kappa^4}{5!}\mspd{\Psi,\mspd{\mspd{\Psi,\qxi},\Psi,\eLambdax}}
\nonumber\\
&
+\frac{8\kappa^4}{5!}\mspd{\Psi,\mspd{\Psi,\qxi},\mspd{\Psi,\eLambdax}}
-\frac{4\kappa^4}{5!}\mspd{\mspd{\Psi,\mspd{\Psi,\qxi}},\Psi,\eLambdax},\\
\delta^{[4]}_{\tilde{\Lambda}_{1/2}}\Xi =&\
-\frac{2\kappa^4}{5!}\mspd{\Psi,\qpsi,(\qxi)^2,\eLambdax}
+\frac{\kappa^4}{5!}\mspd{\Psi,\mspd{\Psi,(\qxi)^2,\eLambdax}}
\nonumber\\
&
+\frac{3\kappa^4}{5!}\mspd{\Psi,\qxi,\mspd{\Psi,\qxi,\eLambdax}}
+\frac{3\kappa^4}{5!}\mspd{\Psi,(\qxi)^2,\mspd{\Psi,\eLambdax}}
\nonumber\\
&
-\frac{7\kappa^4}{5!}\mspd{\mspd{\Psi,\qxi},\Psi,\qxi,\eLambdax}
-\frac{14\kappa^4}{6!}\mspd{\mspd{\Psi,(\qxi)^2},\Psi,\eLambdax}
\nonumber\\
&
+\frac{2\kappa^4}{4!}\mspd{\Xi,\mspd{\Psi,\qpsi,\qxi,\eLambdax}}
-\frac{2\kappa^4}{4!}\mspd{\Xi,\mspd{\Psi,\mspd{\Psi,\qxi,\eLambdax}}}
\nonumber\\
&
-\frac{\kappa^4}{3!}\mspd{\Xi,\mspd{\Psi,\qxi,\mspd{\Psi,\eLambdax}}}
+\frac{\kappa^4}{3!}\mspd{\Xi,\mspd{\mspd{\Psi,\qxi},\Psi,\eLambdax}},\\
B_{\delta_{\tilde{\Lambda}_{1/2}}}^{[6]} =&\
\frac{3\kappa^5}{6!}\mspd{\Psi,(\qpsi)^2,(\qxi)^2,\eLambdax}
-\frac{\kappa^5}{5!}\mspd{\Psi,\mspd{\Psi,\qpsi,(\qxi)^2,\eLambdax}}
\nonumber\\
&
-\frac{3\kappa^5}{6!}\mspd{\Psi,\qpsi,\mspd{\Psi,(\qxi)^2,\eLambdax}}
-\frac{21\kappa^5}{6!}\mspd{\Psi,\qxi,\mspd{\Psi,\qpsi,\qxi,\eLambdax}}
\nonumber\\
&
-\frac{9\kappa^5}{6!}\mspd{\Psi,\qpsi,\qxi,\mspd{\Psi,\qxi,\eLambdax}}
-\frac{7\kappa^5}{6!}\mspd{\Psi,(\qxi)^2,\mspd{\Psi,\qpsi,\eLambdax}}
\nonumber\\
&
-\frac{9\kappa^5}{6!}\mspd{\Psi,\qpsi,(\qxi)^2,\mspd{\Psi,\eLambdax}}
+\frac{3\kappa^5}{6!}\mspd{\mspd{\Psi,\qpsi},\Psi,(\qxi)^2,\eLambdax}
\nonumber\\
&
+\frac{21\kappa^5}{6!}\mspd{\mspd{\Psi,\qxi},\Psi,\qpsi,\qxi,\eLambdax}
+\frac{9\kappa^5}{6!}\mspd{\mspd{\Psi,\qpsi,\qxi},\Psi,\qxi,\eLambdax}
\nonumber\\
&
+\frac{7\kappa^5}{6!}\mspd{\mspd{\Psi,(\qxi)^2},\Psi,\qpsi,\eLambdax}
+\frac{9\kappa^5}{6!}\mspd{\mspd{\Psi,\qpsi,(\qxi)^2},\Psi,\eLambdax}
\nonumber\\
&
+\frac{3\kappa^5}{6!}\mspd{\Psi,\mspd{\Psi,\mspd{\Psi,(\qxi)^2,\eLambdax}}}
+\frac{9\kappa^5}{6!}\mspd{\Psi,\mspd{\Psi,\qxi,\mspd{\Psi,\qxi,\eLambdax}}}
\nonumber\\
&
+\frac{9\kappa^5}{6!}\mspd{\Psi,\mspd{\Psi,(\qxi)^2,\mspd{\Psi,\eLambdax}}}
+\frac{2\kappa^5}{5!}\mspd{\Psi,\qxi,\mspd{\Psi,\mspd{\Psi,\qxi,\eLambdax}}}
\nonumber\\
&
+\frac{4\kappa^5}{5!}\mspd{\Psi,\qxi,\mspd{\Psi,\qxi,\mspd{\Psi,\eLambdax}}}
+\frac{8\kappa^5}{6!}\mspd{\Psi,(\qxi)^2,\mspd{\Psi,\mspd{\Psi,\eLambdax}}}
\nonumber\\
&
-\frac{21\kappa^5}{6!}\mspd{\Psi,\mspd{\mspd{\Psi,\qxi},\Psi,\qxi,\eLambdax}}
-\frac{7\kappa^5}{6!}\mspd{\Psi,\mspd{\mspd{\Psi,(\qxi)^2},\Psi,\eLambdax}}
\nonumber\\
&
-\frac{8\kappa^5}{5!}\mspd{\Psi,\qxi,\mspd{\mspd{\Psi,\qxi},\Psi,\eLambdax}}
-\frac{2\kappa^5}{5!}\mspd{\Psi,\mspd{\Psi,\qxi},\mspd{\Psi,\qxi,\eLambdax}}
\nonumber\\
&
-\frac{8\kappa^5}{6!}\mspd{\Psi,\mspd{\Psi,(\qxi)^2},\mspd{\Psi,\eLambdax}}
-\frac{4\kappa^5}{5!}\mspd{\Psi,\qxi,\mspd{\Psi,\qxi},\mspd{\Psi,\eLambdax}}
\nonumber\\
&
+\frac{2\kappa^5}{5!}\mspd{\mspd{\Psi,\mspd{\Psi,\qxi}},\Psi,\qxi,\eLambdax}
+\frac{8\kappa^5}{6!}\mspd{\mspd{\Psi,\mspd{\Psi,(\qxi)^2}},\Psi,\eLambdax}
\nonumber\\
&
+\frac{4\kappa^5}{5!}\mspd{\mspd{\Psi,\qxi,\mspd{\Psi,\qxi}},\Psi,\eLambdax}
\nonumber\\
&
+\frac{4\kappa^5}{5!}\mspd{\mspd{\Psi,\qxi},\mspd{\Psi,\qxi},\Psi,\eLambdax},
\end{align}
which transforms the pseudo-action as:
\begin{align}
  &\delta^{[6]}_{\tilde{\Lambda}_{1/2}}S_{NS}+\delta^{[4]}_{\tilde{\Lambda}_{1/2}}S_{R[2]}
+\delta^{[2]}_{\tilde{\Lambda}_{1/2}}S_{R[4]}+\delta^{[0]}_{\tilde{\Lambda}_{1/2}}S_{R[6]}
\nonumber\\
=&\
\frac{\kappa^2}{12}
\langle\eLambdax,\mspd{\qxi,\mspd{\Psi,
\left(\frac{\kappa^2}{3!}\mspd{\Psi,\epsi,\qxi}\right)}}\rangle
\nonumber\\
&\hspace{60mm}
-\frac{\kappa^2}{12}\langle\eLambdax,\mspd{
\left(\frac{\kappa^2}{3!}\mspd{\Psi,\epsi,\qxi}\right),\mspd{\Psi,\qxi}}\rangle
\nonumber\\
&
-\frac{\kappa^4}{5!}\langle\eLambdax,\mspd{\qxi,\mspd{\Psi,\qpsi,\epsi,\qxi}}\rangle
+\frac{\kappa^4}{5!}\langle\eLambdax,\mspd{\epsi,\mspd{\Psi,\qpsi,(\qxi)^2}}\rangle
\nonumber\\
&
-\frac{\kappa^4}{6!}
\langle\eLambdax,\mspd{\Psi,\qxi,\mspd{\qpsi,\epsi,\qxi}}\rangle
+\frac{\kappa^4}{6!}
\langle\eLambdax,\mspd{\Psi,\epsi,\mspd{\qpsi,(\qxi)^2}}\rangle
\nonumber\\
&
-\frac{2\kappa^4}{6!}
\langle\eLambdax,\mspd{\qpsi,\qxi,\mspd{\Psi,\epsi,\qxi}}\rangle
+\frac{2\kappa^4}{6!}
\langle\eLambdax,\mspd{\qpsi,\epsi,\mspd{\Psi,(\qxi)^2}}\rangle
\nonumber\\
&
-\frac{3\kappa^4}{6!}
\langle\eLambdax,\mspd{(\qxi)^2,\mspd{\Psi,\qpsi,\epsi}}\rangle
+\frac{3\kappa^4}{6!}
\langle\eLambdax,\mspd{\epsi,\qxi,\mspd{\Psi,\qpsi,\qxi}}\rangle
\nonumber\\
&
-\frac{3\kappa^4}{6!}
\langle\eLambdax,\mspd{\Psi,\qpsi,\qxi,\mspd{\epsi,\qxi}}\rangle
+\frac{3\kappa^4}{6!}
\langle\eLambdax,\mspd{\Psi,\qpsi,\epsi,\mspd{(\qxi)^2}}\rangle
\nonumber\\
&
-\frac{9\kappa^4}{6!}
\langle\eLambdax,\mspd{\qpsi,(\qxi)^2,\mspd{\Psi,\epsi}}\rangle
+\frac{9\kappa^4}{6!}
\langle\eLambdax,\mspd{\qpsi,\epsi,\qxi,\mspd{\Psi,\qxi}}\rangle
\nonumber\\
&
+\frac{2\kappa^4}{6!}
\langle\eLambdax,\mspd{\Psi,\mspd{\qxi,\mspd{\Psi,\epsi,\qxi}}}\rangle
-\frac{2\kappa^4}{6!}
\langle\eLambdax,\mspd{\Psi,\mspd{\epsi,\mspd{\Psi,(\qxi)^2}}}\rangle
\nonumber\\
&
-\frac{\kappa^4}{5!}
\langle\eLambdax,\mspd{\qxi,\mspd{\Psi,\mspd{\Psi,\epsi,\qxi}}}\rangle
+\frac{\kappa^4}{5!}
\langle\eLambdax,\mspd{\epsi,\mspd{\Psi,\mspd{\Psi,(\qxi)^2}}}\rangle
\nonumber\\
&
-\frac{2\kappa^4}{5!}
\langle\eLambdax,\mspd{\Psi,\mspd{\Psi,\qxi,\mspd{\epsi,\qxi}}}\rangle
+\frac{2\kappa^4}{5!}
\langle\eLambdax,\mspd{\Psi,\mspd{\Psi,\epsi,\mspd{(\qxi)^2}}}\rangle
\nonumber\\
&
-\frac{\kappa^4}{5!}
\langle\eLambdax,\mspd{\Psi,\mspd{(\qxi)^2,\mspd{\Psi,\epsi}}}\rangle
+\frac{\kappa^4}{5!}
\langle\eLambdax,\mspd{\Psi,\mspd{\epsi,\qxi,\mspd{\Psi,\qxi}}}\rangle
\nonumber\\
&
-\frac{2\kappa^4}{5!}
\langle\eLambdax,\mspd{\qxi,\mspd{\Psi,\qxi,\mspd{\Psi,\epsi}}}\rangle
-\frac{\kappa^4}{5!}
\langle\eLambdax,\mspd{\qxi,\mspd{\Psi,\epsi,\mspd{\Psi,\qxi}}}\rangle
\nonumber\\
&
+\frac{3\kappa^4}{5!}
\langle\eLambdax,\mspd{\epsi,\mspd{\Psi,\qxi,\mspd{\Psi,\qxi}}}\rangle
\nonumber\\
&
-\frac{4\kappa^4}{6!}
\langle\eLambdax,\mspd{\Psi,\qxi,\mspd{\qxi,\mspd{\Psi,\epsi}}}\rangle
+\frac{2\kappa^4}{6!}
\langle\eLambdax,\mspd{\Psi,\qxi,\mspd{\epsi,\mspd{\Psi,\qxi}}}\rangle
\nonumber\\
&
+\frac{2\kappa^4}{6!}
\langle\eLambdax,\mspd{\Psi,\epsi,\mspd{\qxi,\mspd{\Psi,\qxi}}}\rangle
\nonumber\\
&
-\frac{2\kappa^4}{6!}
\langle\eLambdax,\mspd{\Psi,\qpsi,\mspd{\Psi,\mspd{\epsi,\qxi}}}\rangle
+\frac{2\kappa^4}{6!}
\langle\eLambdax,\mspd{\Psi,\epsi,\mspd{\Psi,\mspd{(\qxi)^2}}}\rangle
\nonumber\\
&
-\frac{\kappa^4}{5!}
\langle\eLambdax,\mspd{(\qxi)^2,\mspd{\Psi,\mspd{\Psi,\epsi}}}\rangle
+\frac{\kappa^4}{5!}
\langle\eLambdax,\mspd{\epsi,\qxi,\mspd{\Psi,\mspd{\Psi,\qxi}}}\rangle
\nonumber\\
&
+\frac{8\kappa^4}{6!}
\langle\eLambdax,\mspd{\mspd{\Psi,\qxi},\mspd{\Psi,\epsi,\qxi}}\rangle
-\frac{8\kappa^4}{6!}
\langle\eLambdax,\mspd{\mspd{\Psi,\epsi},\mspd{\Psi,(\qxi)^2}}\rangle
\nonumber\\
&
-\frac{\kappa^4}{5!}
\langle\eLambdax,\mspd{\Psi,\mspd{\Psi,\qxi},\mspd{\epsi,\qxi}}\rangle
+\frac{\kappa^4}{5!}
\langle\eLambdax,\mspd{\Psi,\mspd{\Psi,\epsi},\mspd{(\qxi)^2}}\rangle
\nonumber\\
&
-\frac{3\kappa^4}{5!}
\langle\eLambdax,\mspd{\qxi,\mspd{\Psi,\epsi},\mspd{\Psi,\qxi}}\rangle
\nonumber\\
&
+\frac{3\kappa^4}{5!}
\langle\eLambdax,\mspd{\epsi,\mspd{\Psi,\qxi},\mspd{\Psi,\qxi}}\rangle.
\end{align}
The first two terms give the correction to the constraint 
in (\ref{tf tildelambda}).
\begin{align}
 &\nonumber\\
 &\nonumber\\
 &\nonumber
\end{align}%


\begin{thebibliography}{99}


\bibitem{Kunitomo:2014qla} 
  H.~Kunitomo,
  arXiv:1412.5281 [hep-th].

\bibitem{Berkovits:1995ab}
  N.~Berkovits,
  Nucl.\ Phys.\ B {\bf 450}, 90 (1995);
   {\bf 459}, 439 (1996) [erratum] [arXiv:9503099 [hep-th]].

\bibitem{Berkovits:2001im} 
  N.~Berkovits,
 J.~High~Energy~Phys.\ \textbf{0111}, 047 (2001)
  [arXiv:0109100 [hep-th]].

\bibitem{Michishita:2004by} 
  Y.~Michishita,
 J.~High~Energy~Phys.\ \textbf{0501}, 012 (2005)
  [arXiv:0412215 [hep-th]].

\bibitem{Michishita:Riken}
 Y.~Michishita,
  Talk given at \lq\lq String Field Theory 07\rq\rq, RIKEN, 6-7 October 2007.
(Available at http://www.riken.jp/lab-www/theory/sft/sft07\_michishita.pdf,
date last accessed June 30, 2015.) 

\bibitem{Michishita:2012ku}
Y.~Michishita,
Bulletin of the Faculty of Education, Kagoshima University.
Natural science, {\bf 63}: 17-43, (2012), (in Japanese) 
(Available at http://ir.kagoshima-u.ac.jp/bitstream/10232/14240/1/AN00408518\_v63\_p17-43.pdf,
date last accessed , June 30, 2015).

\bibitem{Okawa:2004ii} 
  Y.~Okawa and B.~Zwiebach,
  J.~High~Energy~Phys.\ \textbf{0407}, 042 (2004)
  [hep-th/0406212].

\bibitem{Berkovits:2004xh} 
  N.~Berkovits, Y.~Okawa and B.~Zwiebach,
 J.~High~Energy~Phys.\ \textbf{0411}, 038 (2004)
  [arXiv:0409018 [hep-th]].

\bibitem{Kunitomo:2013mqa}
  H.~Kunitomo,
  Prog.~Theor.~Exp.~Phys.\ \textbf{2014}, 043B01 (2014)
  [arXiv:1312.7197 \mbox{[hep-th]}].

\bibitem{Kunitomo:2014hba}
  H.~Kunitomo,
 Prog.~Theor.~Exp.~Phys.\ \textbf{2014}, 093B07 (2014) 
[arXiv:1407.0801 \mbox{[hep-th]}].

\bibitem{Saadi:1989tb} 
  M.~Saadi and B.~Zwiebach,
  Annals Phys.\  {\bf 192}, 213 (1989).

\bibitem{Kugo:1989aa} 
  T.~Kugo, H.~Kunitomo and K.~Suehiro,
  Phys.\ Lett.\ B {\bf 226}, 48 (1989).

\bibitem{Kugo:1989tk}
  T.~Kugo and K.~Suehiro,
  Nucl.\ Phys.\ B {\bf 337} (1990) 434.

\bibitem{Zwiebach:1992ie} 
  B.~Zwiebach,
  Nucl.\ Phys.\ B {\bf 390}, 33 (1993)
  [arXiv:9206084 [hep-th]].


\bibitem{LeClair:1988sp} 
  A.~LeClair, M.~E.~Peskin and C.~R.~Preitschopf,
  Nucl.\ Phys.\ B {\bf 317}, 411 (1989).

\bibitem{LeClair:1988sj} 
  A.~LeClair, M.~E.~Peskin and C.~R.~Preitschopf,
  Nucl.\ Phys.\ B {\bf 317}, 464 (1989).


\bibitem{Moeller:2004yy} 
  N.~Moeller,
  JHEP {\bf 0411}, 018 (2004)
  [hep-th/0408067].

\bibitem{Friedan:1985ge}
  D.~Friedan, E.~J.~Martinec and S.~H.~Shenker,
  Nucl.\ Phys.\ B {\bf 271}, 93 (1986).

\bibitem{Sen:2014pia}
  A.~Sen,
  arXiv:1408.0571 [hep-th].

\bibitem{Kroyter:2012ni}
  M.~Kroyter, Y.~Okawa, M.~Schnabl, S.~Torii and B.~Zwiebach,
  J.\ High\ Energy\ Phys.\ {\bf 1203}, 030 (2012) 
  [arXiv:1201.1761 [hep-th]].




\end{thebibliography}
\end{document}